\documentclass[prd,aps,preprint, tightenlines, showpacs,nofootinbib,superscriptaddress]{revtex4-1}

\usepackage{epsfig}
\usepackage{amsmath}
  
\begin{document}

\title{Sudakov Double Logarithms Resummation in Hard Processes in Small-$x$ Saturation Formalism}

\author{A. H. Mueller}
\affiliation{Department of Physics, Columbia University, New York, NY 10027, USA}

\author{Bo-Wen Xiao}
\affiliation{Institute of Particle Physics, Central China Normal University, Wuhan 430079, China}

\author{Feng Yuan}
\affiliation{Nuclear Science Division, Lawrence Berkeley National
Laboratory, Berkeley, CA 94720, USA}

\begin{abstract}
In this manuscript, we present a complete study of the Sudakov double 
logarithms resummation for various hard processes in $eA$ and $pA$ 
collisions in the small-$x$ saturation formalism. 
We first employ a couple of slightly different formalisms to perform the one-loop 
analysis of the Higgs boson production process in $pA$ collisions, and demonstrate 
that Sudakov-type logarithms arise as the leading correction and that they
can be systematically resummed in addition to the usual small-$x$ resummation. 
We further study the Sudakov double logarithms for other processes such as 
heavy quark pair production and back-to-back dijet production in $eA$ and $pA$ 
collisions through detailed calculation of the corresponding one-loop diagrams. 
As the most important contribution from the one-loop correction, 
the Sudakov factor should play an important role in the phenomenological study of 
saturation physics in the $pA$ programs at RHIC and the LHC. 
\end{abstract}
\maketitle

\section{Introduction}

Saturation physics concerns the high energy QCD formalism which can describe the dense 
partons, in particlular, the gluons inside hadrons~\cite{Gribov:1984tu,Mueller:1985wy,McLerran:1993ni,eic}. This formalism employs the 
small-$x$ evolution equations, namely, the BFKL equation~\cite{Balitsky:1978ic} 
as well as its non-linear extension~\cite{Balitsky:1995ub,JalilianMarian:1997jx,Iancu:2000hn}, 
the BK-JIMWLK equation~\cite{Balitsky:1995ub,JalilianMarian:1997jx,Iancu:2000hn} to 
resum the small-$x$ large logarithms $\alpha_s \ln ({1}/{x})$. 

As proposed in Refs.~\cite{Dominguez:2010xd, Dominguez:2011wm}, the 
di-jet productions in $pA$ collisions plays a crucial role in terms of distinguishing 
two fundamental unintegrated gluon distributions, namely, the Weizsacker-Williams 
(WW) gluon distribution and dipole gluon distribution, and searching for the saturation 
phenomenon in high energy experiments. To be more precise, the dijet configuration 
with transverse momenta $k_{1\perp}$ and $k_{2\perp}$ that we are especially 
interested in is the almost back-to-back dijet configuration, in which we can define 
two interesting variables, i.e., the dijet transverse momentum imbalance  
$k_\perp \equiv k_{1\perp}+k_{2\perp}$ and the dijet relative transverse momentum 
$P_\perp \equiv \frac{1}{2}|k_{1\perp}-k_{2\perp}|$ with $P_\perp^2 \gg k_\perp^2$. 
In particular, it has been argued that the dijet processes in $pA$ collisions
could be the smoking gun for the discovery of saturation 
phenomena~\cite{Dominguez:2010xd, Dominguez:2011wm,JalilianMarian:2004da,Blaizot:2004wv,Marquet:2007vb,Albacete:2010pg,Stasto:2011ru,Deak:2011ga, Kutak:2012rf,Lappi:2012nh,Dominguez:2012ad,Iancu:2013dta}.

An important advantage of the dijet production as probe to the gluon 
saturation is that the differential cross section can be directly related to the
un-integrated gluon distributions of the nucleus in the small-$x$ 
factorization. For example, we can write down a generic expression for the 
dijet production cross section depending on $k_\perp$ as
\begin{equation}
\frac{d\sigma}{dy_1dy_2 dP_\perp^2  d^2k_\perp}\propto H(P_\perp^2)\int d^2x_\perp d^2y_\perp e^{ik_\perp\cdot (x_\perp-y_\perp)}
{\widetilde{W}}_{x_A}(x_\perp,y_\perp) \ , \label{generic}
\end{equation}
where $H(P_\perp)$ represents the hard factor only depending on the 
hard momentum scale $P_\perp$ and the rapidities of the two jets, $\widetilde{W}_{x_A}$ 
represents the Wilson lines associated with the un-integrated gluon distributions
of nucleus. The dependence on incoming parton distribution of nucleon
and other kinematic variables are omitted in the above expression for
simplicity. When the impact parameter dependence is neglected for large nucleus targets, $\widetilde{W}_{x_A}$ only depends on the difference
of $x_\perp$ and $y_\perp$ (see the detailed expressions in Refs.~\cite{Dominguez:2010xd, Dominguez:2011wm}) in the large $N_c$ limit. 
In the sense, in the correlation limit of dijet production mentioned above, $P_\perp$
dependence is decouple from $k_\perp$ dependence, and the differential
cross section can be directly calculated from the un-integrated gluon distribution
of the nucleus.

Nevertheless, only the leading order calculation is presented in 
Refs.~\cite{Dominguez:2010xd, Dominguez:2011wm} for various dijet processes, 
which may not be sufficient for drawing definite conclusions. In other words, in order to 
probe saturation phenomena by comparing with experimental data from $pA$ 
collisions, the one-loop correction to the leading order results for various dijet processes 
seems to be indispensable. The major objective of this manuscript is to study the 
one-loop correction to dijet processes and other similar processes, identify the 
characteristic feature of all possible divergences and demonstrate that the most 
important correction is the Sudakov type double logarithmic terms. The appearance 
of Sudakov double logarithms, i.e., $\bar \alpha_s \ln ^2 \left({P_\perp^2}/{k_\perp^2}\right)$, 
requires another type of resummation, namely, the Sudakov 
resummation~\cite{Sudakov:1954sw, Collins:1984kg}, since 
$\bar \alpha_s \ln ^2\left({P_\perp^2}/{k_\perp^2}\right)$ 
can be of order $1$ when $P_\perp^2\gg k_\perp^2$. 

Therefore, it is important to know whether or not we can consistently compute 
and perform the Sudakov resummation in the saturation formalism, in the 
presence of small-$x$ resummation. Through an explicit calculation of 
the production of Higgs particle with mass $M$ and transverse momentum 
$k_\perp$ in $pA$ collisions, we show that one can consistently resum
Sudakov type double logarithms and small-$x$ single logarithms simultaneously. 
The Higgs boson production process in $pA$ collisions is one of the simplest process which 
allows us to compute Sudakov factors, including the double logarithmic term 
$\bar \alpha_s \ln ^2 \left({M^2}/{k_\perp^2}\right)$ and single logarithmic term 
$\bar \alpha_s \ln \left({M^2}/{k_\perp^2}\right)$ exactly, since this process can 
be effectively viewed as a $2\to 1$ process ($gg\to h (k_\perp, M)$) without final 
state interactions at leading order. Thus, the one-loop correction to the Higgs boson inclusive 
production is relatively simpler to compute as compared to the dijet processes. 

The result for Higgs boson production in $pA$ collisions has been summarised and 
published earlier in Ref~\cite{Mueller:2012uf}, in which we also comment on the 
phenomenological application of the Sudakov resummation in the saturation 
formalism to the dijet processes. This manuscript not only contains many more 
details on the the calculation of the Higgs boson production, but also provides the calculation 
for all the possible dijet productions in $pA$ collisions and in deep inelastic 
scattering (DIS). 

Furthermore, we find that there is another interesting and simple example, i.e., the 
heavy quark pair production in DIS, which only has final state interactions. This 
process can be viewed as the reverse process of the Higgs boson productions in terms of 
the color flow.  One can also easily show that this process contains a similar Sudakov 
double logarithmic term at the one-loop order but with an additional $\frac{1}{2}$ factor. 

In addition, the Sudakov logarithms typically have a factor $\mathcal{C}$ depending 
on the color flow, which has been absorbed into the definition of 
$\bar \alpha_s \equiv \frac{\alpha_s\mathcal{C}}{2\pi}$ in the above expression. 
The color factor $\mathcal{C}$ is process dependent, which implies that it has to 
be determined case by case for various dijet production processes. 

Last but not least, the complete evaluation of the one-loop corrections is 
too complicated to do, even for the Higgs boson production. However, the leading 
power contributions are much easier to obtain. Especially, the Sudakov double 
logarithms, which is the most important leading power contribution at one-loop 
order, can be computed. Therefore, in the following 
section, we first develop some useful technical tools through the detailed calculation 
of the one-loop correction to Higgs boson production in $pA$ collisions and heavy 
quark pair production in DIS. After that, we evaluate the one-loop order graphs for 
various dijet production processes in $pA$ collisions and in DIS, and determine the 
corresponding coefficients $\mathcal{C}$ for the Sudakov double logarithmic terms. 
The computation of the single Sudakov logarithms and other subleading 
contributions are less straightforward and more subtle, similar to that 
for the threshold resummation for hard processes such as heavy
quark pair production and dijet production,
where a matrix form has to be used~\cite{Kidonakis:1997gm}. We leave this for future studies. 

The bottom line is that the Sudakov double logarithmic terms and small-$x$ type 
logarithms can be resumed simultaneously and independently in the saturation 
formalism. This conclusion holds for many processes, such as Higgs boson production, 
heavy quark pair productions and back-to-back dijet production in $pA$ collisions as well as in DIS.

Therefore, after the Sudakov double logarithms resummation, the differential cross section of Eq.~(\ref{generic}) can be modified as
\begin{equation}
\frac{d\sigma}{dy_1 dy_2 dP_\perp^2 d^2k_\perp} \propto H(P_\perp^2)\int d^2x_\perp d^2y_\perp e^{ik_\perp\cdot R_\perp}
e^{-{\cal S}_{sud}(P_\perp, R_\perp)}{\widetilde{W}}_{x_A}(x_\perp,y_\perp) \ ,
\end{equation}
where the Sudakov factor can be written as
\begin{equation}
{\cal S}_{sud}=\frac{\alpha_s}{\pi}{\cal C}\int_{c_0^2/R_\perp^2}^{P_\perp^2}\frac{d\mu^2}{\mu^2}\ln\frac{P_\perp^2}{\mu^2}  \ .
\end{equation}
where $c_0$ is defined in the following calculations and $R_\perp=x_\perp-y_\perp$.
The ${\cal C}$ coefficient is identified in the Sudakov double logarithmic 
factor. The main object of this paper to evaluate ${\cal C}$ for all the partonic
processes in $eA$ and $pA$ collisions by a detailed analysis of gluon radiation
at one-loop order.

The physics behind the above arguments and the following analysis in this paper
can be understood as follows. The gluon radiation comes from 
three different regions, for a particular partonic channel in $pA$ collisions:
(1) collinear gluon parallel to the incoming nucleon; (2) collinear gluon
parallel to the incoming nucleus; (3) soft gluon radiation. For example,
for one-gluon radiation contribution to the Born diagram of the above
hard processes in $pA$ collisions, we can parameterize the radiated
gluon momentum as
\begin{equation}
q=\alpha_qp_1+\beta_qp_2+q_\perp \ ,
\end{equation}
where $p_1$ and $p_2$ denote the four momenta of incoming 
partons in the partonic process. Since the phase space
of the radiated gluon will be integrated out to obtain the 
final differential cross section, we need to consider the three
different momentum regions discussed above. 
The momentum region of (1) corresponds to $\alpha_q\sim {\cal O}(1)$,
whereas $\beta_q\ll 1$, which contributes to the collinear divergence associated
with the parton distribution from the nucleon. For region (2), the dominant collinear
gluon radiation parallel to the nucleus requires not only $\beta_q$ of order $1$, 
but also close to $1$, i.e., $1-\beta_q\ll 1$. Effectively, because of $q_\perp\sim k_\perp\ll M(P_\perp)$,
this leads to $\alpha_q\to 0$. This corresponds to the rapidity divergence at one-loop
order, which can be absorbed into the renormalization of the un-integrated 
gluon distribution of the nucleus in the small-$x$ limit. Region (3) concerns
the Sudakov double logarithms where $\alpha_q\sim\beta_q\ll 1$. The 
gluon radiation in this kinematic region depends on the overall color flow
in the hard partonic processes. That is why the Sudakov double logarithms
have simple counting rule as briefly mentioned above and will be extensively
analyzed in the following sections. The kinematics of the three regions
are well separated, and at the leading power of $k_\perp/M(P_\perp)$, 
they can be factorized into various factors. The resummation for the 
collinear gluon parallel to the nucleon can be performed by DGLAP
evolution of the integrated parton distributions; for the collinear gluons
parallel to the nucleus the resummation can be performed by BK-JIMWLK evolution;
for the soft gluon radiation the resummation be done with the Sudakov 
method (the so-called Collins-Soper-Sterman evolution~\cite{Collins:1984kg}).

The rest of this paper is organized as follows. In Sec. II, we present a detailed derivation
for Higgs boson production in $pA$ collisions in the small-$x$ factorization 
formalism, where we show that the Sudakov resummation can be performed
consistently with the small-$x$ resummation. In particular, the three different 
regions of gluon radiation mentioned above will be analyzed, and
the factorization is demonstrated at one-loop order. In this section, 
we will also present the basic techniques for our computation, which 
we will apply later for more complicated hard processes.
In Sec. III, we extend our discussions to heavy quark pair production in
deep inelastic scattering of $eA$ collisions, where again it is
the WW gluon distribution from the nucleus that contributes at the leading
power of heavy quark mass. Sec. IV is devoted to the discussion of
dijet production in DIS processes. In Sec. V, we summarize the generic
structure of soft gluon radiation for two-particle productions, and the leading
double logarithms are identified for various partonic processes.
The analysis is focused in the collinear factorization framework 
with dilute-dilute scattering processes. In Sec. VI, we derive the double
logarithms for the quark jet-photon production in $pA$ collisions,
where the dipole gluon distribution from the nucleus contributes. 
In this section, we will also identify the small-$x$ evolution 
for the dipole gluon distribution from the nucleus. By comparing 
with that for the Higgs boson production where the WW gluon 
distribution contributes, we understand the small-$x$ resummation
is consistently resummed in our framework. The one-loop calculations
clearly demonstrate that for different gluon distributions (dipole
vs WW), different evolution equations will apply. Of course, in the
dilute limit, they will be identical, i.e., both reduce to the BFKL.
In Sec. VII, the analysis of the double logarithms is extended to various
dijet production processes in $pA$ collisions. The leading double 
logarithmic terms will be identified.
We summarize our paper in Sec. VIII.

\section{Higgs Boson Production in $pA$ Collisions}

\begin{figure}[tbp]
\begin{center}
\includegraphics[width=7cm]{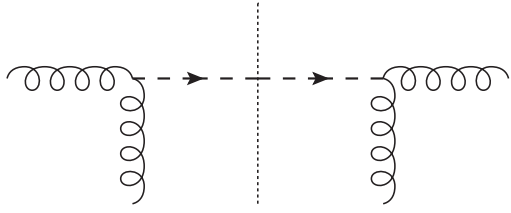}
\end{center}
\caption[*]{Leading order contribution to the Higgs boson production in $pA$ collisions.}
\label{higgsLO}
\end{figure}

Similar to the complete one-loop calculation in Ref.~\cite{Chirilli:2011km}, as demonstrated 
in Ref.~\cite{Mueller:2012bn}, we can adopt the so-called dilute-dense factorization 
(also known as the hybrid factorization) to simplify the calculation in this manuscript, 
since the proton projectile is considered to be dilute as compared to the dense large 
nucleus target. Therefore, we use collinear gluon distribution for the incoming gluons 
(the horizontal gluon as shown in Fig.~\ref{higgsLO}) from the proton projectile. Due to 
the high gluon density in the large nucleus target, multiple gluon interactions (represented 
by the vertical gluon as shown in the Fig.~\ref{higgsLO}) have to be taken in account. 
Thus, we need to use the unintegrated gluon distribution for the incoming gluon from the 
target nucleus. Because Higgs particles do not carry any color, the final state interaction 
is absent. It is then easy to see that the relevant un-integrated gluon distribution should 
be the so-called WW gluon distribution~\cite{Dominguez:2010xd, Dominguez:2011wm}. 

Calculated from Fig.~\ref{higgsLO}, the leading order cross section for producing Higgs (or heavy scalar particles) with forward rapidity $y$ and transverse momentum $k_\perp$ in $pA$ collisions\cite{CU-TP-441a, Mueller:1999wm}, can be written as 
\begin{equation}
\frac{\text{d}\sigma_{\textrm{LO}}}{\text{d} y \text{d}^2 k_\perp} = \sigma_0 x_pg_p(x_p) \int\frac{d^2x_\perp d^2x^{\prime}_\perp}{(2\pi)^2}\,e^{-ik_\perp\cdot(x_\perp-x^{\prime}_\perp)}S^{WW}(x_\perp,x^\prime_\perp)
\end{equation}
where $\sigma_0=\frac{ g_\phi^2}{4g^2(N_c^2-1)\left(1-\epsilon\right)}$. $S^{WW}(x_\perp,x^\prime_\perp)$, which represents the WW gluon distribution in the coordinate space and resums the initial state multiple interactions, is defined as
\begin{equation}
S^{WW}(x_\perp,x^\prime_\perp)=-
\left\langle\text{Tr}\left[\partial^iU(x_\perp)\right]U^\dagger(x^{\prime}_%
\perp)\left[\partial^iU(x^{\prime}_\perp)\right]U^\dagger(x_\perp)\right%
\rangle_{x_g}, \label{sww}
\end{equation}
where the fundamental Wilson line
\begin{equation}
U(x_\perp)=\mathcal{P}\exp\left\{ig_S\int_{-\infty}^{+\infty} \text{d}%
x^+\,T^cA_c^-(x^+,x_\perp)\right\} \ ,
\end{equation}
with $A_c^-(x^+,x_\perp)$ being the gluon field solution of the classical
Yang-Mills equation inside the large nucleus target. As mentioned earlier, we can use the 
collinear gluon distribution $x_pg_p(x_p)$ where $x_p =\tau\equiv \frac{M e^{y}}{\sqrt{s}}$ 
with $M \gg k_\perp$ being the mass of the Higgs particle. According to kinematics, one 
can also find that $x_g=\frac{M e^{-y}}{\sqrt{s}}$ which gives the longitudinal momentum 
fraction of the incoming gluon distribution with respect to that of the target nucleus per 
nucleon. The minus sign in $S^{WW}(x_\perp,x^\prime_\perp)$ ensures that the above 
correlation function is positive since the derivatives bring down two factors of $-i g$ from 
the exponent of the Wilson lines. The notation $\langle\cdots\rangle_{x_g}$ in Eq.~(\ref{sww})
indicates the random colour charge average over the large nucleus wave function and implies 
the small-$x$ evolution of the relevant correlators inside. Here $\epsilon=\frac{4-D}{2}$ is the 
dimensional regularization parameter. The factor of $\frac{1}{1-\epsilon}$ comes from the 
average of the polarization for the incoming gluon according to the convention in the 
dimensional regularization. This factor is universal for all diagrams at higher order, 
therefore we always factorize this factor into the leading order cross section when we 
perform the one-loop calculation in terms of the dimensional regularization. 

Before we start the detailed evaluation for the one-loop amplitudes for the Higgs boson 
production, let us briefly discuss the usual divergences that we are anticipating at 
the one-loop order from the point of view of the leading order cross section. First of 
all, when the radiated gluon is collinear to the proton projectile, or more precisely, 
the initial state gluon which comes from the proton, we find the collinear divergence 
which is corresponding to the renormalization of the incoming collinear gluon distribution 
$x_pg_p(x_p)$. The renormalization of this collinear gluon distribution is governed by the 
well-known DGLAP evolution equation. Furthermore, there should be the so-called rapidity 
divergence, which appears when the rapidity of the radiated gluon is reaching $-\infty$. The appearance of the rapidity divergence 
is simply due to the high energy limit which puts both the projectile and targets on the light cone. It comes from the region where the radiated gluon becomes collinear to the target nucleus. It is natural to associate this type of divergence with the renormalization of the wave function of the target nucleus, which is represented by the WW correlator $S^{WW}(x_\perp,x^\prime_\perp)$. As known before, the WW correlator satisfies the following small-$x$ evolution equation\cite{Dominguez:2011gc}
\begin{eqnarray}
\frac{\partial}{\partial Y}S^{WW}(x_\perp,x^\prime_\perp) &=& \int {\bf K}_{\rm DMMX}\otimes S^{WW}(x_\perp,x^\prime_\perp) \notag \\
&=&-\frac{\alpha_s N_c}{2\pi^2}\int d^2 z_{\perp} \frac{(x_\perp-x_\perp^\prime)^2}{%
(x_\perp-z_\perp)^2(z_\perp-x_\perp^\prime)^2}S^{WW}(x_\perp,x^\prime_\perp)  +\cdots 
 \label{wwe}
\end{eqnarray}
The exact form of the kernel ${\bf K}_{\rm DMMX}$ is rather complicated, which is the reason that only the first term is listed above while the rest are abbreviated. Nevertheless, we have checked that we can reproduce every term of this evolution equation in our calculation when the WW correlator $S^{WW}(x_\perp,x^\prime_\perp)$ is involved. 

There are three different ways to compute the Sudakov factor associated with the Higgs boson productions in $pA$ collisions when we evaluate the one-loop contributions to this process. The first approach is the momentum space evaluation which is already presented in Ref.~\cite{Mueller:2012uf}. The second way to evaluate the one-loop contribution is based on the dipole model in the coordinate space representation which will be provided as follows.  Last, we give a heuristic derivation of the Sudakov factor in this process in which the physical meaning is more transparent. 

\subsection{The derivation of the Sudakov factor in the dipole model}
\label{HiggsDipole}
Following the dipole formalism which allows us to resum multiple scatterings easily into Wilson lines in either fundamental representations or adjoint representations, we can compute cross sections in coordinate space, which is in general expressed in terms of splitting functions and scattering amplitudes. There is an underlying assumption in our calculation which is the scattering energy of the collision is so high that the size of the target nucleus $L$ is always much smaller than the coherence time of the fluctuated gluon. Therefore, we just need to consider the case that the multiple interaction with the target nucleus either happens before or after the gluon splitting, but not during the gluon splitting. 

To perform the calculation for the one-loop Higgs boson productions in $pA$ collisions in coordinate space, we need to first derive the splitting functions involved in the calculations. There are three types of splitting functions ($g\to gg$, $g\to g h$ and $h\to gg$) which appear in this calculation. For the coupling vertex between two gluons with momenta $k_1, k_2$ and the Higgs particle, we employ the following effective Feynman rule $ig_{\phi}\delta_{ab}(k_1\cdot k_2 g_{\mu\nu}- k_{1\nu}k_{2\mu})$ which comes from the effective Lagrangian 
\begin{equation}
\mathcal{L}_{\textrm{eff}}=-\frac{1}{4}g_\phi \Phi F_{\mu\nu}^a F^{a \mu\nu }.
\end{equation}
This effective coupling has been used in many different calculations, e.g., see Refs.~\cite{Dawson:1990zj, CU-TP-441a, Mueller:1999wm}. 

The $g\to gg$ splitting function, which is already known\cite{Dominguez:2011wm, Iancu:2013dta}, can be written as 
\begin{eqnarray}
\Psi_{g\to gg}\left(\xi, k_\perp\right)&=&\sqrt{\frac{2\xi(1-\xi)}{p^+}}\frac{1%
}{k_\perp^2}\notag \\
&&\left(\frac{1}{\xi} k_\perp \cdot
\epsilon_\perp^{(2)}\epsilon_\perp^{(1)\ast} \cdot \epsilon_\perp^{(3)}+%
\frac{1}{1-\xi} k_\perp \cdot \epsilon_\perp^{(3)}\epsilon_\perp^{(1)\ast}
\cdot \epsilon_\perp^{(2)}-k_\perp \cdot
\epsilon_\perp^{(1)\ast}\epsilon_\perp^{(2)} \cdot
\epsilon_\perp^{(3)}\right),
\end{eqnarray}
where $\epsilon_\perp^{(1)\ast}$ is the transverse polarization vector of the incoming gluon, whereas $\epsilon_\perp^{(2),(3)}$ are the polarizations of the outgoing gluons. $\xi, (1-\xi)$ are the longitudinal momentum fractions of outgoing gluons. After the Fourier transform, we can obtain the splitting function in the coordinate space
\begin{eqnarray}
\Psi_{g\to gg}\left(\xi, u_\perp\right)&=&\sqrt{\frac{2\xi(1-\xi)}{p^+}%
}\frac{2\pi i}{u_\perp^2}\notag \\
&&\left(\frac{1}{\xi} u_\perp \cdot
\epsilon_\perp^{(2)}\epsilon_\perp^{(1)\ast} \cdot \epsilon_\perp^{(3)}+%
\frac{1}{1-\xi} u_\perp \cdot \epsilon_\perp^{(3)}\epsilon_\perp^{(1)\ast}
\cdot \epsilon_\perp^{(2)}-u_\perp \cdot
\epsilon_\perp^{(1)\ast}\epsilon_\perp^{(2)} \cdot
\epsilon_\perp^{(3)}\right).
\end{eqnarray}
After summing over all polarizations, the squared splitting function in transverse coordinate space reads\cite{Dominguez:2011wm}
\begin{equation}
\sum \Psi_{g\to gg}^*(\xi,u_\perp^{\prime})\Psi_{g\to gg}(\xi,u_\perp)=(2\pi)^2 \frac{4}{p^+}\left[\frac{\xi}{1-\xi}+\frac{1-\xi}{\xi}+\xi(1-\xi)\right]\frac{u_\perp^{\prime}\cdot u_\perp}{u_\perp^{\prime 2} u_\perp^{ 2}}.
\end{equation}

For the $g\rightarrow g h $ splitting function, using the effective coupling between the Higgs particle and gluons, we obtain the vertex and propagator contributions in momentum space
\begin{equation}
\Psi _{g\rightarrow g h }\left( \xi ,k_{\perp }\right) =\sqrt{\frac{\xi }{%
2(1-\xi )p^{+}}}\frac{1}{k_{\perp }^{2}+(1-\xi )M^{2}}\left( \frac{1}{2}%
k_{\perp }^{2}\epsilon _{\perp }^{(1)\ast }\cdot \epsilon _{\perp
}^{(2)}-k_{\perp }\cdot \epsilon _{\perp }^{(1)\ast }k_{\perp }\cdot
\epsilon _{\perp }^{(2)}\right) ,  \label{LO}
\end{equation}%
where $M,\xi$ are the mass and the longitudinal momentum fraction of the Higgs boson $h$. In the coordinate space, the above expression can be cast into 
\begin{equation}
\Psi _{g\rightarrow g h }\left( \xi ,u_{\perp }\right) =-\sqrt{\frac{\xi }{%
2(1-\xi )p^{+}}}\frac{2\pi }{u_{\perp }^{2}}K(\epsilon _{f}u_{\perp })\left( 
\frac{1}{2}\epsilon _{\perp }^{(1)\ast }\cdot \epsilon _{\perp }^{(2)}-\frac{%
1}{u_{\perp }^{2}}u_{\perp }\cdot \epsilon _{\perp }^{(1)\ast }u_{\perp
}\cdot \epsilon _{\perp }^{(2)}\right).
\end{equation}
The splitting kernel defined above
\begin{eqnarray}
K(\epsilon _{f}u_{\perp })
= 2\epsilon _{f}u_{\perp }\text{K}%
_{1}(\epsilon _{f}u_{\perp })+\epsilon _{f}^{2}u_{\perp }^{2}
\text{K}_{0}(\epsilon _{f}u_{\perp }),
\end{eqnarray}
where $\epsilon _{f}^{2}=(1-\xi )M^{2}$ and $\text{K}_{0,1}$ are modified Bessel functions of the second kind. If one sets $\epsilon _{f}^{2}=0$, one can easily find $K(0)=2$ and reproduce the
splitting function used in Ref.~\cite{Mueller:1999wm}.

For the virtual graphs, we also need to use the $h \rightarrow gg$ splitting function, which in momentum space can be written as
\begin{equation}
\Psi _{h \rightarrow gg}\left( \xi ,k_{\perp }\right) =\sqrt{\frac{1}{%
2\xi (1-\xi )p^{+}}}\frac{1}{k_{\perp }^{2}-\xi (1-\xi )M^{2}}\left( 
\frac{1}{2}k_{\perp }^{2}\epsilon _{\perp }^{(2)}\cdot \epsilon _{\perp
}^{(3)}-k_{\perp }\cdot \epsilon _{\perp }^{(2)}k_{\perp }\cdot \epsilon
_{\perp }^{(3)}\right) .
\end{equation}
In the coordinate space, this function can be cast into 
\begin{equation}
\Psi _{h \rightarrow gg}\left( \xi ,u_{\perp }\right) =-\sqrt{\frac{1}{%
2\xi (1-\xi )p^{+}}}\frac{2\pi }{u_{\perp }^{2}}K(\epsilon _{f}^{\prime
}u_{\perp })\left( \frac{1}{2}\epsilon _{\perp }^{(2)}\cdot \epsilon _{\perp
}^{(3)}-\frac{1}{u_{\perp }^{2}}u_{\perp }\cdot \epsilon _{\perp
}^{(2)}u_{\perp }\cdot \epsilon _{\perp }^{(3)}\right) ,
\end{equation}%
where $\epsilon _{f}^{\prime 2}=-\xi (1-\xi )M^{2}$.  

\begin{figure}[tbp]
\begin{center}
\includegraphics[width=12cm]{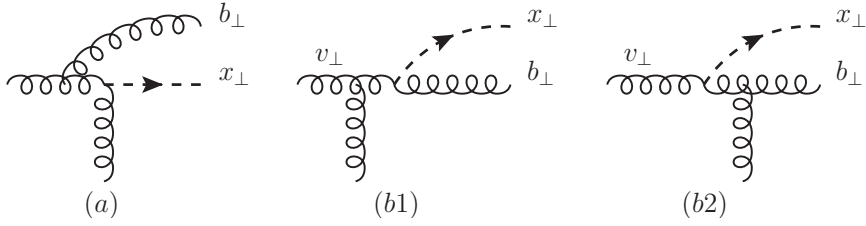}
\end{center}
\caption[*]{The type (a) and (b) graph at the amplitude level. The graph (a) also includes the situation in which small-$x$ gluons exchange between the unobserved gluon and target nucleus.}
\label{a}
\end{figure}
With the above splitting functions, we can now write down the cross sections for the one-loop Higgs boson productions in $pA$ collisions graph by graph. First of all, let us consider the type $(a)$ graph as shown in Fig.~\ref{a}(a) which only involves the $g\to gg$ splitting function and thus the Higgs boson can only be produced after the splitting occurs. After integrating over the phase space of the unobserved gluon, the square of the type $(a)$ diagram gives 
\begin{eqnarray}
\frac{\text{d}\sigma_a^{pA\to h X}}{\text{d} y\text{d}^2 k_\perp}&=&p^+%
\alpha_s N_c \sigma_0 \int_{\tau}^1 \text{d}\xi xg(x) \int \frac{\text{d}%
^{2}x_{\perp}}{(2\pi)^{2}}\frac{\text{d}^{2}x_{\perp}^{\prime }}{(2\pi )^{2}}%
\frac{\text{d}^{2}b_{\perp}}{(2\pi)^{2}}\notag \\
&&\times e^{-ik_{\perp
}\cdot(x_{\perp}-x^{\prime }_{\perp})}  \sum \Psi^{\ast}_{g\to g g}\left(\xi,
u^{\prime}_\perp\right)\Psi_{g\to gg}\left(\xi, u_\perp\right) S^{WW}(x_\perp,x^\prime_\perp) 
\label{r1}
\end{eqnarray}
where the last line represents the WW gluon distribution in the coordinate space due to the resummation of the initial state interactions. According to the definition of the splitting function, we always define $u_\perp=x_\perp -b_\perp$ and $u_\perp^\prime=x_\perp^\prime -b_\perp$. The kinematics requires that $1\geq x=\frac{\tau}{\xi}\geq\tau\equiv \frac{Me^y}{\sqrt{s}}$, where we define $\xi$ to be the longitudinal momentum fraction of the split gluon, which eventually turns into Higgs boson at the transverse coordinate $x_\perp$ (or at $x_\perp^\prime$ in the complex conjugate of the amplitude) after interacting with the target nucleus, with respect to the horizontal incoming gluon. This implies that the unobserved gluon, which is at the position $b_\perp$\footnote{The integration over the phase space of the unobserved gluon, which generates the two dimensional delta function $\delta^{(2)}(b_\perp-b_\perp^\prime)$, identifies its transverse position in the complex conjugate amplitude to be same as that in the amplitude.}, carries the momentum fraction $1-\xi$. 

Since the multiple interactions between the unobserved gluon and target nucleus cancel between the amplitude and complex conjugate amplitude for the above case, Eq.~(\ref{r1}) is directly proportional to $S^{WW}(x_\perp,x^\prime_\perp)$. However, such cancellation does not happen for the interference graphs between the type $(a)$ and type $(b)$ diagrams, which we shall see immediately in the following calculation. 

It is not hard to see that this type of contribution has both the rapidity divergence and collinear divergence. The rapidity divergence, which is the contribution in the $\xi \to 1$ limit, leads to the real graph contribution of the first terms of Eq.~(\ref{wwe}). The collinear divergence comes from the configuration that the radiated gluon is collinear to the incoming gluon from the proton projectile. 

The contribution from the sum of type (b) graphs, as shown in Fig.~\ref{a} $(b1)$ and $(b2)$,  can be written as 
\begin{eqnarray}
\frac{\text{d}\sigma_b^{pA\to h X}}{\text{d} y\text{d}^2 k_\perp}&=&p^+%
4\alpha_s (N_c^2-1) \sigma_0 \int_{\tau}^1 \text{d}\xi xg(x) \int \frac{\text{d}%
^{2}x_{\perp}}{(2\pi)^{2}}\frac{\text{d}^{2}x_{\perp}^{\prime }}{(2\pi )^{2}}%
\frac{\text{d}^{2}b_{\perp}}{(2\pi)^{2}}\notag \\ 
&& \times e^{-ik_{\perp
}\cdot(x_{\perp}-x^{\prime }_{\perp})} \sum \Psi^{\ast}_{g\to g
h}\left(\xi, u^{\prime}_\perp\right)\Psi_{g\to g h}\left(\xi,
u_\perp\right)  \notag \\
&&\times \left[\tilde{S}^{(2)}(v_\perp,v_\perp^{\prime}) +1-\tilde{S}^{(2)}(b_\perp,v_\perp) -\tilde{S}^{(2)}(v^\prime_\perp,b_\perp)\right] \label{typeb}
\end{eqnarray}
where the four scattering amplitudes in the last line come from four different combination of the initial and final state multiple interactions. Here $v_\perp = \xi x_\perp + (1-\xi) b_\perp$ represents the coordinate of the incoming gluon. The gluon dipole amplitude $\tilde{S}^{(2)}(x_\perp,y_\perp)$ is defined as $\frac{1}{N_{c}^2-1}\left\langle 
\text{Tr}W(x_{\perp})W^{\dagger }(y_{\perp })\right\rangle _{x_a}$,  where $W(x_\perp)$
is the Wilson line in the adjoint representation. Using the identity 
\begin{equation}
\tilde{S}_Y^{(2)}(x_\perp,y_\perp)=\frac{1}{N_{c}^2-1}\left[\left\langle 
\text{Tr}U(x_{\perp})U^{\dagger }(y_{\perp })\text{Tr}U(y_{\perp})U^{\dagger
}(x_{\perp })\right\rangle _{Y}-1\right],
\end{equation}
it is straightforward to show that the contribution from type $(b)$ graphs in
the $\xi \to 1$ limit generates the corresponding part of the small-$x$ evolution equation for the WW correlator. 

For the interference diagrams between the type $(a)$ and $(b)$ graphs, we can cast the cross section into
\begin{eqnarray}
\frac{\text{d}\sigma_{(a)\times (b)}^{pA\to h X}}{\text{d} y\text{d}^2 k_\perp}&=&-2ip^+\alpha_s \sigma_0 \int_{\tau}^1 \text{d}\xi xg(x) \int \frac{\text{d}%
^{2}x_{\perp}}{(2\pi)^{2}}\frac{\text{d}^{2}x_{\perp}^{\prime }}{(2\pi )^{2}}%
\frac{\text{d}^{2}b_{\perp}}{(2\pi)^{2}}\notag \\
&& \times e^{-ik_{\perp
}\cdot(x_{\perp}-x^{\prime }_{\perp})} \sum \Psi^{\ast}_{g\to g h}\left(\xi,
u^{\prime}_\perp\right)\Psi_{g\to gg}\left(\xi, u_\perp\right)  \notag \\
&&\left\{\text{Tr}U^\dagger(x_\perp)\left[\epsilon_{\perp}^{(2)\ast } \cdot \partial U(x_\perp)\right]U^\dagger(b_\perp)U(v^{\prime}_\perp) \text{Tr} U^\dagger(v^{\prime}_\perp)U(b_\perp)\right. \nonumber \\
&&\left. - \text{Tr}U^\dagger(x_\perp)\left[\epsilon_{\perp}^{(2)\ast } \cdot \partial U(x_\perp)\right]U^\dagger(v^{\prime}_\perp)U(b_\perp) \text{Tr} U^\dagger(b_\perp)U(v^{\prime}_\perp)\right\}\notag \\
&&+2ip^+\alpha_s \sigma_0 \int_{\tau}^1 \text{d}\xi xg(x) \int \frac{\text{d}%
^{2}x_{\perp}}{(2\pi)^{2}}\frac{\text{d}^{2}x_{\perp}^{\prime }}{(2\pi )^{2}}%
\frac{\text{d}^{2}b_{\perp}}{(2\pi)^{2}}\notag \\
&&\times e^{-ik_{\perp
}\cdot(x_{\perp}-x^{\prime }_{\perp})} \sum \Psi^{\ast}_{g\to g g}\left(\xi,
u^{\prime}_\perp\right)\Psi_{g\to g h}\left(\xi, u_\perp\right)  \notag \\
&&\left\{\text{Tr}U^\dagger(x^{\prime}_\perp)\left[\epsilon_{\perp}^{(2)} \cdot \partial U(x^{\prime}_\perp)\right]U^\dagger(b_\perp)U(v_\perp) \text{Tr} U^\dagger(v_\perp)U(b_\perp)\right. \nonumber \\
&&\left. - \text{Tr}U^\dagger(x^{\prime}_\perp)\left[\epsilon_{\perp}^{(2)} \cdot \partial U(x^{\prime}_\perp)\right]U^\dagger(v_\perp)U(b_\perp) \text{Tr} U^\dagger(b_\perp)U(v_\perp)\right\}.\label{atimesb}
\end{eqnarray}
Similarly, this type of contribution contains rapidity divergence but no collinear divergence. As we have mentioned earlier and shown in Eq.~(\ref{atimesb}), the multiple interactions between the radiated gluon and the target nucleus does not cancel for the interference diagrams, which is the reason that the Wilson lines are quite complicated in Eq.~(\ref{atimesb}).

\begin{figure}[tbp]
\begin{center}
\includegraphics[width=12cm]{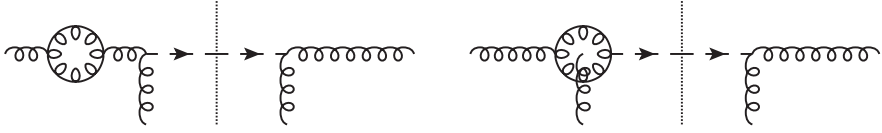}
\end{center}
\caption[*]{Two types of virtual graphs with the gluon loop.}
\label{vh}
\end{figure}

Let us now turn to the virtual graphs. There are two types of virtual diagrams as shown in Fig.~\ref{vh}. The left virtual graph in Fig.~\ref{vh} is nothing but the gluon self-energy correction to the leading order diagram. It can be simply cast into 
\begin{eqnarray}
\frac{\text{d}\sigma_{va}^{pA\to h X}}{\text{d} y\text{d}^2 k_\perp}&=&-\frac{1}{2} p^+ \alpha_s N_c \sigma_0 x_p g(x_p)\int_{0}^1 \text{d}\xi \int \frac{\text{d}%
^{2}v_{\perp}}{(2\pi)^{2}}\frac{\text{d}^{2}v_{\perp}^{\prime }}{(2\pi )^{2}}%
\frac{\text{d}^{2}u_{\perp}}{(2\pi)^{2}}\notag \\
&&\times e^{-ik_{\perp
}\cdot(v_{\perp}-v^{\prime }_{\perp})} \sum \Psi^{\ast}_{g\to g g}\left(\xi,
u_\perp\right)\Psi_{g\to gg}\left(\xi, u_\perp\right) S^{WW}(v_\perp,v^\prime_\perp),  \label{vl} 
\end{eqnarray}
where the factor of $\frac{1}{2}$ is the symmetry factor coming from two identical gluons in the
closed gluon loop. It is straightforward to see that this graph has both the rapidity divergence and the collinear singularity. 

The right virtual graph in Fig.~\ref{vh} yields 
\begin{eqnarray}
\frac{\text{d}\sigma_{v b}^{pA\to h X}}{\text{d} y\text{d}^2 k_\perp}&=&-ip^+\alpha_s \sigma_0 x_p g(x_p) \int_{0}^1 \text{d}\xi \int \frac{\text{d}%
^{2}v_{\perp}}{(2\pi)^{2}}\frac{\text{d}^{2}v_{\perp}^{\prime }}{(2\pi )^{2}}%
\frac{\text{d}^{2}u_{\perp}}{(2\pi)^{2}}\notag\\
&&\times e^{-ik_{\perp
}\cdot(v_{\perp}-v^{\prime }_{\perp})} \sum \Psi^{\ast}_{h \to g g}\left(\xi,
u_\perp\right)\Psi_{g\to gg}\left(\xi, u_\perp\right)  \notag \\
&&\left\{\text{Tr}U^\dagger(v^\prime_\perp)\left[\epsilon_{\perp}^{(1) } \cdot \partial U(v^\prime_\perp)\right]U^\dagger(x_\perp)U(b_\perp) \text{Tr} U^\dagger(b_\perp)U(x_\perp)\right. \nonumber \\
&&\left. - \text{Tr}U^\dagger(v^\prime_\perp)\left[\epsilon_{\perp}^{(1) } \cdot \partial U(v^\prime_\perp)\right]U^\dagger(b_\perp)U(x_\perp) \text{Tr} U^\dagger(x_\perp)U(b_\perp)\right\}\notag \\
&&+ip^+\alpha_s \sigma_0  x_p g(x_p) \int_{0}^1 \text{d}\xi\int \frac{\text{d}%
^{2}v_{\perp}}{(2\pi)^{2}}\frac{\text{d}^{2}v_{\perp}^{\prime }}{(2\pi )^{2}}%
\frac{\text{d}^{2}u^\prime_{\perp}}{(2\pi)^{2}}\notag \\
&&\times e^{-ik_{\perp
}\cdot(v_{\perp}-v^{\prime }_{\perp})} \sum \Psi^{\ast}_{g\to g g}\left(\xi,
u^\prime_\perp\right)\Psi_{h \to gg}\left(\xi, u^\prime_\perp\right)  \notag \\
&&\left\{\text{Tr}U^\dagger(v_\perp)\left[\epsilon_{\perp}^{(1)\ast} \cdot \partial U(v_\perp)\right]U^\dagger(x^\prime_\perp)U(b^\prime_\perp) \text{Tr} U^\dagger(b^\prime_\perp)U(x^\prime_\perp)\right. \nonumber \\
&&\left. - \text{Tr}U^\dagger(v_\perp)\left[\epsilon_{\perp}^{(1)\ast} \cdot \partial U(v_\perp)\right]U^\dagger(b^\prime_\perp)U(x^\prime_\perp) \text{Tr} U^\dagger(x^\prime_\perp)U(b^\prime_\perp)\right\}. \label{vr}
\end{eqnarray}
Of course, this graph contains the rapidity divergence and contributes to the small-$x$ evolution of the WW correlator. The collinear divergence of this diagram is a little bit hard to see but it can show up later in our calculation. 

\begin{figure}[tbp]
\begin{center}
\includegraphics[width=12cm]{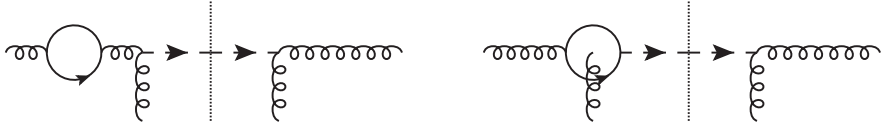}
\end{center}
\caption[*]{Two types of virtual graphs with the gluon loop.}
\label{vh2}
\end{figure}

For the virtual diagrams with quark loops as shown in Fig.~\ref{vh2}. The left virtual graph in Fig.~\ref{vh2} gives a contribution, which is very similar to the gluon loop as shown in Eq.~(\ref{vl}), as follows
\begin{eqnarray}
\frac{\text{d}\sigma_{vq}^{pA\to h X}}{\text{d} y\text{d}^2 k_\perp}&=&- p^+ \alpha_s T_f N_f \sigma_0 x_p g(x_p)\int_{0}^1 \text{d}\xi \int \frac{\text{d}%
^{2}v_{\perp}}{(2\pi)^{2}}\frac{\text{d}^{2}v_{\perp}^{\prime }}{(2\pi )^{2}}%
\frac{\text{d}^{2}u_{\perp}}{(2\pi)^{2}}\notag \\
&&\times e^{-ik_{\perp
}\cdot(v_{\perp}-v^{\prime }_{\perp})} \sum \Psi^{\ast}_{g\to q \bar q}\left(\xi,
u_\perp\right)\Psi_{g\to q\bar q}\left(\xi, u_\perp\right) S^{WW}(v_\perp,v^\prime_\perp),  \label{vl2} 
\end{eqnarray}
where $T_f =\frac{1}{2}$ and $N_f$ represents the number of quark flavors. The $g\to q\bar q$ splitting function $\Psi^{\ast}_{g\to q \bar q}\left(\xi,u_\perp\right)$ can be found in Ref.~\cite{Dominguez:2011wm, Chirilli:2011km} and its sum is
\begin{equation}
\sum \Psi^{\ast}_{g\to q \bar q}\left(\xi,
u_\perp\right)\Psi_{g\to q\bar q}\left(\xi, u_\perp\right) = \frac{2}{p^+} (2\pi)^2 \left[\xi^2+(1-\xi)^2\right] \frac{1}{u_\perp^2}.
\end{equation}
The right virtual graph in Fig.~\ref{vh2} does not generate any logarithms in this process, therefore it is neglected in this calculation. 

Now we can evaluate all the one-loop contributions by following the same procedure as developed in Ref.~\cite{Chirilli:2011km}. First of all, we need to subtract the rapidity divergence from the above contributions, since we can show that the right-hand side of the small-$x$ evolution of the WW gluon distribution, as derived in Ref.~\cite{Dominguez:2011gc}, can be found as the coefficient of the rapidity divergence in the $\xi\to 1$ limit when we add all the above contributions together. As a matter of fact, the rapidity of the radiated gluon goes to $-\infty$ when $\xi\to 1$ and this gluon becomes collinear to the target nucleus. Second, we should compute and identify the collinear singularities by using dimensional regularization, and absorb the collinear singularities into the incoming collinear gluon distribution from the proton projectile according to the corresponding DGLAP equation. In the meantime, in order to simplify the calculation, we always work in the leading power approximation in the $k_\perp^2 \ll M^2$ limit, which means that we perform the power expansion in terms of $\frac{k_\perp^2}{M^2}$ and only keep the contribution which is not suppressed by factors of $\frac{k_\perp^2}{M^2}$ as comparing to the leading order cross section. At the end of the day, the Sudakov logarithms should appear from the soft region of the radiated gluon, where $k^+\sim k^-\sim k_\perp$. Within the leading power approximation, the corresponding graphs must contain both the rapidity divergence and collinear divergence in order to have the soft gluon radiation which contributes to the Sudakov factor. 

Using the above technique, we can examine the above one-loop contributions, and find out the Sudakov factor. According to our observation, Eqs~(\ref{typeb}) and (\ref{atimesb}) only contribute to the small-$x$ evolution of the WW correlator in the $\xi \to 1$ limit, but do not contribute to the Sudakov factor, since they are suppressed by factors of $\frac{k_\perp^2}{M^2}$ when $\xi \neq 1$. Therefore, only Eqs.~(\ref{r1}), (\ref{vl}) and (\ref{vr}) contain the leading power contributions to the Sudakov factor. 

Let us first study Eq.~(\ref{r1}) which is calculated from the type $(a)$ real graph only. It is straightforward to show that Eq.~(\ref{r1}) can be cast into
\begin{eqnarray}
\frac{\text{d}\sigma_{a}^{pA\to h X}}{\text{d} y\text{d}^2 k_\perp}
&=& 4\alpha_s N_c \sigma_0 \int \frac{\text{d}%
^{2}x_{\perp}}{(2\pi)^{2}}\frac{\text{d}^{2}x_{\perp}^{\prime }}{(2\pi )^{2}}%
e^{-ik_{\perp
}\cdot R_\perp}S^{WW}(x_\perp,x^\prime_\perp)  \notag \\
&&\times\int\frac{\text{d}^{2}q_{\perp}}{(2\pi)^{2}}e^{-iq_{\perp
}\cdot R_\perp}\frac{1}{q_{\perp}^{ 2}} \int_{\tau}^1 \text{d}\xi xg(x)\left[\frac{\xi}{1-\xi}+\frac{1-\xi}{\xi}+\xi(1-\xi)\right], \label{r1a}
\end{eqnarray}
where $R_\perp=x_{\perp}-x^{\prime }_{\perp}$ and $q_\perp$ can be interpreted as the transverse momentum of the unobserved gluon.
The last line of the above contribution can be rewritten according to the definition of the plus-function as follows 
\begin{eqnarray}
&&\int\frac{\text{d}^{2}q_{\perp}}{(2\pi)^{2}}e^{-iq_{\perp
}\cdot R_\perp}\frac{1}{q_{\perp}^{ 2}} \int_{\tau}^1 \text{d}\xi xg(x)\left[\frac{\xi}{(1-\xi)_+}+\frac{1-\xi}{\xi}+\xi(1-\xi)\right] \notag \\
&& +x_p g(x_p)\int\frac{\text{d}^{2}q_{\perp}}{(2\pi)^{2}}e^{-iq_{\perp
}\cdot R_\perp}\frac{1}{q_{\perp}^{ 2}} \int_{0}^1 \text{d}\xi \frac{1}{1-\xi}. \label{r1b}
\end{eqnarray}
It is clear that one can identify the first term, which only contains the collinear singularity and is proportional to 
\begin{equation}
\frac{1}{4\pi}\left[\frac{\xi}{(1-\xi)_+}+\frac{1-\xi}{\xi}+\xi(1-\xi)\right]\left(-\frac{1}{\epsilon}+\ln\frac{c_0^2}{\mu^2 R_\perp^2}\right) \label{pggreal}
\end{equation}
in the $\overline{\textrm{MS}}$ scheme, as part of the DGLAP splitting function $\mathcal{P}_{gg}$ which contributes to the renormalization of the collinear gluon distribution $xg(x)$. 

The second term is a little bit more subtle, since it contains both the rapidity divergence and collinear singularity. From the kinematics of the leading order graph, we know 
\begin{equation}
x_p x_g s=M^2+k_\perp^2 \simeq M^2. \label{locon}
\end{equation} 
The high energy limit means that we take the limit $s\to \infty$ and $x_g\to 0$, but keep their product finite to ensure the above kinematical constraint. For the real diagrams at the one-loop order, according to the energy-momentum conservation, we should have the constraint 
\begin{equation}
x_p x_g^\prime s=\frac{M^2}{\xi}+\frac{q_\perp^2}{1-\xi},
\end{equation}
where $x_g^\prime$ represents the sum of the longitudinal momentum fraction that the vertical gluons carry with respect to the target nucleus at the one-loop order. It is clear that $x_g^\prime$ does not have to be the same as $x_g$ which is defined according to the leading order kinematics. When we take $\xi\to 1$ limit, we can find
\begin{equation}
\xi <1-\frac{q_\perp^2}{x_p s},
\end{equation}
which sets upper limit of the $\xi$ integration. If we take the high energy limit for this configuration, namely $s\to \infty$, we can integrate over $\xi$ up to $1$ as shown above. Let us take $s$ to be large but finite for now and compute the $\xi$ integral in the second term of Eq.~(\ref{r1b}) which gives
\begin{equation}
 \int_{0}^{1-\frac{q_\perp^2}{x_p s}} \text{d}\xi \frac{1}{1-\xi}= \ln\left( \frac{x_p s}{q_\perp^2} \right) =\ln \frac{1}{x_g} +\ln\frac{M^2}{q_\perp^2},
\end{equation}
where Eq.~(\ref{locon}) has been used in arriving at the above expression. In fact, the above two logarithms come from the kinematical region $1-\frac{q_\perp^2}{M^2}<\xi <1-\frac{q_\perp^2}{x_p s}$ and $0<\xi <1-\frac{q_\perp^2}{M^2}$, respectively. It is clear that the first logarithm $\ln \frac{1}{x_g}$, accompanied by a factor of $\alpha_s$, is exactly the same small-$x$ logarithm which is resummed by solving the small-$x$ evolution equation. Furthermore, as we are going to show in the following discussion, the second logarithm yields the Sudakov factor. The separation of these two types of large logarithms are clear and therefore both the small-$x$ resummation and the Sudakov resummation can be done simultaneously and independently in the saturation formalism. In addition, if we take the high energy limit again, which yields $x_g\to 0$ as $s\to \infty$, the logarithmic term $\ln \frac{1}{x_g}$ becomes divergent and should be absorbed into the renormalization of the WW correlator according to the corresponding small-$x$ evolution equation. 

Now the integral in question, which is divergent, is
\begin{equation}
\mu^{2\epsilon}\int\frac{\text{d}^{2-2\epsilon}q_{\perp}}{(2\pi)^{2-2\epsilon}}e^{-iq_{\perp
}\cdot R_\perp}\frac{1}{q_{\perp}^{ 2}}\ln\frac{M^2}{q_\perp^2}, \label{realsudakov}
\end{equation}
where we have shifted the dimension of the integral from $2$ to $2-2\epsilon$ according to the common practice dimensional regularization. With the $\overline{\textrm{MS}}$ subtraction scheme, we find that the above integral yields 
\begin{equation}
\frac{1}{4\pi}\left[\frac{1}{\epsilon^2}-\frac{1}{\epsilon}\ln\frac{M^2}{\mu^2}
+\frac{1}{2}\left(\ln\frac{M^2}{\mu^2}\right)^2-\frac{1}{2}\left(\ln\frac{M^2 R_\perp^2}{c_0^2}\right)^2
-\frac{\pi^2}{12}\right] \label{realc}
\end{equation}
where $c_0=2e^{-\gamma_E}$ and $\gamma_E$ is the Euler constant. The detail of the evaluation is provided in the appendix. 

Following the same procedure, the virtual contributions can be evaluated similarly in dimensional regularization as well. It is straightforward to find that first type virtual graph as given in Eq.~(\ref{vl}), after subtracting the rapidity divergent contribution, can be cast into the following integral
\begin{equation}
\frac{1}{2}\int_{0}^1\text{d}\xi \frac{\left[1-\xi(1-\xi)\right]^2-1}{\xi (1-\xi)}\int\frac{\text{d}^{2}u_{\perp}}{(2\pi)^{2}}\frac{1}{u_\perp^2}.
\end{equation}
The last integral can be also written as $\int\frac{\text{d}^{2}q_{\perp}}{(2\pi)^{2}}\frac{1}{q_\perp^2}$ in the momentum space. In dimensional regularization convention, this scale invariant integral is required to be zero since it has both the infrared (IR) and ultraviolet (UV) divergence. In practice, we can write it as 
\begin{equation}
\int\frac{\text{d}^{2}q_{\perp}}{(2\pi)^{2}}\frac{1}{q_\perp^2}=\frac{1}{4\pi} \left(-\frac{1}{\epsilon_{\textrm{IR}}}+\frac{1}{\epsilon_{\textrm{UV}}}\right)
\end{equation}
where $\epsilon_{\textrm{IR}}=\epsilon=\epsilon_{\textrm{UV}}$. Taking the $d\xi$ integration, which gives $-\frac{11}{6}$, into account, we can find that Eq.~(\ref{vl}) is then proportional to
\begin{equation}
-\frac{11}{12}\frac{1}{4\pi}\left[\left(-\frac{1}{\epsilon}+\ln \frac{Q^2}{\mu^2}\right)+\left(\frac{1}{\epsilon_{\textrm{UV}}}-\ln \frac{Q^2}{\mu^2}\right)\right].
\end{equation}
The quark loop contribution contains no rapidity divergence and requires no subtraction. Through a similar calculation, we find that, after dividing over the common factor, Eq.~(\ref{vl2}) yields,
\begin{equation}
\frac{N_f}{6N_c}\frac{1}{4\pi}\left[\left(-\frac{1}{\epsilon}+\ln \frac{Q^2}{\mu^2}\right)+\left(\frac{1}{\epsilon_{\textrm{UV}}}-\ln \frac{Q^2}{\mu^2}\right)\right],
\end{equation}
where $Q^2$ is the scale which the strong coupling is measured. It is natural to set $Q$ to be identical to the Higgs boson mass $M$.
Adding these two contributions together we obtain 
\begin{equation}
-\frac{1}{4\pi}\left[\beta_0\left(-\frac{1}{\epsilon}+\ln \frac{M^2}{\mu^2}\right)+\beta_0\left(\frac{1}{\epsilon_{\textrm{UV}}}-\ln \frac{M^2}{\mu^2}\right)\right],
\end{equation}
where $\beta_0=\frac{11}{12}-\frac{N_f}{6N_c}$. The UV divergent part of the above contribution, after factorizing out the leading order cross section, reads
\begin{equation}
-\frac{\alpha_s}{\pi}N_c \beta_0 \left(\frac{1}{\epsilon_{\textrm{UV}}}-\ln \frac{M^2}{\mu^2}\right).
\end{equation}
It can be naturally interpreted as the colour charge renormalization. By noting the $\alpha_s$ dependence in $\sigma_0\propto \frac{1}{\alpha_s}$ that we have defined in the LO calculation, and comparing to the one-loop order result that we have obtained above, we find that the renormalized coupling can be written as 
\begin{equation}
\frac{1}{\alpha_s(M^2)}=\frac{1}{\alpha_s^0}\left[1-\frac{\alpha_s^0}{\pi}N_c \beta_0 \left(\frac{1}{\epsilon_{\textrm{UV}}}-\ln \frac{M^2}{\mu^2}\right)\right]=\frac{1}{\alpha_s(\mu^2)}+\frac{N_c \beta_0}{\pi} \ln \frac{M^2}{\mu^2},
\end{equation}
where $\frac{1}{\alpha_s(\mu^2)}\equiv \frac{1}{\alpha_s^0}-\frac{N_c \beta_0}{\pi} \frac{1}{\epsilon_{\textrm{UV}}}$. The above equation exactly gives the one-loop QCD running coupling and renormalization equation for $\alpha_s$ by differentiating over $\mu^2$ and requiring that both sides of the equation is independent of choice of the scale $\mu^2$.

The infrared divergent contribution, combined with the real graph contribution as in Eq.~(\ref{pggreal}), gives rise to
\begin{equation}
-\frac{1}{\epsilon} \frac{\alpha_s}{\pi}N_c \int_{\tau}^1 d\xi x g(x) \mathcal{P}_{gg} (\xi)
\end{equation}
where $\mathcal{P}_{gg} (\xi)$ is the full $g\to gg$ DGLAP type splitting function which is defined as 
\begin{equation}
\mathcal{P}_{gg} (\xi)=\frac{\xi}{(1-\xi)_+}+\frac{1-\xi}{\xi}+\xi(1-\xi)+\beta_0 \delta(1-\xi).
\end{equation}
It is then clear that this part of contribution corresponds to the renormalization of the collinear gluon distribution $xg(x)$. As for the single logarithmic terms in Eq.~(\ref{pggreal}) and $\ln \frac{M^2}{\mu^2}$ from above, we can choose the factorization scale $\mu^2=\frac{c_0^2}{R_\perp^2}$ and obtain the single logarithmic contribution to the Sudakov factor as follows
\begin{equation}
\frac{\alpha_s}{\pi} N_c \beta_0 \ln \frac{M^2R_\perp^2}{c_0^2}.
\end{equation}

At last, we need to evaluate the virtual contribution from Eq.~(\ref{vr}). The complete evaluation of Eq.~(\ref{vr}) is quite complicated. However, the leading power contribution, which is without any suppression of $\frac{k_\perp^2}{M_H^2}$, can be obtained easily by expanding $x_\perp , b_\perp$ at the vicinity of $v_\perp$. Therefore, the leading power contribution can be written as 
\begin{eqnarray}
\frac{\text{d}\sigma_{v b}^{pA\to \phi X}}{\text{d} y\text{d}^2 k_\perp}&=& 4ip^+ \alpha_s N_c \sigma_0 x_pg(x_p) \int_{0}^1 \text{d}\xi \int \frac{\text{d}%
^{2}v_{\perp}}{(2\pi)^{2}}\frac{\text{d}^{2}v_{\perp}^{\prime }}{(2\pi )^{2}}%
\frac{\text{d}^{2}u_{\perp}}{(2\pi)^{2}}e^{-ik_{\perp
}\cdot(v_{\perp}-v^{\prime }_{\perp})}  \notag \\
&&  \times \sum \Psi^{\ast}_{h \to g g}\left(\xi,
u_\perp\right)\Psi_{g\to gg}\left(\xi, u_\perp\right) \notag \\
&& \times u_\perp^j \left\{\text{Tr}U^\dagger(v^\prime_\perp)\left[\epsilon_{\perp}^{(1) } \cdot \partial U(v^\prime_\perp)\right]U^\dagger(v_\perp)\partial^j U(v_\perp) \right\}.
\end{eqnarray}
Using the definition of the relevant splitting functions which have been calculated earlier which yields
\begin{equation}
\sum \Psi^{\ast}_{\phi \to g g}\left(\xi,
u_\perp\right)\Psi_{g\to gg}\left(\xi, u_\perp\right)=i\frac{(2\pi)^2}{p^+u_\perp^4} K(\epsilon _{f}^{\prime}u_{\perp }) \frac{1}{2\xi (1-\xi)} u_\perp \cdot \epsilon_{\perp}^{(1)\ast },
\end{equation}
and taking into account the angular integration of $u_\perp$ which turns $u^i_\perp u^j_\perp$ into $\frac{1}{d} u_\perp^2$,\footnote{It is important to keep in mind that we are doing calculation in the dimensional regularization which changes the number of transverse dimension from $2$ to $d=2+2\epsilon$ in the coordinate space according to the convention used in this manuscript.} one can cast the above result into
\begin{eqnarray}
\frac{\text{d}\sigma_{v b}^{pA\to \phi X}}{\text{d} y\text{d}^2 k_\perp}&=&- \alpha_s N_c \sigma_0  x_pg(x_p) \int \frac{\text{d}%
^{2}v_{\perp}\text{d}^{2}v_{\perp}^{\prime }}{(2\pi)^{2}}e^{-ik_{\perp
}\cdot(v_{\perp}-v^{\prime }_{\perp})} \int_{0}^1 \frac{\text{d}\xi}{\xi (1-\xi)} \notag \\
&& \frac{2}{d}\int \frac{\text{d}^{2}u_{\perp}}{(2\pi)^{2}} \frac{K(\epsilon _{f}^{\prime}u_{\perp }) }{u_\perp^2}\left\{\text{Tr}U^\dagger(v^\prime_\perp)\left[ \partial^i U(v^\prime_\perp)\right]U^\dagger(v_\perp)\left[\partial^i U(v_\perp) \right]\right\}.
\end{eqnarray}
Following the procedure illustrated above, we should subtract the rapidity divergence, which contributes to the small-$x$ evolution of the WW correlator, and factorize out the tree level cross section. After that, the integrals in question is
\begin{equation}
\frac{2}{d}\int_{0}^1 \frac{\text{d}\xi}{\xi (1-\xi)}\int\frac{\text{d}^{2}u_{\perp}}{(2\pi)^{2}} \frac{K(\epsilon _{f}^{\prime}u_{\perp })-2 }{u_\perp^2}, \label{virtualsud}
\end{equation}
with $K(\epsilon _{f}^\prime u_{\perp })= 2\epsilon _{f}^\prime u_{\perp }\text{K}%
_{1}(\epsilon _{f}^\prime u_{\perp })+\epsilon _{f}^{\prime 2}u_{\perp }^{2}
\text{K}_{0}(\epsilon _{f}u_{\perp })$.
The above divergent integral can be either evaluated in the momentum space or in the coordinate space in the dimensional regularization. The detail of the evaluation can be found in the appendix.  Using the $\overline{\textrm{MS}}$ scheme, we can find Eq.~(\ref{virtualsud}) gives
\begin{eqnarray}
\frac{1}{\pi} \left(-\frac{1}{\epsilon^2}+\frac{1}{\epsilon}\ln \frac{M^2}{\mu^2}-\frac{1}{2}\ln^2\frac{M^2}{\mu^2} +\frac{\pi^2}{2}+\frac{\pi^2}{12}\right).
\end{eqnarray}

At the end of the day, by putting all the above real and virtual contributions together, which cancels all the divergent terms, and factorizing out the leading order contribution, we arrive at the following Sudakov factor at the one-loop order
\begin{equation}
\frac{\alpha_s}{\pi} N_c \left(\beta_0 \ln \frac{M^2R_\perp^2}{c_0^2}-\frac{1}{2}\ln^2 \frac{M^2R_\perp^2}{c_0^2} +\frac{\pi^2}{2}\right), \label{oneloopsud}
\end{equation}
with $\mu^2=\frac{c_0^2}{R_\perp^2}$.

Our calculations at one-loop
order in the above demonstrate that the Sudakov factors are well-separated from the small-$x$ and DGLAP type of logarithms. To resum the Sudakov large logarithms, we 
follow the Collins-Soper-Sterman procedure~\cite{Collins:1984kg}. In particular, we can 
write down an evolution equation with respect to the hard scale $M^2$. 
By solving the differential equation, we can resum the differential 
cross section as follows,
\begin{eqnarray}
\frac{d\sigma^{\rm ({\rm resum})}}{dyd^2k_\perp}|_{k_\perp\ll M}&=&\sigma_0
\int \frac{d^2x_\perp d^2x_\perp'}{(2\pi)^2}e^{ik_\perp\cdot R_\perp}e^{-{\cal S}_{\textrm{sud}}(M^2,R_\perp^2)}
S^{WW}_{Y=\ln 1/x_g}(x_\perp,x_\perp')\nonumber\\
&&\times x_p g_p(x_p,\mu^2=\frac{c_0^2}{R_\perp^2})\left[1+\frac{\alpha_s}{\pi}\frac{\pi^2}{2}N_c\right]\ ,\label{resum}
\end{eqnarray}
where the Sudakov form factor contains all order resummation 
\begin{eqnarray}
{\cal S}_{\textrm{sud}}(M^2,R_\perp^2)=\int_{C_1^2/R_\perp^2}^{C_2^2M^2}\frac{d\mu^2}{\mu^2}\left[A\ln\frac{C_2^2M^2}{\mu^2}+B\right] \ , 
\end{eqnarray}
where $C_{1,2}$ are parameters in order of 1.
The hard coefficients $A$ and $B$ can be
calculated perturbatively: $A=\sum\limits_{i=1}^\infty A^{(i)}\left(\frac{\alpha_s}{\pi}\right)^i$. 
From the explicit results for the one-loop calculations, we find that they are $A^{(1)}=N_c$ and $B^{(1)}=-\beta_0 N_c$, 
where we have chosen the so-called canonical variables 
for $C_1=c_0$ and $C_2=1$. 

As compared to Ref.~\cite{Ji:2005nu} which takes the $\alpha_s$ correction of 
$g_\phi$ into account, we find the above coefficients are consistent
with the known results~\cite{Berger:2002ut} except for $B^{(1)}$ which 
differs by a factor of 2. This difference is due to the fact that the virtual gluon and quark loops associated with 
the vertical small-$x$ gluon lines of Fig.~\ref{vh} are usually considered in the small-$x$ evolution of the relevant scattering amplitude at NLO, therefore are not included in this small-$x$ factorization for the calculation of the Sudakov factor, see, e.g., the discussions in Ref.~\cite{Kovchegov:2006wf}. We believe that this is due to generic difference between the small-$x$ saturation formalism and the normal transverse momentum dependent factorization formalism, which usually applies to the large $x$ regime where parton densities are dilute. We expect that this calculation can be applied to processes, such as two-photon productions\cite{Qiu:2011ai} and heavy quarkonium productions in $pA$ collisions\cite{Sun:2012vc}. 

In the process of the above one-loop calculation, we can derive four different renormalization group equations, namely, the DGLAP equation, the small-$x$ evolution equation, the colour charge renormalization group equation and the CSS type evolution equation. The above result resums all the leading large logarithms of $\alpha_s \ln 1/x$ and $\alpha_s \ln^2 \left(M^2/k_\perp^2\right)$ as well as the DGLAP type logarithms.
Provided that the next-to-leading order evolution for the WW
gluon distribution and and all other possible NLO corrections will be calculated in the future, together with the additional $A^{(2)}$ coefficient which is universal, the above result can be extended to include the resummation of next-to-leading logarithms. 

\subsection{Heuristic Derivation}
Here we provide a more heuristic derivation for the Sudakov factor for the forward Higgs boson production in $pA$ collisions. For the sake of simplicity, we use cut-off instead of dimensional regularization to regularize divergences. As far as the Sudakov factor is concerned, the following discussion is correct and adequate, but some assumptions may not be exactly accurate. 

It is well-known that large logarithms can be generated due to incomplete real and virtual graph cancellation when the phase space for additional gluon emission is limited. In other words, in principle, there is normally no large logarithms for inclusive observables, since the complete real and virtual cancellation are expected in this case. For the forward Higgs boson productions in $pA$ collisions, in order to produce a Higgs boson with fixed transverse momentum $k_\perp$, loosely speaking, we should restrict the transverse momentum of the emitted gluon to be less than $k_\perp$. This is corresponding to put a constraint on the real emission while the phase space of the virtual graphs are unrestricted. Since Sudakov factors can be physically interpreted as the probability for emitting no gluons with transverse momenta greater than $k_\perp$, we should be able to extract the Sudakov factor by simply studying the different constraints on real and virtual graphs. If we were to integrated over the Higgs boson transverse momentum $k_\perp$, we would not get any Sudakov factor. 

The Sudakov factor can be easily obtained through the following probability conservation argument. First, at one loop order, it is conceivable that the sum of the probability ($\mathcal{P}_< (k_\perp)$) that no gluon with transverse momentum greater than $k_\perp$ is emitted, and the probability ($\mathcal{P}_> (k_\perp)$) that one gluon with transverse momentum larger than $k_\perp$ is radiated should be unity, namely, $\mathcal{P}_<(k_\perp)+\mathcal{P}_>(k_\perp)=1$. Usually it is easier to compute $\mathcal{P}_>(k_\perp)$, since it only involves real emissions and it does not have IR divergence, while the computation of $\mathcal{P}_<(k_\perp)$ is slightly harder since it involves both the real and virtual graphs and IR divergence is cancelled between real and virtual graphs. For the UV divergence, we always work under the assumption that real and virtual diagrams cancel when $k_\perp > M$ in this part of discussion, thus the Higgs mass $M$ naturally serves as the UV cutoff. At the one-loop order, the Sudakov factor is nothing but $\mathcal{P}_< (k_\perp)$ or $1-\mathcal{P}_> (k_\perp)$ according to its physical interpretation. Therefore, there are two equivalent ways to compute the Sudakov factor at the one-loop order. The first way is to consider only the real gluon emissions with transverse momentum greater than $k_\perp$ but less than $M$, which gives $\mathcal{P}_> (k_\perp)$. The alternative method is to compute only the virtual graphs with transverse momentum $l_\perp$ integrated in the region $k_\perp <l_\perp <M$ under the assumption that cancellations happen everywhere else. This yields $-\mathcal{P}_> (k_\perp)$ which also allows us to obtain the Sudakov factor. 

\begin{figure}[tbp]
\begin{center}
\includegraphics[width=13cm]{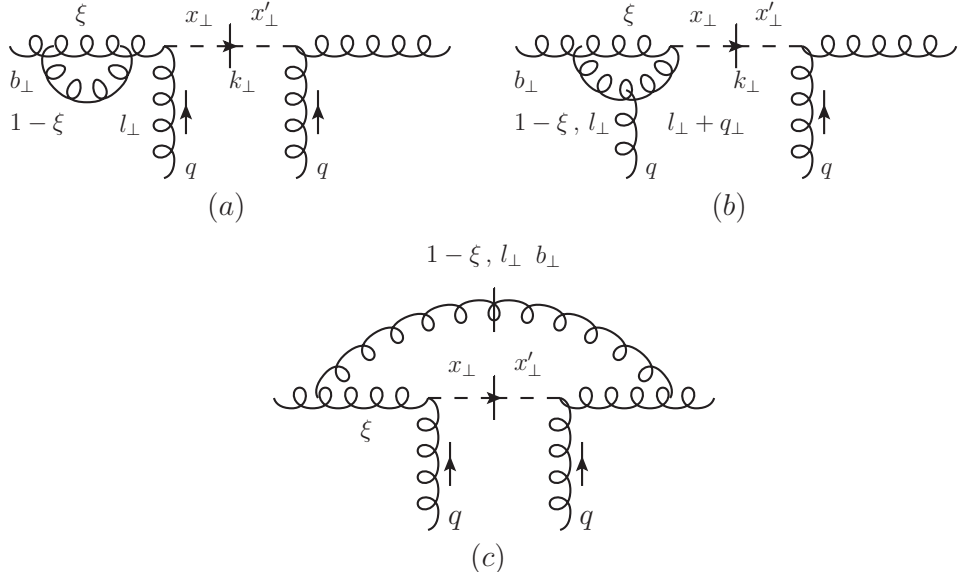}
\end{center}
\caption[*]{Three relevant graphs in the heuristic derivation.}
\label{hh}
\end{figure}

\begin{figure}[tbp]
\begin{center}
\includegraphics[width=9cm]{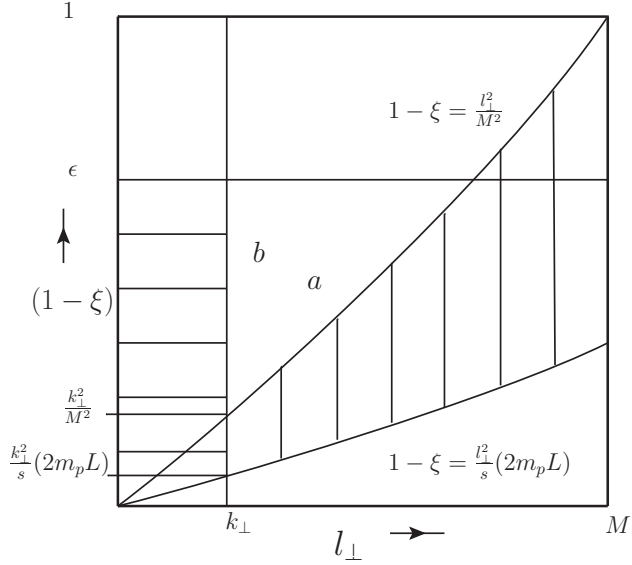}
\end{center}
\caption[*]{The kinematic regions of the emitted gluon in the target rest frame, where $L$ is the longitudinal size of the target nucleus, $s$ is the center of mass energy of the collision, $m_p$ is the rest mass of nucleons. Graphs $(a)$ and $(c)$ of Fig.~\ref{hh} cancel approximately in the horizontally shaded area, while graphs $(a)$ and $(b)$ cancel in the vertically shaded region between the line $1-\xi=\frac{l_\perp^2}{M^2}$ and the line $1-\xi=\frac{l_\perp^2}{s}(2m_pL)$. In this heuristic discussion, the parameter $\epsilon$ is just a small number between $0$ and $1$, and should not be confused with the parameter in the dimensional regularization which we used in the previous discussion.}
\label{ki}
\end{figure}

If one is willing to believe that certain real-virtual cancellation must occur, without explicitly verifying those cancellations, the calculation given in Sec.~\ref{HiggsDipole} can be considerably simplified with the double logarithmic term emerging almost trivially. We begin by explaining the cancellations that occur in terms of the three graphs of Fig.~\ref{hh}, the only graphs, in light cone gauge, that we shall need for this discussion. However, we remind the reader that much of the confidence that we have in this heuristic presentation stems from the more complete calculation which we have just presented. Fig.~\ref{ki} illustrates the kinematic regions of the unmeasured gluon (real or virtual) having transverse momentum $l_\perp$ and longitudinal momentum fraction $(1-\xi)x$ of the incident proton's momentum. The transverse momentum of the Higgs boson is $k_\perp$ and its longitudinal momentum fraction is $\tau$. For graphs $(a)$ and $(b)$ of Fig.~\ref{hh}, $\tau =x$ while for graph $(c)$ of Fig.~\ref{hh}, $\tau=x\xi$. We shall start with a focus on the region $1-\xi <\epsilon$ where $\epsilon$ is some small number; that is the gluon $(l_\perp, 1-\xi)$ is a soft gluon. 

If the transverse momentum of the Higgs boson were not measured, or if $\frac{k_\perp}{M} \sim 1$, there would be only single DGLAP and small-$x$ logarithms, and all Sudakov logarithms would disappear due to a real-virtual cancellation. But this must mean that when $\frac{l_\perp}{k_\perp} \ll 1$, there should also be a real-virtual cancellation. This corresponds to a cancellation between graphs $(a)$ and $(c)$ of Fig.~\ref{hh} with graph $(b)$ being small when $\frac{l_\perp}{k_\perp} \simeq \frac{l_\perp}{q_\perp} \ll 1$. In light cone perturbation theory this cancellation is almost manifest: Graph $(c)$ is the lowest order Higgs boson production amplitude times the probability that the initial gluon has an additional gluon $(l_\perp, 1-\xi)$ in its wave function, Graph $(a)$, along with the corresponding complex conjugate amplitude, must be the negative of graph $(c)$ by probability conservation. Then in Fig.~\ref{ki} the region shaded with horizontal lines, $\frac{l_\perp}{k_\perp}<1$, does not give a Sudakov contribution due to a cancellation of graphs $(a)$ and $(c)$, a real-virtual cancellation. To the right of line $b$ in Fig.~\ref{ki}, that is when $\frac{l_\perp}{k_\perp}>1$, graph $(c)$ ceases to contribute because the Higgs boson and the gluon $(l_\perp, 1-\xi)$ will have balancing transverse momentum. Graph $(b)$ only becomes important when the life time of the gluon $(l_\perp+q_\perp, 1-\xi)$ becomes as long as the formation time for the Higgs boson which occurs when 
\begin{equation}
\frac{(1-\xi)xp^+}{(l_\perp+q_\perp)^2}=\frac{xp^+}{M^2},
\end{equation}
as indicated by line $a$ in Fig.~\ref{ki} when $q_\perp$ is negligible as compared to $l_\perp$. Thus the region between lines $a$ and $b$ in Fig.~\ref{ki} is the Sudakov region and is evaluated solely from graph $(a)$ in Fig.~\ref{hh}. To the right of line $a$ of Fig.~\ref{ki}, graph $(a)$ and $(b)$ cancel. Thus the Sudakov double logarithmic contribution is 
\begin{equation}
2\left(-\frac{1}{2}\right)\frac{g^2N_c}{(2\pi)^3}\int_{k_\perp^2}^{M^2} \frac{\pi d l_\perp^2}{l_\perp^2} \int_0^{1-\frac{l_\perp^2}{M^2}}\frac{2 d\xi}{1-\xi} =-\frac{\alpha_s N_c}{2\pi}\ln^2 \frac{M^2}{k_\perp^2}. \label{hsudd}
\end{equation}
The integrand in Eq.~(\ref{hsudd}) is the negative of the probability that a bare gluon emits an additional gluon $(l_\perp, 1-\xi)$ in forming the dressed gluon wave function. At the double logarithmic level Eq.~(\ref{hsudd}) agrees with Eq.~(\ref{oneloopsud}).

Single logarithmic contributions are located along lines $a$ and $b$ and for $1-\xi < \epsilon$ in Fig.~\ref{ki}. Our strategy for evaluating the single logarithmic Sudakov terms is to integrate, for fixed $1-\xi$, between $l_\perp =0$ and $l_\perp=l^{(0)}_\perp$, with $l_\perp^{(0)}$ lying between lines $a$ and $b$, using graphs $(a)$ and $(c)$ and then integrate graphs $(a)$ and $(b)$ between $l_\perp^{(0)}$ and $l_\perp=M$. We find it convenient, however, to do these evaluations in transverse coordinate space. 

\subsubsection{The contribution across line b}
The correction to the Born graph for Higgs boson production due to graphs $(a)$ and $(c)$ in the region $x_0^2< b_\perp^2 <\infty$ with $1/x_0 =l^{(0)}_\perp$ is straightforwardly given by 
\begin{equation}
C_1=-\frac{\alpha_s N_c}{\pi^2}\int \Theta (b_\perp^2-x_0^2) d^2 b_\perp \left[\frac{1}{b_\perp^2}-\frac{b_\perp\cdot (b_\perp -x_\perp^\prime)}{b_\perp^2 (b_\perp -x_\perp^\prime)^2}\right],\label{c1}
\end{equation}
where we have taken $x_\perp =0$ in graphs $(a)$ and $(c)$ to simplify the calculation. The first term on the right hand side of Eq.~(\ref{c1}) is from graph $(a)$ while the second corresponds to graph $(c)$. Using
\begin{equation}
\int_0^{2\pi} \frac{d\phi_{b_\perp}}{2\pi} \frac{b_\perp -x_\perp^\prime}{(b_\perp -x_\perp^\prime)^2}=-\frac{x_\perp^\prime}{x_\perp^{\prime 2}} \Theta (x_\perp^{\prime 2} -b_\perp^2) =\frac{R_\perp}{R_\perp^2}\Theta (R_\perp^2 -b_\perp^2),
\end{equation}
one easily gets, with $R_\perp =x_\perp -x_\perp^\prime$, 
\begin{equation}
C_1 =-\frac{\alpha_s N_c}{\pi} \ln \frac{R_\perp^2}{x_0^2}.
\end{equation}
The integration over $\xi$ will be done later. 

\subsubsection{The contribution across line $a$}

The integration of graphs $(a)$  and $(b)$ across line $a$, between $b_\perp^2=0$ and $b_\perp^2=x_{0\perp}^2$ (where $x_\perp$ has been taken to be zero again) is not so easy. We shall begin by writing the graphs in light cone perturbation theory in momentum space and then transform to transverse coordinate space. Then
\begin{equation}
C_2 = -\frac{\alpha_s N_c}{\pi^2} \int d^2 l_\perp \left[\frac{1}{l_\perp^2}-\frac{2l_\perp\cdot (l_\perp+q_\perp) q_\perp\cdot (l_\perp+q_\perp)-l_\perp\cdot q_\perp (l_\perp+q_\perp)^2}{l_\perp^2 q_\perp^2 \left[(l_\perp+q_\perp)^2+m^2\right]}\right], \label{c2}
\end{equation}
where 
\begin{equation}
m^2=-M^2(1-\xi).\label{m}
\end{equation}
We do the calculation with $M^2$, in Eq.~(\ref{m}), having a negative value and at the end analytically continue $M^2$ back to its positive physical value. Thus while doing the calculation we suppose $m^2$ in Eq.~(\ref{c2}) is positive. Eq.~(\ref{c2}) can be written as
\begin{equation}
C_2 = -\frac{\alpha_s N_c}{\pi^2} \int d^2 l_\perp \frac{l_{ i}}{l_\perp^2}\left[\frac{l_{ i}}{l_\perp^2}+\frac{q_j}{q_\perp^2}\frac{\delta_{ij} (l_\perp+q_\perp)^2-2 (l+q)_i (l+q)_j }{\left[(l_\perp+q_\perp)^2+m^2\right]}\right].\label{c22}
\end{equation}
In Eqs.~(\ref{c2}) and (\ref{c22}), the first term on the right hand side comes from graph $(a)$ while the second term comes from graph $(b)$. 

Using 
\begin{equation}
\frac{l_i}{l_\perp^2}=\frac{i}{2\pi}\int d^2 z_\perp e^{-il_\perp \cdot z_\perp} \frac{z_i}{z_\perp^2}
\end{equation} 
and 
\begin{equation}
\frac{1}{(l_\perp+q_\perp^2)^2+m^2}=\frac{1}{2\pi} \int d^2 b_\perp K_{0} (mb_\perp) e^{i(q_\perp+l_\perp)\cdot b_\perp},
\end{equation}
one gets
\begin{equation}
C_2 =-\frac{\alpha_s N_c}{\pi^2} \int d^2 l_\perp d^2 z_\perp \frac{i}{2\pi}e^{-il_\perp \cdot z_\perp} \frac{z_i}{z_\perp^2} \mathcal{I}, \label{c23}
\end{equation}
with 
\begin{equation}
\mathcal{I}=\int d^2 b_\perp \left[-\frac{i}{2\pi}e^{il_\perp\cdot b_\perp}\frac{b_i}{b_\perp^2}-\frac{q_i}{2\pi q_\perp^2}\left(\delta_{ij}\delta_{kl}-2\delta_{il}\delta_{jk}\right)K_{0}(mb_\perp)\nabla_k \nabla_l e^{i(q_\perp+l_\perp)\cdot b_\perp}\right]. \label{inte}
\end{equation}
One integration by parts in Eq.~(\ref{inte}) is safe, so 
\begin{equation}
\mathcal{I}=\int d^2 b_\perp \left[-\frac{i}{2\pi}e^{il_\perp\cdot b_\perp}\frac{b_i}{b_\perp^2}+\frac{q_i}{2\pi q_\perp^2}\left(\delta_{ij}\delta_{kl}-2\delta_{il}\delta_{jk}\right)\nabla_k K_{0}(mb_\perp) \nabla_l e^{i(q_\perp+l_\perp)\cdot b_\perp}\right]. \label{inte2}
\end{equation}
The integration in Eq.~(\ref{inte2}) is convergent at $b_\perp =0$ but not sufficiently convergent to do a second integration by parts. Introduce a cutoff for $b_\perp \rho$ and then integrate by parts, one gets 
\begin{equation}
\mathcal{I} =\lim_{\rho\to 0} \left(\mathcal{I}_{1}+\mathcal{I}_{2}\right),
\end{equation}
where 
\begin{equation}
\mathcal{I}_1=\int d^2 b_\perp \Theta (b_\perp-\rho) \frac{q_j}{2\pi q_\perp^2}\left(\delta_{ij}\delta_{kl}-2\delta_{il}\delta_{jk}\right)\nabla_l\left(\nabla_k K_{0}(mb_\perp) e^{i(q_\perp+k_\perp)\cdot b_\perp}\right), \label{inte3}
\end{equation}
and
\begin{equation}
\mathcal{I}_2=\int d^2 b_\perp \Theta (b_\perp-\rho)e^{il_\perp\cdot b_\perp} \left[-\frac{i}{2\pi}\frac{b_i}{b_\perp^2}-\frac{q_i}{2\pi q_\perp^2}e^{iq_\perp\cdot b_\perp} \left(\delta_{ij}\delta_{kl}-2\delta_{il}\delta_{jk}\right)\nabla_k \nabla_lK_{0}(mb_\perp) \right]. \label{inte4}
\end{equation}

It is straightforward to partially integrate Eq.~(\ref{inte3}) to find 
\begin{equation}
\mathcal{I}_1 =\int_{b_\perp=\rho} d\phi_{b_\perp}\frac{b_k b_l}{b_\perp^2}\frac{q_i}{2\pi q_\perp^2}\left(\delta_{ij}\delta_{kl}-2\delta_{il}\delta_{jk}\right) e^{i(q_\perp+k_\perp)\cdot b_\perp},\label{inte5}
\end{equation} 
Substituting Eq.~(\ref{inte5}) into Eq.~(\ref{c23}) and first doing the $d^2 l_\perp$ integration gives
\begin{equation}
C_{2,1}=-\frac{\alpha_s N_c}{\pi}. \label{c21f}
\end{equation}
Next substitute Eq.~(\ref{inte4}) into Eq.~(\ref{c23}) and integrate over $l_\perp$ and $z_\perp$, one finds 
\begin{equation}
C_{2,2}=-\frac{\alpha_s N_c}{\pi^2} \int d^2 b_\perp  \left[\frac{1}{b_\perp^2}-\frac{i b_i q_j}{b_\perp^2 q_\perp^2}e^{iq_\perp\cdot b_\perp} \left(\delta_{ij}\nabla^2-2\nabla_i \nabla_j\right) K_{0}(mb_\perp) \right]. 
\end{equation}
Anticipating that the final integration over $d^2 q_\perp$ will give no additional angular dependence, we may average over direction of $q_\perp$ to get the leading power contribution 
\begin{equation}
C_{2,2}=-\frac{\alpha_s N_c}{\pi^2} \int d^2 b_\perp  \left[\frac{1}{b_\perp^2}+\frac{i b_i b_j}{2 b_\perp^2} \left(\delta_{ij}\nabla^2-2\nabla_i \nabla_j\right) K_{0}(mb_\perp) \right]. \label{inte6}
\end{equation}
Now recall that the $b_\perp$ integral should be limited to $0<b_\perp^2 <x_0^2$ for $C_{2}$. Doing the integral in Eq.~(\ref{inte6}) over the region $0<b_\perp^2 <x_0^2$ gives
\begin{equation}
C_{2,2}=-\frac{\alpha_s N_c}{\pi} \left[ \ln\frac{x_0^2m^2}{4}+2\gamma_E-1\right]. \label{c22f}
\end{equation}
with $\gamma_E$ the Euler constant. Adding Eq.~(\ref{c21f}) and Eq.~(\ref{c22f}) finally yields
\begin{equation}
C_{2}=-\frac{\alpha_s N_c}{\pi} \left[ \ln\frac{x_0^2m^2}{4}+2\gamma_E\right]. \label{c2f}
\end{equation}

Now turn to the $\xi$-integration. The $\xi$-integral goes from $1-\xi=\frac{k_\perp^2}{M^2}$, the intersection of lines $a$ and $b$ in Fig.~\ref{ki} up to $\epsilon$. (The region $(1-\xi)>\epsilon$ will shortly be considered separately.) Then, going from momentum space limits to coordinate space, and recalling that at the moment $M^2<0$, 
\begin{eqnarray}
\overline{C}&=&\int_\frac{-4}{M^2R_\perp^2}^\epsilon d(1-\xi) \left(C_1+C_2\right) \notag\\
&=&-\frac{\alpha_s N_c}{\pi}\int_\frac{-4}{M^2R_\perp^2}^\epsilon d(1-\xi) \left[ \ln\frac{R_\perp^2}{x_0^2}+\ln\frac{x_0^2m^2}{4}+2\gamma_E\right]. \label{ctotal}
\end{eqnarray}
We note that all $x_0$ dependence cancels. Using Eq.~(\ref{m}) in Eq.~(\ref{ctotal}),
\begin{equation}
\overline{C}=-\frac{\alpha_s N_c}{2\pi} \left[ \left(\ln\frac{-M^2R_\perp^2}{4}-\ln\frac{1}{\epsilon}\right)^2+2\gamma_E\ln\frac{-M^2R_\perp^2 \epsilon}{4} \right]. \label{cf}
\end{equation}
Now continue $M^2$ to positive values, take the real part of $\overline{C}$ and drop non-logarithmic terms. (The answer in Eq.~(\ref{cf}) comes partly from a complex conjugate amplitude. Taking the real part corresponds to opposite direction of continuation in the amplitude and complex conjugate amplitude.) One finds 
\begin{equation}
\overline{C}=-\frac{\alpha_s N_c}{2\pi} \left[ \ln^2\frac{M^2R_\perp^2}{4}+2\left(2\gamma_E-\ln\frac{1}{\epsilon}\right)\ln\frac{M^2R_\perp^2}{4} -\pi^2\right]. \label{cff}
\end{equation}
The upper limit $\epsilon$ in Eq.~(\ref{ctotal}) is somewhat arbitrary, and we shall see that the $\epsilon$-dependence in Eq.~(\ref{cff}) will be exactly cancelled when we evaluate the contribution of the $1-\xi>\epsilon$ region. The lower limit of the integral in Eq.~(\ref{ctotal}) is also somewhat arbitrary. We might, more generally, write this lower limit as $-\frac{M^2R_\perp^2}{4} \zeta$ with $\zeta$ a constant. One can check that with such a lower limit the result in Eq.~(\ref{cff}) remains unchanged, that is the single and double logarithmic contributions are insensitive to the details of the exact point where lines $a$ and $b$ of Fig.~\ref{ki} cross. 

Now turn to the $1-\xi>\epsilon$ contribution. When $l_\perp^2 <k_\perp^2$, there are single DGLAP logarithms which are included in the evolution of the gluon distribution of the proton. (Graphs $(a)$ and $(c)$ do not cancel when $1-\xi >\epsilon$ and $l_\perp<k_\perp$ because the $x$-value for the gluon distribution corresponding to graph $(a)$ and $(b)$ is $x=\tau$, while for graph $(c)$ it is $x=\frac{\tau}{\xi}$ and for $1-\xi >\epsilon$, these different $x$-values can destroy the real-virtual cancellation.) Thus only graph $(a)$ is important and leads to a contribution 
\begin{equation}
\overline{C}^\prime=-\frac{\alpha_s N_c}{\pi} \int_{\frac{4}{R_\perp^2}}^{M^2}\frac{d l_\perp^2}{l_\perp^2}\int_{\epsilon}^{1/2} d (1-\xi) \left[\frac{1-\xi}{\xi}+\frac{\xi}{1-\xi}+\xi(1-\xi)\right] ,
\end{equation}
or
\begin{equation}
\overline{C}^\prime=-\frac{\alpha_s N_c}{\pi}\ln\frac{M^2R_\perp^2}{4}\left(\ln\frac{1}{\epsilon}-\frac{11}{12}\right). \label{cpf}
\end{equation}
Adding Eq.~(\ref{cff}) and Eq.~(\ref{cpf}) along with quark loops corresponding to graph $(a)$ gives Eq.~(\ref{oneloopsud}). Including a resummation of Sudakov logarithms, small-$x$ single logarithmic terms and multiple scattering in a nucleus target leads to Eq.~(\ref{resum}). Finally we note that, in terms of the kinematics shown in Fig.~\ref{ki}, the single logarithmic region is located at $l_\perp \sim k_\perp$  and $\frac{k_\perp^2}{s}(2m_p L) <1-\xi <\frac{k_\perp^2}{M^2}$. The highest momentum small-$x$ gluons just match onto the softest Sudakov gluons, but there is no overlapping of their kinematic domains.  

\begin{figure}[tbp]
\begin{center}
\includegraphics[width=8cm]{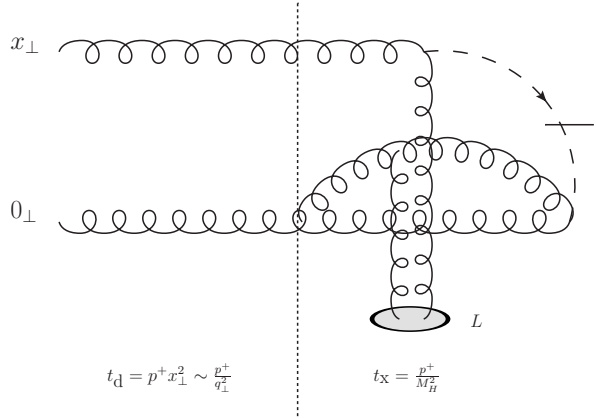}
\end{center}
\caption[*]{The dipole graph corresponding to the graph shown in Fig.~\ref{hh} $(b)$.}
\label{time}
\end{figure}

In addition, we can also use the dipole picture as show in Fig.~\ref{time} to understand the separation of the small-$x$ resummation and the Sudakov resummation. In this dipole graph, the incoming gluon is either located at the coordinate $0_\perp$ or at $x_\perp$ and the outgoing Higgs particle is produced after the scattering with the target nucleus with longitudinal size $L$. We always assume that the scattering energy is so high that $\frac{p^+}{ M^2}\gg L$ with $p^+=\frac{s}{2m_p}$ in the target rest frame. Because the incoming gluon is massless while the outgoing Higgs particle has a mass $M$ which is much larger than its transverse momentum $q_\perp$, the time region before and after (roughly separated by the dotted line) the scattering are asymmetric. If the lifetime of the virtual fluctuation $t_{\textrm{f}}=\frac{l^+}{l_\perp^2}$, with $l^+=(1-\xi)p^+$ is in the region $L< t_{\textrm{f}} < t_{\textrm{x}}=\frac{p^+}{M_H^2}$, we can attribute it to the small-$x$ fluctuation which gives the small-$x$ evolution. If the virtual fluctuation has a lifetime in the region $ t_{\textrm{x}}=\frac{p^+}{M_H^2}< t_{\textrm{f}} < t_{\textrm{d}}=\frac{p^+}{q_\perp^2}$, then it can be characterized as the Sudakov fluctuation which only contribute to the Sudakov factor. Therefore, in time, the Sudakov fluctuation happens before the small-$x$ fluctuations so that there can be no interference between them. 


\section{The Sudakov factor for heavy quark pair production}
\label{hqp}
This section is devoted to the calculation of the Sudakov double logarithm for heavy quark pair productions in DIS. We will not provide much detail for this calculation since it bears a lot of resemblance to that of the Higgs boson productions in $pA$ collisions. The main objective is to compute the coefficient of the Sudakov double logarithm, which has an additional factor of $\frac{1}{2}$ as compared to the double logarithmic term in Higgs boson production. 

To avoid additional complication, we assume that the transverse momentum of each produced heavy quark $k_{\perp i}$ is not too much larger than the saturation momentum, and so is the virtuality of the photon $Q^2$. Therefore, the only large scale in this problem is the heavy quark mass $m_q$. 

\begin{figure}[tbp]
\begin{center}
\includegraphics[width=7cm]{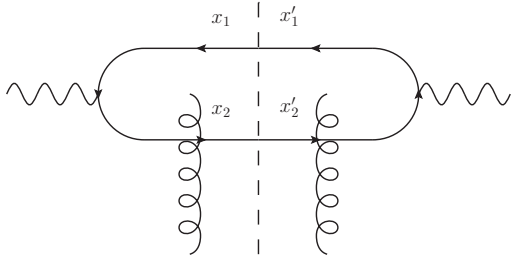}
\end{center}
\caption[*]{The LO Born diagram for producing heavy quark pairs in DIS in small-$x$ limit.}
\label{LOh}
\end{figure}

The leading order cross section for heavy quark pair productions in DIS, as shown in Fig.~\ref{LOh}, can be written as \cite{Dominguez:2011wm}
\begin{eqnarray}
\frac{d\sigma ^{\gamma_{T,L}^{\ast }A\rightarrow q\bar{q}}}{d^3k_1d^3k_2}
&=&N_{c}\alpha _{em}e_{q}^{2}\delta(p^+-k_1^+-k_2^+) \int
\frac{\text{d}^{2}x_1}{(2\pi)^{2}}\frac{\text{d}^{2}x_1^{\prime }}{(2\pi )^{2}}
\frac{\text{d}^{2}x_2}{(2\pi)^{2}}\frac{\text{d}^{2}x_2^{\prime }}{(2\pi )^{2}} \notag \\
&&\times e^{-ik_{1\perp }\cdot(x_1-x_1^{\prime })} e^{-ik_{2\perp }\cdot (x_2-x_2^{\prime })}
\sum_{\lambda\alpha\beta} \psi_{\alpha\beta 0}^{T, L \lambda}(x_1-x_2)
\psi_{\alpha\beta  0}^{T, L\lambda*}(x_1^{\prime }-x_2^{\prime })  \notag \\
&&\times \left[1+S^{(4)}_{x_g}(x_1,x_2;x_2^{\prime },x_1^{\prime})
-S^{(2)}_{x_g}(x_1,x_2)-S^{(2)}_{x_g}(x_2^{\prime },x_1^{\prime })\right] \ ,\label{xsdis}
\end{eqnarray}
where 
\begin{align}
\psi^{T\,\lambda}_{\alpha\beta 0}(p^+,z,r)&=2\pi\sqrt{\frac{2}{p^+}}\begin{cases}i\epsilon_f^{\prime}K_1(\epsilon_f^{\prime}|r|)\tfrac{r\cdot\epsilon^{(1)}_\perp}{|r|}[\delta_{\alpha+}\delta_{\beta+}(1-z)+\delta_{\alpha-}\delta_{\beta-}z]\\
\quad+\delta_{\alpha -}\delta_{\beta +}m_qK_0(\epsilon_f^{\prime}|r|), & \lambda=1,\\
i\epsilon_f^{\prime}K_1(\epsilon_f^{\prime}|r|)\tfrac{r\cdot\epsilon^{(2)}_\perp}{|r|}[\delta_{\alpha-}\delta_{\beta-}(1-z)+\delta_{\alpha+}\delta_{\beta+}z]\\
\quad+\delta_{\alpha +}\delta_{\beta -}m_qK_0(\epsilon_f^{\prime}|r|), & \lambda=2,\end{cases}\\
\psi^L_{\alpha\beta 0}(p^+,z,r)&=2\pi\sqrt{\frac{4}{p^+}}z(1-z)QK_0(\epsilon_f^{\prime}|r|)\delta_{\alpha\beta}\;,
\end{align}
and $\epsilon^{\prime 2}_f=z(1-z)Q^2+m^2_q$. Due to the large quark mass $m_q$ as compared to the saturation momentum, one can easily find that $u_\perp \ll v_\perp$ with $u=x_1-x_2$ and $v=zx_1+(1-z)x_2$. Therefore, in the heavy quark mass limit, the last line of Eq.~(\ref{xsdis}) can be further simplified as 
\begin{equation}
-u_iu'_j\frac{1}{N_c}\langle\text{Tr}\left[\partial_iU(v)\right]U^\dagger(v')\left[\partial_jU(v')\right]
U^\dagger(v)\rangle_{x_g}\ .
\end{equation}
Similar to the Higgs boson productions in $pA$ collisions, the LO cross section is proportional to the WW gluon distribution since this process only has final state interactions, while there is only initial state interactions in the Higgs boson productions. 

Now we study the one-loop correction of this process by emitting one extra gluon from the quark line. The important difference that we have to take into account is the quark mass which modifies the light cone denominator and therefore the splitting function. The light cone denominator now becomes
\begin{equation}
\frac{1}{E_q +E_{g}-E_{q 0}}=\frac{\xi(1-\xi) p^+}{l_\perp^2+\xi^2m_q^2},
\end{equation}
where $l_\perp$ is the relative transverse momentum between the quark and radiated gluon and $\xi$ is the longitudinal momentum fraction of the gluon as respect to the original quark. It is crucial to note that the coefficient in front of the quark mass square $m_q^2$ is now $\xi^2$ which yields a factor of $\frac{1}{2}$ as compared to the Higgs boson production where the coefficients are either $-\xi(1-\xi)$ or $\xi$ for the Sudakov double logarithmic term. This particular effect associated with the gluon radiation from a heavy quark is sometimes known as the 'dead-zone' effect.

\begin{figure}[tbp]
\begin{center}
\includegraphics[width=6cm]{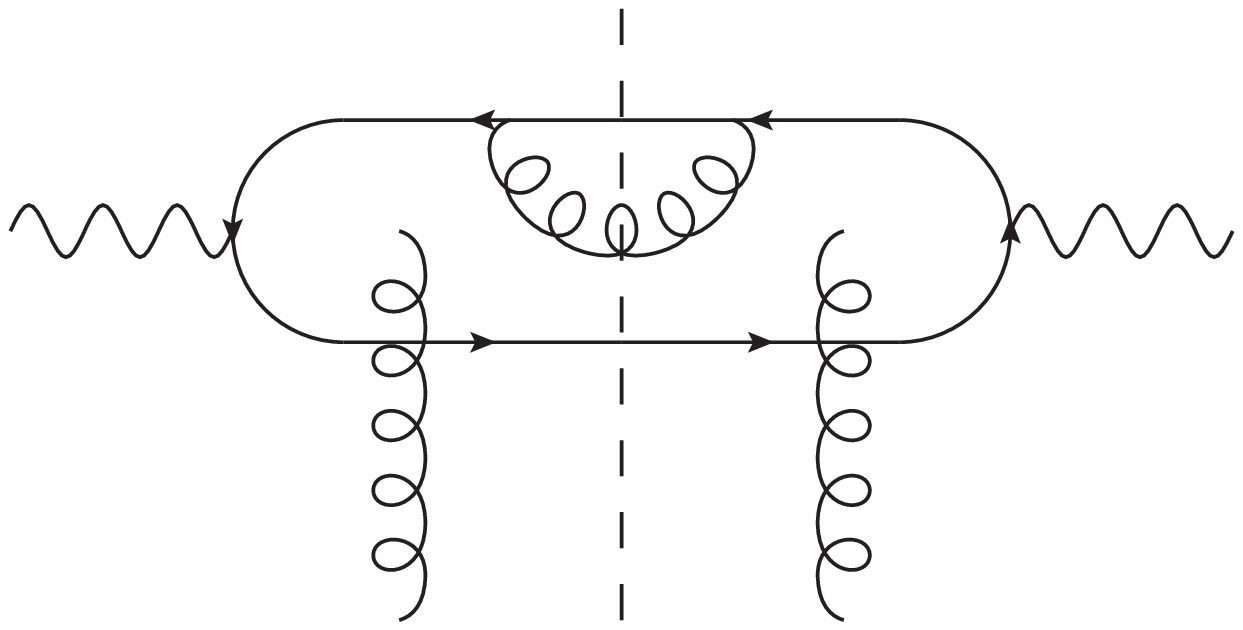}\includegraphics[width=6cm]{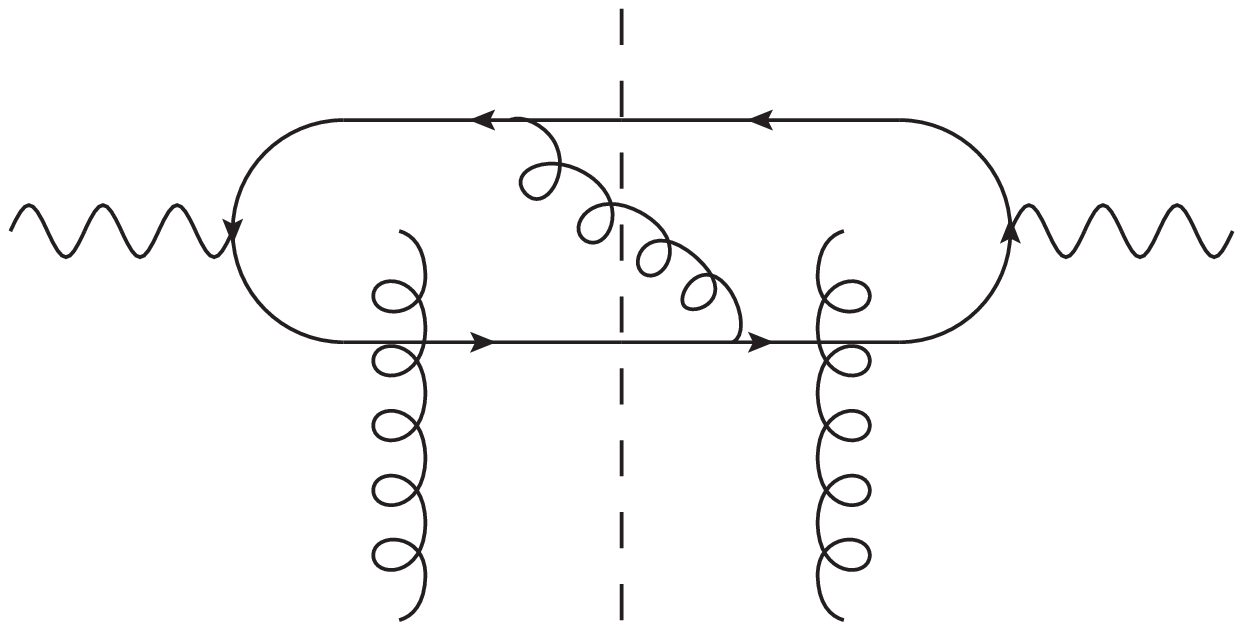}
\end{center}
\caption[*]{The left graph yields a Sudakov double logarithm with a color factor $C_F$ while the right graph is power suppressed or suppressed by powers of $N_c$. Eventually the combination of these two graphs give the color factor $\frac{N_c}{2}$ for the Sudakov double logarithm. By including the other two similar real graphs, one can show that the effective color factor is actually $N_c$.}
\label{sudr}
\end{figure}

For the corresponding real contributions from Fig.~\ref{sudr} after factorizing out the contribution from the Born diagram, one obtains
\begin{equation}
\alpha_s N_c \times 4 \int \frac{\textrm{d} ^2l_\perp}{(2\pi)^2}\int_{\frac{l_\perp^2}{s}}^1\frac{\textrm{d} \xi}{\xi} \left[ \frac{l_\perp^2}{(l_\perp^2+\xi^2 m_q^2)^2}\right] e^{-i l_\perp\cdot R_\perp},
\end{equation}
with $R_\perp =x_1-x_1^\prime \simeq v-v^\prime$. In the spirit of the power expansion, here we have used the fact that the $q\bar q$ pair has a very small transverse size and therefore $v\simeq x_1 \simeq x_2$. It is easy to check the coefficient by comparing with the corresponding contribution to the small-$x$ evolution equation as indicated in Eq.~\ref{wwe} after transforming into coordinate space. Again, by following the procedure developed in the Higgs boson production calculation, after subtracting off the energy evolution contribution which is proportional to $\ln\frac{1}{x_g}\simeq \ln \frac{s}{m_q^2}$, together with the identities that we have derived in the Appendix and factorizing out the Born diagram contribution, one finds that the contribution from Fig.~\ref{sudr} becomes
\begin{eqnarray}
&&2\alpha_s N_c\int \frac{\textrm{d} ^2l_\perp}{(2\pi)^2}\frac{1}{l_\perp^2}\ln\frac{m_q^2}{l_\perp^2} e^{-i l_\perp\cdot R_\perp} \notag \\
&=& \frac{\alpha_sN_c}{2\pi} \left[\frac{1}{\epsilon^2} - \frac{1}{\epsilon} \ln \frac{m_q^2}{\mu^2}+\frac{1}{2} \ln^2 \frac{m_q^2}{\mu^2}-\frac{1}{2} \ln^2 \frac{m_q^2R^2_\perp}{c_0^2}+\frac{\pi^2}{12} \right],
\end{eqnarray}
in dimensional regularization with the $\overline{\textrm{MS}}$ scheme.

\begin{figure}[tbp]
\begin{center}
\includegraphics[width=6cm]{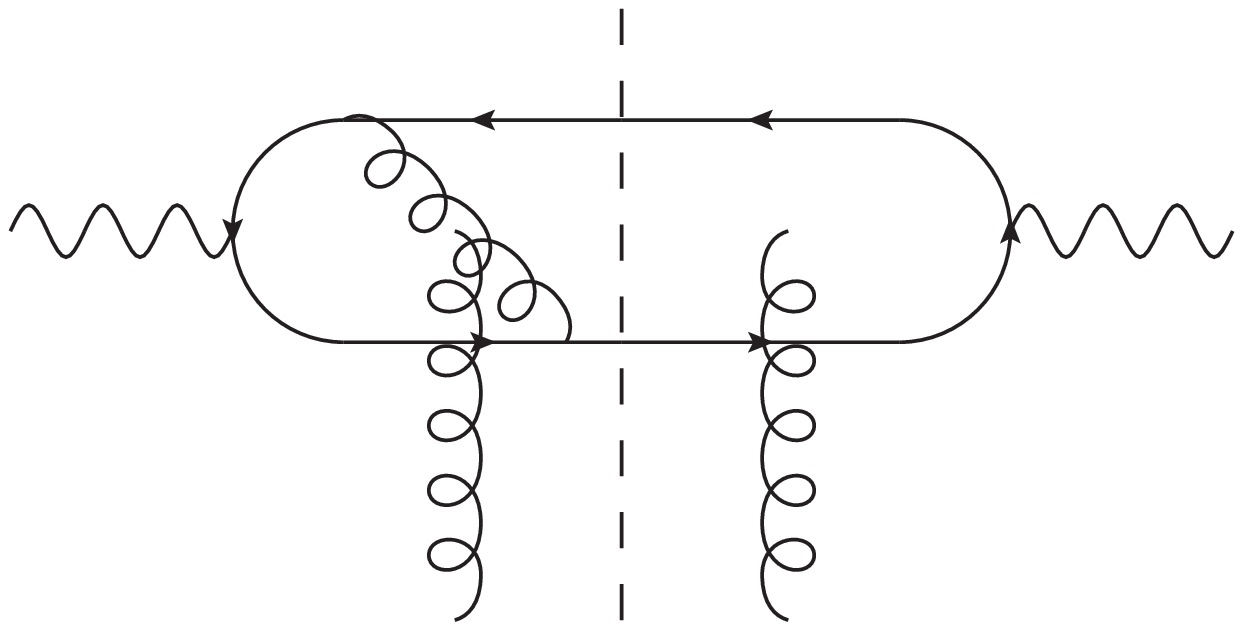}\includegraphics[width=6cm]{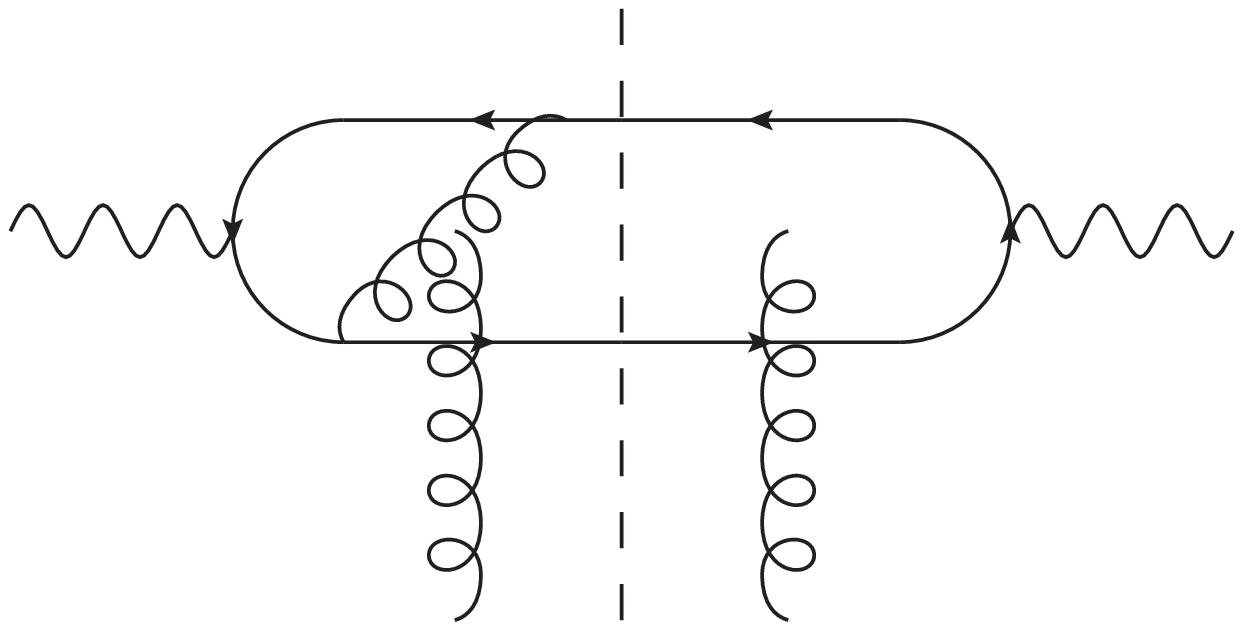}
\includegraphics[width=6cm]{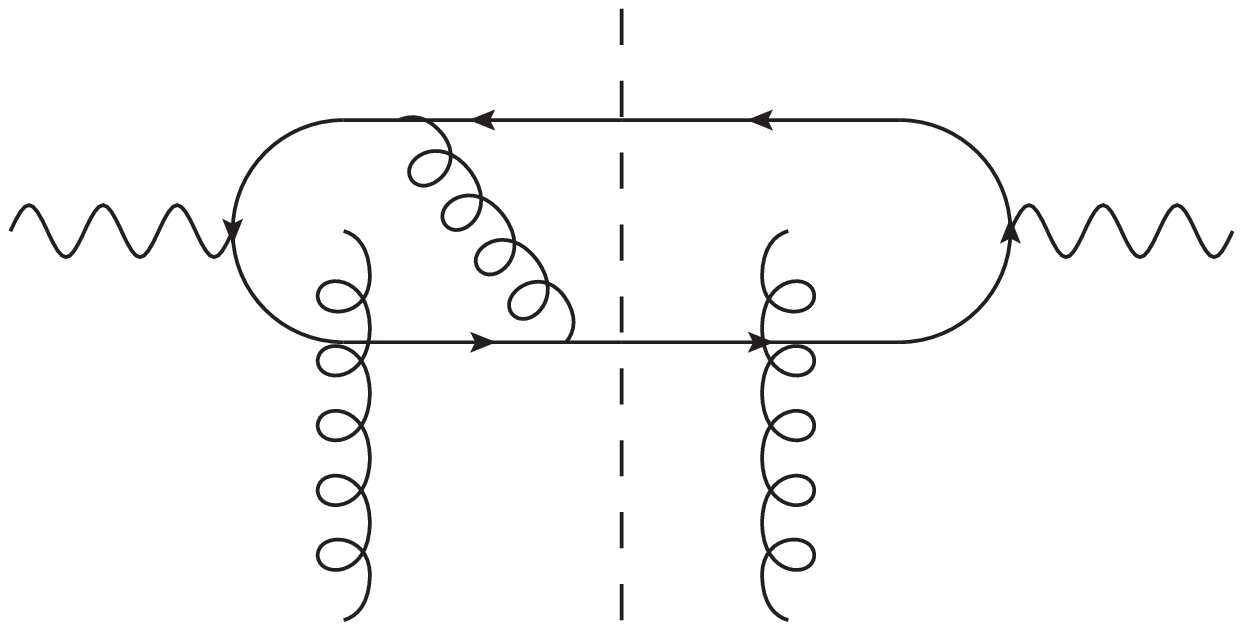}\includegraphics[width=6cm]{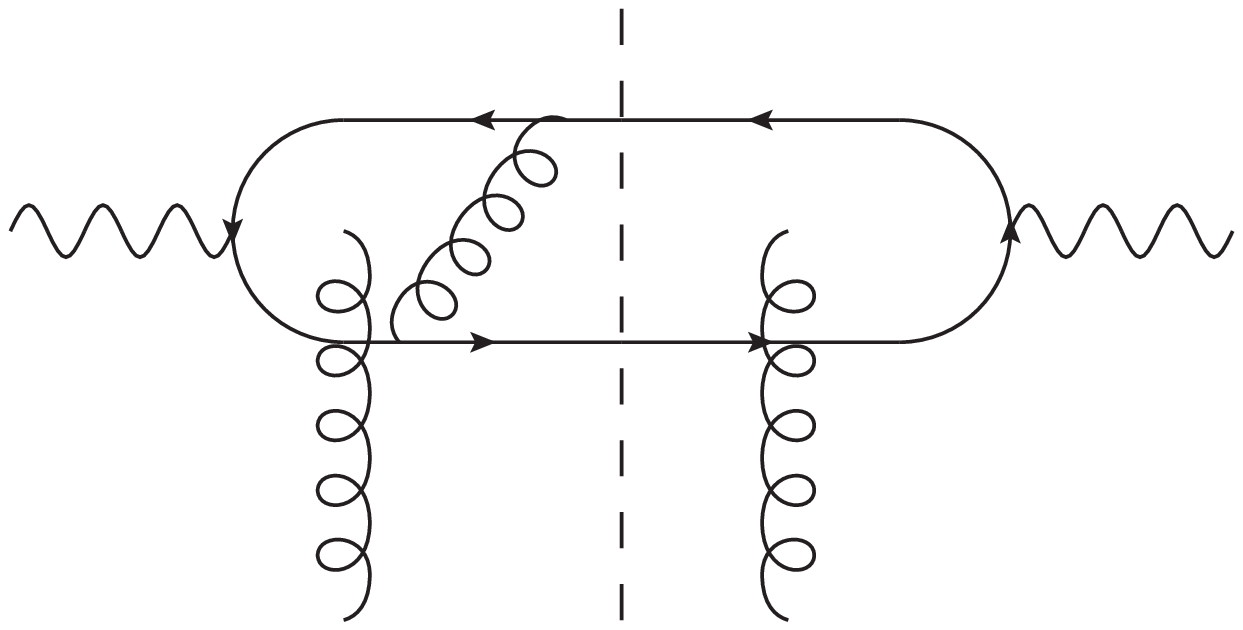}
\end{center}
\caption[*]{Four virtual graphs which contribute to the Sudakov double logarithm. The leading $N_c$ contributions of the lower two graphs are power suppressed, and therefore are neglected. The large $N_c$ corrections of these four diagrams cancel each other, and give an effective color factor $N_c$ when the mirror diagrams are taken into account.}
\label{noww}
\end{figure}

Let us now consider the virtual diagrams as shown in Fig.~\ref{noww}, which actually resembles the right virtual graph in Fig.~\ref{vh} in terms of the color structure. It is not surprising that the calculation with respect to Fig.~\ref{noww} is very similar to that of the right graph in Fig.~\ref{vh}. By following the same procedure especially the power expansion, and keeping in mind that $u=x_1-x_2$ and $u^\prime =x_1^\prime -x_2^\prime$ are very small as compared to other scales, we obtain the following contribution after factorizing out the Born contribution 
\begin{eqnarray}
&&\alpha_s \frac{N_c}{2} \times 2 \times 4\int_0^1\frac{\textrm{d} \xi}{\xi} \int \frac{\textrm{d} ^2l_\perp}{(2\pi)^2}\left[ \frac{l_\perp^2}{(l_\perp^2+\xi^2 m_q^2)^2}-\frac{1}{l_\perp^2}\right] \notag \\
&=&-\frac{\alpha_sN_c}{2\pi}\left[\frac{1}{\epsilon^2} - \frac{1}{\epsilon} \ln \frac{m_q^2}{\mu^2}+\frac{1}{2} \ln^2 \frac{m_q^2}{\mu^2}+\frac{\pi^2}{12}+\cdots \right].
\end{eqnarray}
Here the first factor of $2$ is to take into account the complex conjugate amplitudes while the factor of 4 is from the factor $2(1+(1-\xi)^2)$ in the splitting function in the small $\xi$ limit. 

\begin{figure}[tbp]
\begin{center}
\includegraphics[width=6cm]{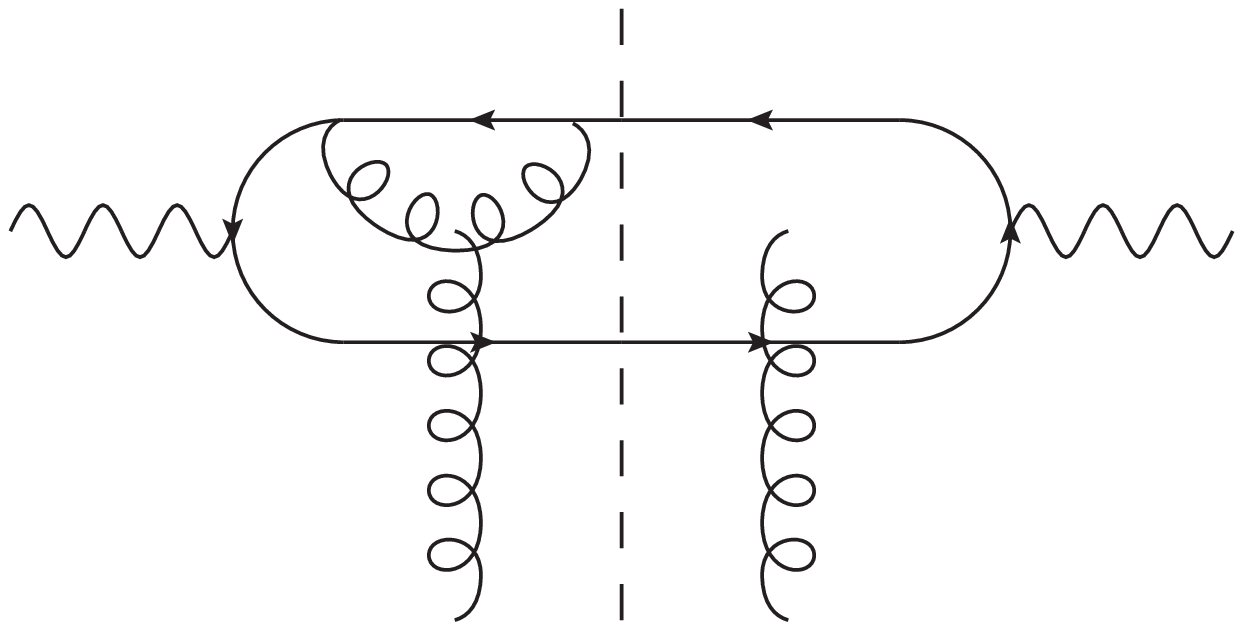}\includegraphics[width=6cm]{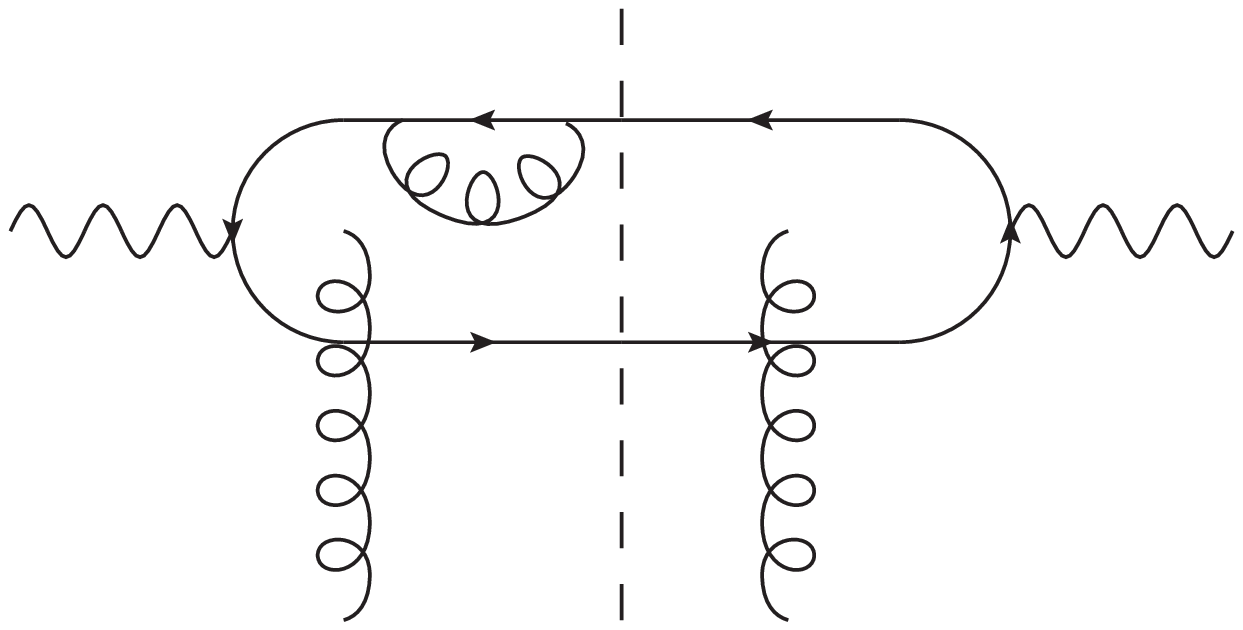}
\end{center}
\caption[*]{Two typical self-energy type of virtual graphs.}
\label{self}
\end{figure}

We assume that there is no purely initial state radiative corrections, because the $q\bar q$ pair coming from the virtual photon is color neutral and has a very small transverse size. As to the quark self energy diagrams as shown in Fig.~\ref{self}, we believe that the sum of these diagrams only contains single logarithms. 
 
At the end of the day, by adding the real and virtual contributions together, one obtains the Sudakov double logarithm term 
\begin{equation}
-\frac{\alpha_s N_c}{4\pi}\ln^2 \frac{m_q^2R^2_\perp}{c_0^2},
\end{equation}
where $R_\perp$ is conjugate to the typical transverse momentum of the produced heavy quark pair. It is interesting to note that the coefficient of this Sudakov double logarithmic term is half of that in the Higgs boson production case due to their generic difference in terms of QCD interaction and dynamics. Due to the complexity of the problem, we will leave the evaluation of the single logarithmic term and the constant correction for future studies.

\section{The Sudakov factor for dijet prodution in DIS}

This section of the paper is devoted to the derivation of the Sudakov double logarithmic term for the dijet productions in DIS from the corresponding one-loop calculation. A collinear factorization approach to this problem can be found in Ref.~\cite{Banfi:2008qs}.

The DIS dijet processes at low $x$ limit is very important, since it provides us the direct access to the well-known WW gluon distribution which has never been measured before. The back-to-back dijet configuration, with jets' transverse momentum $k_{1\perp}$ and $k_{2\perp}$, of this process contains two separated scales, i.e., the large jet transverse momentum $P_\perp\equiv \frac{1}{2}|k_{1\perp}-k_{2\perp}|$ and the small dijet momentum inbalance $k_\perp\equiv |k_{1\perp}+k_{2\perp}|$. We define the total transverse momentum carried by the gluons from the target hadron as $q_\perp$, which is identical to the dijet transverse momentum inbalance $k_\perp$ at the leading order. We assume that the photon virtuality $Q^2$ is of the same order of $P_\perp^2$ to simplify the calculation. In general, this problem is harder than the Higgs boson production and heavy quark pair production cases.
Nevertheless, we find that there are a lot similarities between this calculation and the heavy quark pair productions in DIS, and use similar techniques to obtain the Sudakov double logarithms. In addition, we find that all the discussion in Sec.~\ref{hqp} on the color factor still holds. In addition, the same sets of real and virtual graphs contribute to the Sudakov double logarithm. 

There are two differences between the dijet process and the heavy quark pair production. The first difference is quite obvious. In dijet production, the largest scale is now the jet transverse momentum $P_\perp$ while the mass the quarks are taken to be zero. 
The other difference comes from the jet cone singularity. In this section, we have to deal with the jet cone singularities when the emitted gluon is collinear to the final state quark or antiquark, while in the heavy quark pair production, this type of singularity is absent. 

The LO calculation for dijet production in DIS is very similar to the one for the heavy quark pair productions as shown in Eq.~(\ref{xsdis}). One just needs to set the quark mass to zero and take the transverse momenta $k_{\perp i}$ to be the large scale in the process. The LO cross section can be written either in coordinate space or in momentum space which are both available in Ref.~\cite{Dominguez:2011wm}.

For the one-loop correction to the Born amplitude, the sum of the real diagrams as illustrated in Fig.~\ref{sudr} is proportional to
\begin{eqnarray}
\frac{d\sigma}{d^3k_1d^3k_2d^3k_3}&\propto&\alpha_{em} \alpha_s N_c\int \frac{\textrm{d}^2x_1\textrm{d}^2x_2\textrm{d}^2x_3 \textrm{d}^2x^\prime_1\textrm{d}^2x^\prime_2\textrm{d}^2x^\prime_3}{(2\pi)^{12}} e^{-ik_1\cdot(x_1-x_1^\prime)}e^{-ik_2\cdot(x_2-x_2^\prime)}e^{-ik_3\cdot(x_3-x_3^\prime)}   \nonumber \\
&& \times \psi_{\gamma^\ast \to q\bar q}(\alpha , x_1-v) \psi^{\ast}_{\gamma^\ast \to q\bar q}(\alpha, x_1^\prime-v^\prime) \psi_{q \to q g}(\xi, x_2-x_3) \psi^{\ast}_{q\to qg}(\xi, x_2^\prime-x^\prime_3)  \nonumber \\
&& \times \left[1+S^{(4)}_{x_g}(x_1,v ; v^{\prime },x_1^{\prime})
-S^{(2)}_{x_g}(x_1,v)-S^{(2)}_{x_g}(v^{\prime },x_1^{\prime })\right] , \label{djreal}
\end{eqnarray}
where $\psi_{\gamma^\ast \to q\bar q}$ represents the initial photon splitting function into quark $q$ (at $x_1$) and antiquark $\bar q$ (at $x_2$) while $\psi_{q\to qg}$ stands for the second splitting function from the quark into a quark and a gluon (at $x_3$). The full definitions for $\psi_{\gamma^\ast \to q\bar q}$ and $\psi_{q\to qg}$ can be found in Ref.~\cite{Dominguez:2011wm}. With the definition $v=(1-\xi) x_2+\xi x_3$ as the coordinate of the quark before the splitting occurs, it is straightforward to check that the above coordinate expression can reproduce the momentum space light-cone denominators. By taking the dijet correlation limit, one can reduce the above scattering amplitudes into the WW form as 
\begin{equation}
-u_iu'_j\frac{1}{N_c}\langle\text{Tr}\left[\partial_iU(w)\right]U^\dagger(w')\left[\partial_jU(w')\right]
U^\dagger(w)\rangle_{x_g}\ ,
\end{equation}
where $u=x_1-v$ and $w=x_1\alpha+v(1-\alpha)$ which stand for the distance between the $q\bar q$ pair and center of mass coordinates of the $q\bar q$ pair right after the photon splitting, respectively. 

Also, one should integrate over the phase space of the radiated gluon ($\xi, k_{\perp 3}$) which sets $x_3=x_3^\prime$. By using the Fourier transform $$\psi_{\gamma^\ast \to q\bar q} (\alpha, u) =\int d^2 l_\perp \psi_{\gamma^\ast \to q\bar q} (\alpha, l_\perp) e^{-il_\perp\cdot u}$$ and $$u^i \psi_{\gamma^\ast \to q\bar q} (\alpha, u) =\int d^2 l_\perp e^{-il_\perp\cdot u} (-i)\partial^i_{l_\perp} \psi_{\gamma^\ast \to q\bar q} (\alpha,l_\perp),$$ together with the approximation $w\simeq x_1$ and $w'\simeq x_1'$ in the $\xi\to 0$ limit (this means that the radiated gluon has very little longitudinal momentum), one can integrate over $x_3$, $x_2$ and $x_2^\prime$ and find that the real contribution in Eq.~(\ref{djreal}) eventually factorizes into
\begin{equation}
\sigma_{\textrm{LO}} \times  4\alpha_s N_c \int_{l_\perp^2/s}^1 \frac{\textrm{d} \xi}{\xi}  \int \frac{\textrm{d} ^2l_\perp}{(2\pi)^2} \frac{1}{(l_\perp-\xi P_\perp)^2} e^{-il_\perp \cdot R_\perp} , \label{realdj}
\end{equation}
where $R_\perp \simeq w-w'\simeq x_1-x_1^\prime \sim 1/q_\perp$ is large. Here $P_\perp=k_{\perp 2}$ (or $k_{\perp 1}$ depending on from which the gluon is radiated) stands for the large jet transverse momentum which is much larger than $q_\perp$. The subtle part of the above derivation is to take $\xi \to 0$ limit while keep $\xi P_\perp$ finite. In Eq.~(\ref{realdj}), we have included the other half of real contributions in which the gluon is first emitted from the quark. 

Now the remaining work is to evaluate the integrals in Eq.~(\ref{realdj}) and extract the Sudakov factor. Although the direct evaluation seems difficult, one can first average over the azimuthal angle of the $P_\perp$, since one has the freedom to choose the orientation of the leading jet as the azimuthal angle $\phi =0$ reference in a measurement. Using the identity 
\begin{equation}
\frac{1}{2\pi} \int_0^{2\pi} d \theta \frac{1}{1+a \cos \theta} =\frac{1}{\sqrt{1-a^2}},  \quad \textrm{with} \quad a<1,
\end{equation}
one can cast the above integral into
\begin{equation}
\int_{l_\perp^2/s}^1 \frac{\textrm{d} \xi}{\xi}  \int \frac{\textrm{d} ^2l_\perp}{(2\pi)^2} \frac{1}{|l_\perp^2-\xi^2 P_\perp^2|} e^{-il_\perp \cdot R_\perp}.
\end{equation}

Clearly the above integration has a collinear singularity at $\xi=\frac{l_\perp}{P_\perp}$ which is expected since this comes from the region where the radiated gluon is collinear to the quark. Let us just regularize this collinear singularity by putting a cutoff in the $\xi$ integral which gives
\begin{equation}
\int_{l_\perp^2/s}^{\frac{l_\perp}{P_\perp} (1-\delta)} \frac{\textrm{d} \xi}{\xi} \frac{1}{l_\perp^2-\xi^2 P_\perp^2}-\int_{\frac{l_\perp}{P_\perp} (1+\delta)}^{1}\frac{\textrm{d} \xi}{\xi}  \frac{1}{l_\perp^2-\xi^2 P_\perp^2}\simeq \frac{1}{l_\perp^2} \left[\ln\frac{1}{x_g}+\frac{1}{2}\ln\frac{1}{4\delta^2}+\frac{1}{2}\ln\frac{P_\perp^2}{l_\perp^2}\right],
\end{equation}
where $\delta \ll 1$ should depend the angular resolution of the jet measurement. The above results contain three terms which correspond to three kinds of different physics. 

The first term $\ln\frac{1}{x_g}$ comes from the light-cone singularity of the problem and gives the small-$x$ resummation. After integrating over $l_\perp$ with lower cutoff $\bar\epsilon /R_\perp$ which depends on the transverse energy resolution of the measurement, the second term eventually yields $\frac{\alpha_s N_c}{\pi} \ln\frac{1}{\bar\epsilon} \ln \frac{1}{\delta^2}$ (This term actually corresponds to the three body final states which should be subtracted from the dijet cross section. Also the color factor should be $C_F$ instead of $N_c/2$ since the right diagram in Fig.~\ref{sudr} does not have such collinear singularity.) Combining the divergent contribution from the region that we have removed from the above integral $[\frac{l_\perp}{P_\perp}(1-\delta),\frac{l_\perp}{P_\perp}(1+\delta)]$ together with the virtual diagram contribution which is also divergent, we should get $-\frac{2\alpha_s C_F}{\pi} \ln\frac{1}{\bar \epsilon} \ln \frac{1}{\delta^2}$ which is the NLO correction of the dijet cross section. This result differs from the three body final state result by a minus sign simply due to probability conservation. At last, it is straightforward to see that the last term yields 
\begin{eqnarray}
&&2 \alpha_s N_c \int \frac{\textrm{d} ^2l_\perp}{(2\pi)^2}\frac{1}{l_\perp^2}\ln\frac{P_\perp^2}{l_\perp^2} e^{-i l_\perp\cdot R_\perp} \notag \\
&=& \frac{\alpha_s N_c}{2 \pi} \left[\frac{1}{\epsilon^2} - \frac{1}{\epsilon} \ln \frac{P_\perp^2}{\mu^2}+\frac{1}{2} \ln^2 \frac{P_\perp^2}{\mu^2}-\frac{1}{2} \ln^2 \frac{P_\perp^2R^2_\perp}{c_0^2}+\cdots \right]. \label{realsud} 
\end{eqnarray}

Similarly, for virtual diagrams shown in Fig.~\ref{noww} and Fig.~\ref{self}, one can do the similar analysis and find out that only the four diagrams in Fig~\ref{noww} contribute to the Sudakov double logarithm which can be approximately cast into
\begin{equation}
\sigma_{\textrm{LO}} \times  4 \alpha_s N_c \int_{0}^1 \frac{\textrm{d} \xi}{\xi}  \int \frac{\textrm{d} ^2l_\perp}{(2\pi)^2} \frac{l_\perp\cdot (l_\perp+\xi P_\perp)}{l_\perp^2 (l_\perp+\xi P_\perp)^2} 
. \label{virtualdj}
\end{equation}
Following the standard procedure, we should remove the rapidity divergence by subtracting a term like $1/l_\perp^2$ from the above virtual contribution which gives
\begin{equation}
-\sigma_{\textrm{LO}} \times  4 \alpha_s N_c \int_{0}^1 \frac{\textrm{d} \xi}{\xi}  \int \frac{\textrm{d} ^2l_\perp}{(2\pi)^2} \frac{\xi P_\perp\cdot (l_\perp+\xi P_\perp)}{l_\perp^2 (l_\perp+\xi P_\perp)^2} 
. \label{virtualdj2}
\end{equation}
Next, we should average over the azimuthal orientation of $P_\perp$ which simplifies the calculation dramatically. The angular average turns $\frac{\xi P_\perp\cdot (l_\perp+\xi P_\perp)}{l_\perp^2 (l_\perp+\xi P_\perp)^2} $ into $\frac{1}{l_\perp^2}$ with the condition that $l_\perp^2<\xi^2P_\perp^2$. Now it is very simple to integrate over $\xi$ for the virtual contribution and obtain
\begin{equation}
-4 \alpha_s N_c  \int \frac{\textrm{d} ^2l_\perp}{(2\pi)^2}\frac{1}{l_\perp^2 } \int_{\frac{l_\perp}{P_\perp}}^1 \frac{\textrm{d} \xi}{\xi}=-2\alpha_s N_c \left. \int \frac{\textrm{d} ^2l_\perp}{(2\pi)^2}\frac{1}{l_\perp^2 }\ln \frac{P_\perp^2}{l_\perp^2}\right|_{|l_\perp|<P_\perp}.\label{virtualdj33}
\end{equation}
Using the dimensional regularization in $\overline{\textrm{MS}}$ scheme with upper cutoff $P_\perp$, the integration over $l_\perp$ gives
\begin{equation}
-\frac{\alpha_s N_c}{2 \pi} \left[\frac{1}{\epsilon^2} - \frac{1}{\epsilon} \ln \frac{P_\perp^2}{\mu^2}+\frac{1}{2} \ln^2 \frac{P_\perp^2}{\mu^2}+\cdots \right].\label{virtualdj3} 
\end{equation}
At the end of the day, by combining the real and virtual contributions together as shown in Eq.~(\ref{realsud}) and Eq.~(\ref{virtualdj3}), respectively, one can find that the divergences cancel and obtain the following Sudakov double logarithm 
\begin{equation}
S_{\textrm{Sud}} (R_\perp, P_\perp)=\frac{\alpha_s N_c}{4 \pi} \ln^2 \frac{P_\perp^2 R^2_\perp}{c_0^2}+\cdots,
\end{equation} 
with $R_\perp \simeq 1/q_\perp \gg 1/P_\perp$. The effective color factor $\mathcal{C}$ of the Sudakov double logarithm for DIS dijet processes is found to be $\frac{N_c}{2}$, which is the same as the heavy quark pair productions in DIS. It is natural to interpret this effective color factor as the product of the $N_c$ factor and the $\frac{1}{2}$ factor. The $N_c$ factor simply comes from the observation that the final state back-to-back $q\bar q$ pair acts like a gluon, which effectively yields the color factor $N_c$. On the other hand, the factor of $\frac{1}{2}$ is more intricate. This factor comes from the dynamics that the additional gluon is radiated from the final state quarks with large transverse momentum $P_\perp$. The similar situation also occurs in the heavy quark pair production, where the gluon is radiated from quarks with large masses. It is interesting to note that the $\xi$-dependent coefficient (either $(1-\xi)$ or $\xi(1-\xi)$) in front of the Higgs boson  mass square is always linear when it vanishes. As a comparison, we find that such coefficient for the heavy quark pair or dijet productions in DIS is always quadratic in terms of function of $\xi$ when it vanishes. This brings this interesting factor of $\frac{1}{2}$ from the integration over $\xi$.

Therefore, including the Sudakov factor obtained from the one loop calculation, the cross section for dijet productions in DIS becomes
\begin{equation}
\frac{d\sigma^{\gamma^\ast A \to q\bar q X}}{dy_1 dy_2 d^2 P_\perp d^2q_\perp} =\sigma_0(P_\perp^2, Q^2, z) \int d^2 r_\perp \int d^2 b_\perp e^{-iq_\perp \cdot b_\perp} S^{WW} (x_\perp, x_\perp^\prime) e^ {-S_{\textrm{Sud}} (b_\perp, P_\perp)},
\end{equation}
where $\sigma_0(P_\perp^2, Q^2, z)$ is calculated from the Born diagrams\cite{Dominguez:2011wm}. Here we define $b_\perp =x_\perp-x_\perp^\prime$ and $r_\perp=\frac{1}{2} (x_\perp+x_\perp^\prime)$. If one neglects the impact parameter $r_\perp$ dependence in the WW corrlator $S^{WW} (x_\perp, x_\perp^\prime)$, then $d^2 r_\perp$ integration becomes trivial and just yields the area of the target $S_\perp$.

\section{General Structure of Soft Gluon Radiation for Two Particle Production}

Two-particle production in $pA$ collisions come from the hard partonic $2\to 2$ 
processes. Before we go into the detailed derivation in the small-$x$ formalism,
in this section, we discuss the general structure in these processes in the collinear
factorization approach calculations, i.e., in the dilute-dilute scattering. Again, we analyze 
one gluon radiation contribution to the leading order Born $2\to 2$ diagrams, in 
particular, focusing on the soft gluon contribution to the leading double logarithms
in these hard processes. 
In Ref.~\cite{Qiu:2007ey}, the collinear gluon radiation in dijet production has been analyzed, and it
was found that the gluon radiation at one-loop order can be written into the collinear
splitting function of the incoming parton distributions. In this paper, we will study the 
gluon radiation in the soft momentum region, and focus on the leading double logarithmic
contribution. In addition, we work in the center of mass frame of the incoming two
particles. The soft gluon momentum can be parameterized by
\begin{equation}
k_g=\alpha_g p_1+\beta_g p_2+k_{g\perp} \ , \label{kg}
\end{equation}
where $\alpha$ and $\beta$ are momentum fractions of the incoming partons
carried by the radiated gluon. $p_1$ and $p_2$ are the four momentum of the incoming nucleon and the incoming nucleus, respectively. Soft gluon radiation corresponds to the kinematics:
$k_{g\perp}\sim \alpha_g p_1\sim \beta_g p_2$. Therefore, we will take the limit
of $\alpha_g,\beta_g\ll 1$. 

All of the results in this section has been well studied in the literature, in 
particular, in a series papers by Sterman et al., in the context of
threshold resummation~\cite{Kidonakis:1997gm}. Although the detailed resummation formalism 
is different, the soft gluon radiation share some common features between 
transverse momentum resummation and the threshold resummation. 
The soft gluon radiation is an example.

\subsection{Eikonal Approximation}

\begin{figure}[tbp]
\begin{center}
\includegraphics[width=12cm]{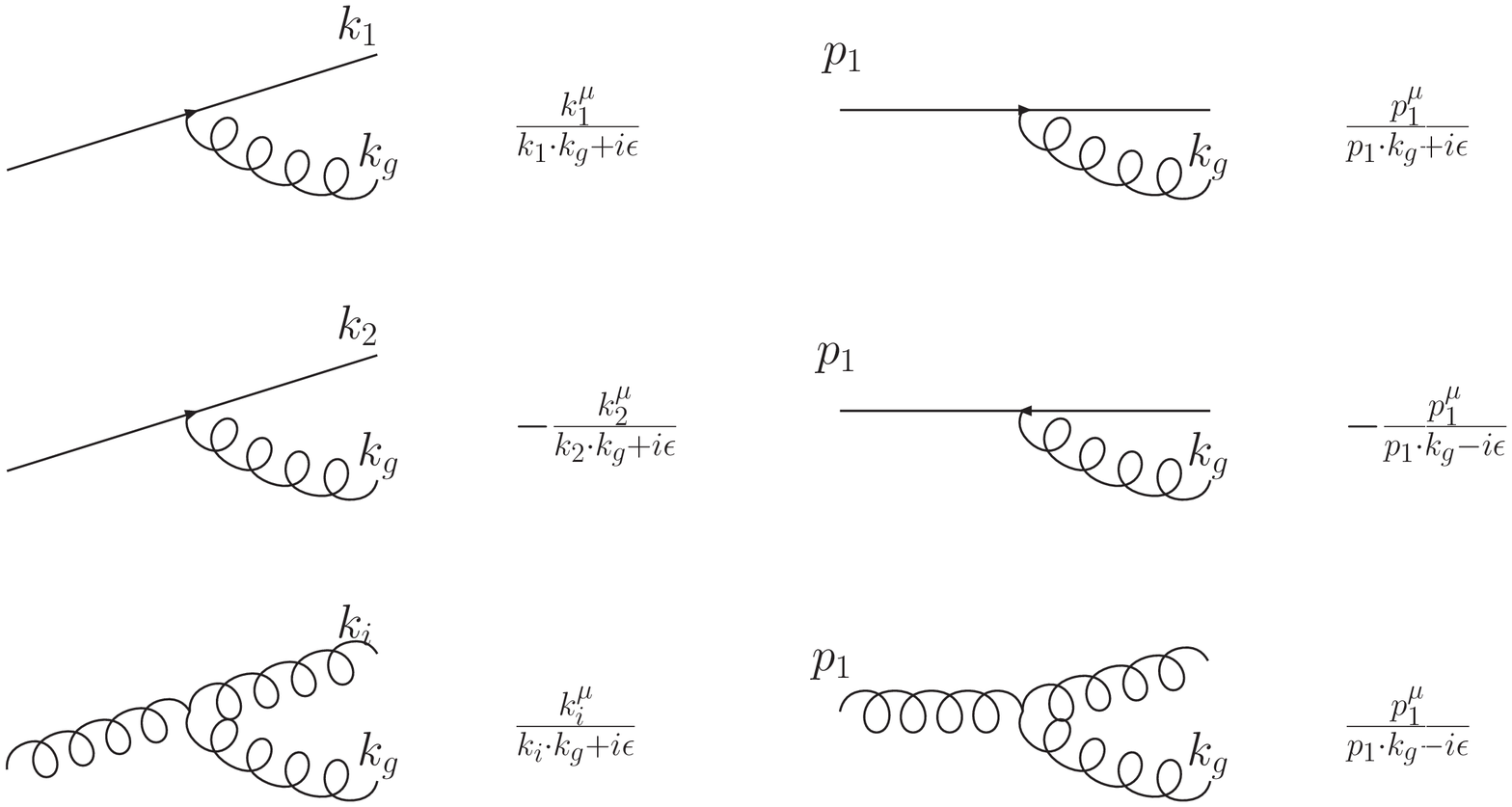}
\end{center}
\caption[*]{Eikonal approximation in the soft gluon radiations.}
\label{eikonal}
\end{figure}

For soft gluon radiations, we can apply
the leading power expansion and derive the dominant contribution by 
the Eikonal approximation. We listed these rules in Fig.~\ref{eikonal}.
For our convenience, we also choose the physical polarization
of the radiated gluon along $p_2$: $\epsilon(k_g)\cdot p_2=0$.
This will simplify the derivation, and in particular, we do not need
to consider the gluon radiation from the gluon line from the nucleus,
which is consistent with the small-$x$ calculations.

For outgoing quark line, we have
\begin{equation}
\frac{2k_1^\mu}{2k_1\cdot k_g+i\epsilon} \ ,
\end{equation}
where $k_1$ represents the momentum of the outgoing quark. For the antiquark, we have
\begin{equation}
-\frac{2k_2^\mu}{2k_2\cdot k_g+i\epsilon} \ ,
\end{equation}
where $k_2$ denotes the momentum of the outgoing antiquark. We notice that the above also hold
for massive quark lines. The only difference is that $k_1^2=m_q^2$,
instead of $k_1^2=0$ for massless case.
For incoming gluon line, we have,
\begin{equation}
\frac{2p_1^\mu}{2p_1\cdot k_g-i\epsilon} \ ,
\end{equation}
where $p_1$ represents the momentum for the incoming gluon For incoming quark line,
\begin{equation}
\frac{2p_1^\mu}{2p_1\cdot k_g+i\epsilon} \ .
\end{equation}
For outgoing gluon line,
\begin{equation}
\frac{2k_2^\mu}{2k_2\cdot k_g+i\epsilon} \ .
\end{equation}
In the above we only list the momentum dependence of these
Eikonal approximation, the associated color factor shall be worked out 
accordingly.

Once we have the above eikonal approximation of the soft gluon radiation in
the scattering amplitude, we can calculate the contribution to the leading double logarithmic 
terms for the soft gluon radiation. For example, from the amplitude squared
of the initial state gluon radiation,
\begin{eqnarray}
\frac{2p_1^\mu}{2p_1\cdot k_g-i\epsilon}\frac{2p_1^\nu}{2p_1\cdot k_g+i\epsilon}
\times \left(-g_{\mu\nu}+\frac{k_g^\mu p_2^\nu+k_g^\nu p_2^\mu}{k_g\cdot p_2}\right) 
=\frac{2p_1\cdot p_2}{p_1\cdot k_gp_2\cdot k_g}=\frac{4}{k_{g\perp}^2} \ ,
\end{eqnarray}
where we have applied the physical polarization for the radiated gluon and the on-shell condition
of Eq.~(\ref{kg}) to derive the last equation.
Integrating out the phase space of the above result will lead to the following 
double logarithm,
\begin{eqnarray}
\int \frac{d^3k_g}{(2\pi)^32E_{k_g}} \frac{4}{k_{g\perp}^2}e^{ik_{g\perp}\cdot R_\perp}
=\frac{1}{\pi}\int\frac{d^2k_\perp}{(2\pi)^2}\frac{1}{k_{g\perp}^2}\ln\frac{P_\perp^2}{k_{g\perp}^2}e^{ik_{g\perp}\cdot R_\perp} \ ,
\end{eqnarray}
where $P_\perp$ represents the hard momentum scale in the two-particle
production processes.
Performing the dimensional regulation, we will obtain,
\begin{equation}
\frac{1}{4\pi^2}\left[\frac{1}{\epsilon^2}-\frac{1}{\epsilon}\ln\frac{P_\perp^2}{\mu^2}-\frac{1}{2}\left(\ln\frac{P_\perp^2R_\perp^2}{c_0^2}\right)^2
+\cdots \right]\ , \label{initialr}
\end{equation}
where 
we have omitted terms which are not important for this part of the discussions. 
The $1/\epsilon$ divergent terms will be cancelled by the virtual diagrams, and
we are left with a double logarithmic term. In particular,
we will identify the above result as the leading double logarithmic 
term in the soft gluon radiation in the hard processes.

Similarly, for the final state radiation,
\begin{eqnarray}
\frac{2k_1^\mu}{2k_1\cdot k_g-i\epsilon}\frac{2k_1^\nu}{2k_1\cdot k_g+i\epsilon}
\times \left(-g_{\mu\nu}+\frac{k_g^\mu p_2^\nu+k_g^\nu p_2^\mu}{k_g\cdot p_2}\right) 
=\frac{2k_1\cdot p_2}{k_1\cdot k_gp_2\cdot k_g} \ ,
\end{eqnarray}
where we have assumed the massless case. However, as shown in the calculations for
dijet production in DIS process in the last section, 
the above contributes only half of what in Eq.~(\ref{initialr}). To summarize, the momentum integrals
have the following counting for the size of the leading double logarithms,
\begin{eqnarray}
\frac{2p_1\cdot p_2}{p_1\cdot k_gp_2\cdot k_g}\Rightarrow 1\ ,\nonumber\\
\frac{2k_1\cdot p_2}{k_1\cdot k_gp_2\cdot k_g}\Rightarrow \frac{1}{2}\ ,\nonumber\\
\frac{2k_2\cdot p_2}{k_2\cdot k_gp_2\cdot k_g}\Rightarrow \frac{1}{2}\ ,\nonumber\\
\frac{2k_1\cdot k_2}{k_1\cdot k_gk_2\cdot k_g}\Rightarrow 0\ .
\end{eqnarray}
The last equation shows that it does not contribute to the leading double logarithmic
terms in the soft gluon radiation, although it may contribute to the single logarithm.

We can also translate the above rules to the amplitude squared contributions. 
For example, the amplitude squared of $p_1^\mu$ contributes to a factor of 1, whereas
that of $k_1^\mu$ to a factor of $1/2$. We list the rules for these calculations,
\begin{eqnarray}
&&p_1^\mu p_1^\nu\Rightarrow 1 \ ,\nonumber\\
&&k_1^\mu k_1^\nu\Rightarrow \frac{1}{2} \ ,\nonumber\\
&&k_2^\mu k_2^\nu\Rightarrow \frac{1}{2} \ ,\nonumber\\
&&p_1^\mu k_1^\nu\Rightarrow \frac{1}{2} \ ,\nonumber\\
&&p_1^\mu k_2^\nu\Rightarrow \frac{1}{2} \ ,\nonumber\\
&&k_1^\mu k_2^\nu\Rightarrow \frac{1}{2} \ .
\end{eqnarray}
Therefore, in the following, we will show how to construct the leading double logarithmic
contributions from soft gluon radiations in the hard processes, by calculating all the soft
gluon radiation amplitude and their interferences. The technique is to sum all the 
leading terms, together with the correct color factors.

\subsection{Reproduce the Leading Double Logarithms in Drell-Yan and Higgs boson production
processes}

Applying the above method, we can easily reproduce the leading double logarithmic terms
in the Drell-Yan and Higgs boson production in hadronic collisions. For Drell-Yan
process, we only have initial state gluon radiation,
\begin{equation}
A_0\left(ig_s\frac{2p_1^\mu}{2p_1\cdot k_g+i\epsilon}\bar v T^a u\right) \ ,
\end{equation}
where $A_0$ represents the leading order amplitude containing spinor structure,
$a$ for the color index for the radiated gluon. We can easily calculate the amplitude
squared as 
\begin{eqnarray}
{\cal M}^2&=&|\overline{\cal M}_0|^2 g_s^2\left[C_F\left(\frac{2p_1\cdot p_2}{p_1\cdot k_gp_2\cdot k_g}\right)\right] \ ,\nonumber\\
 &=&|\overline{\cal M}_0|^2\left(\alpha_s4\pi C_F\right)\frac{4}{k_{g\perp}^2} \ .
\end{eqnarray}
Together with the Fourier transform result in the last subsection, we will find out the
soft gluon radiation at one-loop order contributes to a factor,
\begin{equation}
-\frac{\alpha_sC_F}{2\pi}\ln^2\left(\frac{Q^2R_\perp^2}{c_0^2}\right) \ ,
\end{equation}
where $Q$ is the invariant mass of the lepton pair.
This is the famous double logarithmic term at one-loop order for Drell-Yan processes.

Similarly, for Higgs boson production, we will find out the one-loop soft
gluon radiation contrites,
\begin{equation}
-\frac{\alpha_sC_A}{2\pi}\ln^2\left(\frac{M_H^2R_\perp^2}{c_0^2}\right) \ ,
\end{equation}
where $M_H$ represents the Higgs boson mass. 

To achieve the similar results for the hard processes discussed in this paper
is the main goal of our calculations. In the following, we will first derive the
results in the collinear factorization approach, and summarize the basic
counting rule for the leading double logarithmic terms. After that, we will extend
the discussions to the small-$x$ formalism, and we will demonstrate that the
counting results remain the same. This tells us that the Sudakov resummation
can be performed consistently with the small-$x$ resummation, at the leading
double logarithmic level. Beyond that, we conjecture it shall follow, and may have 
more complicated form.

\subsection{$\gamma g\to q\bar q$}
\begin{figure}[tbp]
\begin{center}
\includegraphics[width=9cm]{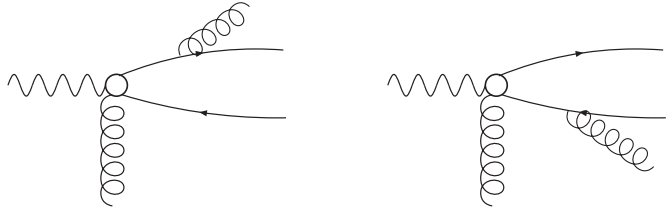}
\end{center}
\caption[*]{Soft gluon radiation in $q\bar q$ production in DIS process.}
\label{fig:disqq}
\end{figure}

There is only one color structure for the Born diagram of this process, $\bar uT^a v$.
So, we can write down the Born amplitude as
\begin{equation}
A_0\bar uT^a v \ ,
\end{equation}
where $A_0$ represents the spinor structure of the amplitude, and $a$ for
the color index for the incoming gluon. With one soft gluon radiation,
we find out that
\begin{equation}
A_0\left(ig_s\frac{2k_1^\mu}{2k_1\cdot k_g+i\epsilon}\bar u T^bT^av
+ig_s\frac{2k_2^\mu}{2k_2\cdot k_g+i\epsilon}\bar u T^aT^bv\right) \ ,
\end{equation}
as shown in Fig.~\ref{fig:disqq}.
The amplitude squared of the soft gluon radiation can be written as,
\begin{eqnarray}
{\cal M}^2=|\overline{\cal M}_0|^2 g_s^2\left[\frac{N_c}{2}\left(\frac{2k_1\cdot p_2}{k_1\cdot k_gp_2\cdot k_g}
+\frac{2k_2\cdot p_2}{k_2\cdot k_g p_2\cdot k_g}\right)+\left(-\frac{1}{2N_c}\right)\frac{2k_1\cdot k_2}{k_1\cdot k_gk_2\cdot k_g}\right] \ ,
\end{eqnarray}
where ${\cal M}_0$ represents the leading Born amplitude. As we discussed above, 
the three terms contribute differently in the leading double logarithmic order.
In particular, the last term does not contribute to the double logarithms, whereas the
first two terms contribute equally one half of the usual double logarithms.
The final result is,
\begin{equation}
-\frac{\alpha_sN_c}{4\pi}\ln^2\left(\frac{P_\perp^2R_\perp^2}{c_0^2}\right) \ .
\end{equation}
It is half of the double logarithms in the Higgs boson production.

This seems to suggest a simple counting rule for the leading double logarithms. 
Every incoming parton contributes to one-half times its associated Casimir color factor, i.e.,
for quark it is half of $C_F$, for gluon it is half of $C_A$. 
This can be understood that all the leading
double logs come from the initial state parton distributions, which can be 
summarized as the counting rule.

\subsection{$qg\to q\gamma$}

\begin{figure}[tbp]
\begin{center}
\includegraphics[width=9cm]{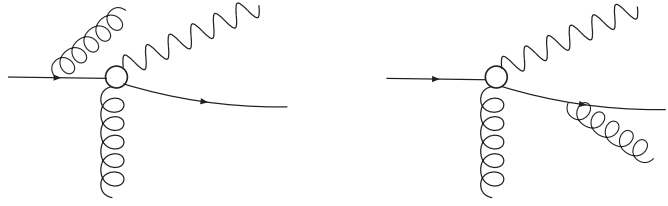}
\end{center}
\caption[*]{Soft gluon radiation in photon-quark production.}
\label{fig:qphoton}
\end{figure}

For this process, again, we only have one color structure in the leading Born
diagram,
\begin{equation}
A_0\bar uT^a u \ ,
\end{equation}
and the soft gluon radiation at one-loop order of Fig.~\ref{fig:qphoton} leads to,
\begin{equation}
A_0\left(ig_s\frac{2p_1^\mu}{2p_1\cdot k_g+i\epsilon}\bar u T^aT^b u
+ig_s\frac{2k_1^\mu}{2k_1\cdot k_g+i\epsilon}\bar u T^bT^a u\right) \ .
\end{equation}
The amplitude squared can be calculated,
\begin{eqnarray}
{\cal M}^2&=&|\overline{\cal M}_0|^2 g_s^2\left[
\left(C_F+\frac{1}{2N_c}\right)\frac{2p_1\cdot p_2}{p_1\cdot k_gp_2\cdot k_g}
+C_F\frac{2k_1\cdot p_2}{k_1\cdot k_gp_2\cdot k_g}\right.\nonumber\\
&&\left.+\left(-\frac{1}{2N_c}\right)\left(\frac{2k_1\cdot p_2}{k_1\cdot k_gp_2\cdot k_g}-
\frac{2k_2\cdot p_2}{k_2\cdot k_g p_2\cdot k_g}\right)\right] \ .
\end{eqnarray}
Following the same analysis of the above, we will find out that 
the first term contributes to a leading double log with coefficient 
$C_A/2$, the second term with half of $C_F$, and the third term 
cancels out. The final result takes the form,
\begin{equation}
-\frac{\alpha_s}{2\pi}\left(\frac{C_A+C_F}{2}\right)\ln^2\left(\frac{P_\perp^2R_\perp^2}{c_0^2}\right) \ .
\end{equation}
Again, it is consistent with the counting rule.

\subsection{$qg\to qg$}

\begin{figure}[tbp]
\begin{center}
\includegraphics[width=12cm]{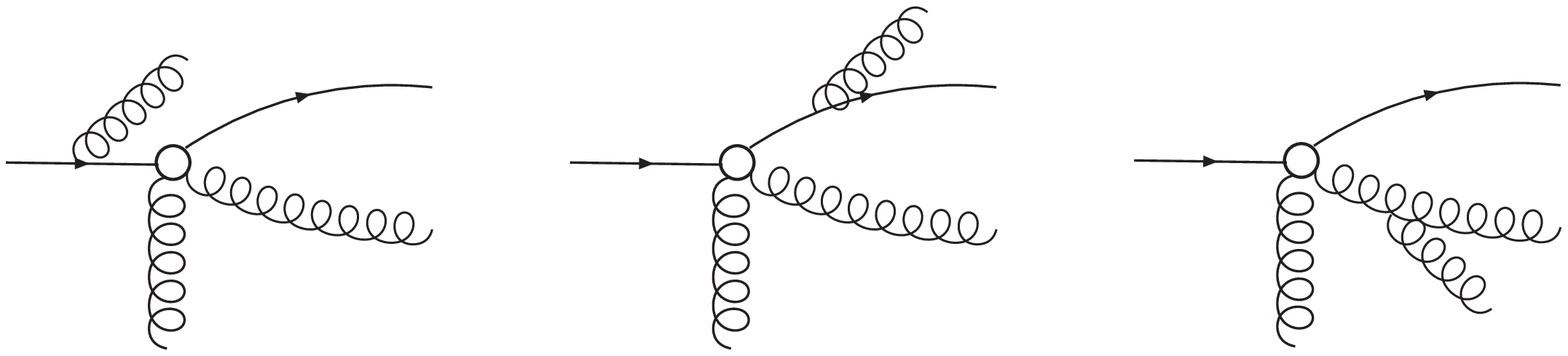}
\end{center}
\caption[*]{Soft gluon radiation in $qg\to qg$ process.}
\label{fig:qg}
\end{figure}

In this process, we have two different color structure at the leading Born
level,
\begin{equation}
A_1\bar u T^aT^bu+A_2\bar uT^bT^au \ ,
\end{equation}
where $a$ and $b$ represent the color indexes for the incoming
and outgoing gluons.
The leading order amplitude squared reads as,
\begin{equation}
|A_0|^2=C_F\left(A_1^2+A_2^2\right)+\left(-\frac{1}{2N_c}\right)2A_1A_2^* \ .
\end{equation}
In the above, we took the straightforward decomposition. In general, 
the scattering amplitude for the hard processes should be decomposed into
orthogonal color bases as that in Ref.~\cite{Kidonakis:1997gm}. 
In this paper, since we focus on the leading double logarithmic contributions,
the above decomposition is good enough.
At one-loop order, we will have gluon radiation as shown in Fig.~\ref{fig:qg}, which come
from the initial quark, the final state quark and gluon. We can write down
the amplitude as
\begin{eqnarray}
&&\frac{2k_1^\mu}{2k_1\cdot k_g}\left[A_1\bar uT^cT^aT^bu+A_2\bar uT^cT^bT^a u\right] \nonumber\\
&+&\frac{-2p_1^\mu}{2p_1\cdot k_g}\left[A_1\bar uT^aT^bT^cu+A_2\bar uT^bT^aT^c u\right] \nonumber\\
&+&\frac{2k_2^\mu}{2k_2\cdot k_g}\left(-if_{cbd}\right)\left[A_1\bar uT^aT^du+A_2\bar uT^dT^a u\right]  \ ,
\end{eqnarray} 
where $c$ represents the color index for the radiated gluon. 

Let us first work out the color factors for different terms,
\begin{eqnarray}
&&k_1^\mu k_1^\nu\Rightarrow C_F |M_0|^2\ ,\nonumber\\
&&k_2^\mu k_2^\nu\Rightarrow C_A |M_0|^2\ ,\nonumber\\
&&p_1^\mu p_1^\nu\Rightarrow C_F |M_0|^2\ ,\nonumber\\
&&k_1^\mu k_2^\nu\Rightarrow \left[\left(-C_F\frac{N_c}{2}\right)A_1^2+\frac{1}{4}\left(A_2^2+2A_1A_2^*\right)\right]\ ,\nonumber\\
&&k_1^\mu p_1^\nu\Rightarrow \left[\left(-\frac{1}{2N_c}\right)\left(-\frac{1}{2N_c}\right)\left(A_1+A_2\right)^2+\frac{1}{4}2A_1A_2^*\right]\ ,\nonumber\\
&&k_2^\mu p_1^\nu\Rightarrow \left[\left(C_F\frac{N_c}{2}\right)A_2^2-\frac{1}{4}\left(A_1^2+2A_1A_2^*\right)\right]\ .
\end{eqnarray}
Adding them together with the weight to the leading double logs, we have
\begin{eqnarray}
&&\left(C_F+\frac{C_F+C_A}{2}\right)|M_0|^2+\left[\left(-C_F\frac{N_c}{2}\right)A_1^2+\frac{1}{4}\left(A_2^2+2A_1A_2^*\right)\right]\nonumber\\
&&-\left[\left(-\frac{1}{2N_c}\right)\left(-\frac{1}{2N_c}\right)\left(A_1+A_2\right)^2+\frac{1}{4}2A_1A_2^*\right]
- \left[\left(C_F\frac{N_c}{2}\right)A_2^2-\frac{1}{4}\left(A_1^2+2A_1A_2^*\right)\right] \nonumber\\
&&=\left(C_F+\frac{C_F+C_A}{2}\right)|M_0|^2-C_F\left(C_F\left(A_1^2+A_2^2\right)+\left(-\frac{1}{2N_c}\right)2A_1A_2^*\right)\nonumber\\
&&=\frac{C_F+C_A}{2}|M_0|^2\ .
\end{eqnarray}
Therefore, at the leading double logarithmic level, the soft gluon radiation contributes to
\begin{equation}
-\frac{\alpha_s}{2\pi}\left(\frac{C_A+C_F}{2}\right)\ln^2\left(\frac{P_\perp^2R_\perp^2}{c_0^2}\right) \ .
\end{equation}

\subsection{$gg\to q\bar q$}
\begin{figure}[tbp]
\begin{center}
\includegraphics[width=12cm]{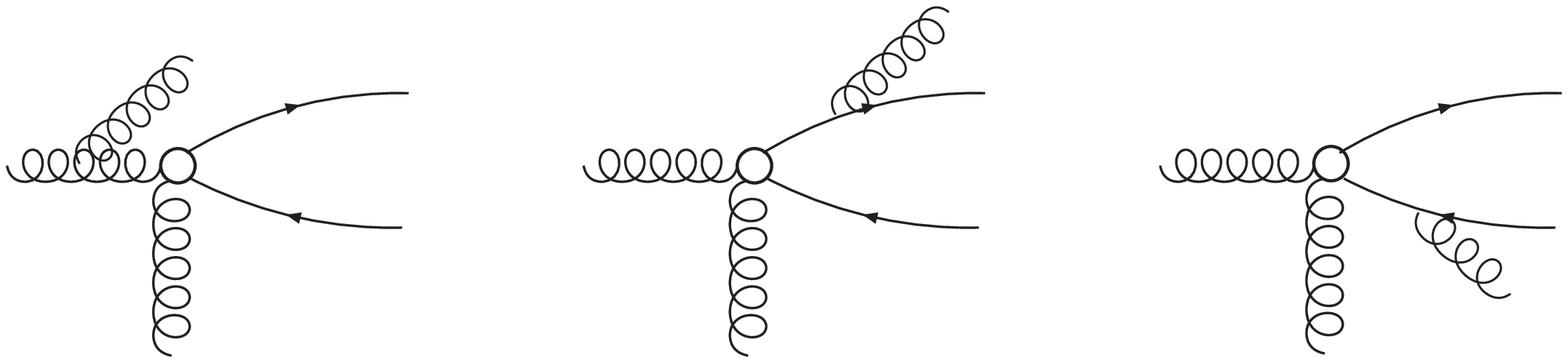}
\end{center}
\caption[*]{Soft gluon radiation in $gg\to q\bar q$ process.}
\label{fig:ggqq}
\end{figure}

Similarly, we have at the leading Born diagram,
\begin{equation}
A_1\bar u T^aT^bv+A_2\bar uT^bT^av \ ,
\end{equation}
with 
\begin{equation}
|A_0|^2=C_F\left(A_1^2+A_2^2\right)+\left(-\frac{1}{2N_c}\right)2A_1A_2^* \ .
\end{equation}
Soft gluon radiation of Fig.~\ref{fig:ggqq} contributes,
\begin{eqnarray}
&&\frac{2k_1^\mu}{2k_1\cdot k_g}\left[A_1\bar uT^cT^aT^bv+A_2\bar uT^cT^bT^a v\right] \nonumber\\
&+&\frac{-2k_2^\mu}{2k_2\cdot k_g}\left[A_1\bar uT^aT^bT^cv+A_2\bar uT^bT^aT^c v\right] \nonumber\\
&+&\frac{2p_1^\mu}{2p_1\cdot k_g}\left(-if_{cad}\right)\left[A_1\bar uT^dT^bv+A_2\bar uT^bT^d v\right]  \ ,
\end{eqnarray} 
where $c$ represents the color index for the radiated gluon. 
The color factors are similar to the above channel, and we have
\begin{eqnarray}
&&k_1^\mu k_1^\nu\Rightarrow C_F |M_0|^2\ ,\nonumber\\
&&k_2^\mu k_2^\nu\Rightarrow C_F |M_0|^2\ ,\nonumber\\
&&p_1^\mu p_1^\nu\Rightarrow C_A |M_0|^2\ ,\nonumber\\
&&k_1^\mu k_2^\nu\Rightarrow\left[\left(-\frac{1}{2N_c}\right)\left(-\frac{1}{2N_c}\right)\left(A_1+A_2\right)^2+\frac{1}{4}2A_1A_2^*\right]
 \ ,\nonumber\\
&&k_1^\mu p_1^\nu\Rightarrow  \left[\left(-C_F\frac{N_c}{2}\right)A_1^2+\frac{1}{4}\left(A_2^2+2A_1A_2^*\right)\right]\ ,\nonumber\\
&&k_2^\mu p_1^\nu\Rightarrow \left[\left(C_F\frac{N_c}{2}\right)A_2^2-\frac{1}{4}\left(A_1^2+2A_1A_2^*\right)\right]\ ,
\end{eqnarray}
Adding them together with the weight to the leading double logs, we have
\begin{eqnarray}
&&\left(C_A+\frac{C_F+C_F}{2}\right)|M_0|^2+\left[\left(-C_F\frac{N_c}{2}\right)A_1^2+\frac{1}{4}\left(A_2^2+2A_1A_2^*\right)\right]\nonumber\\
&&-\left[\left(-\frac{1}{2N_c}\right)\left(-\frac{1}{2N_c}\right)\left(A_1+A_2\right)^2+\frac{1}{4}2A_1A_2^*\right]
- \left[\left(C_F\frac{N_c}{2}\right)A_2^2-\frac{1}{4}\left(A_1^2+2A_1A_2^*\right)\right] \nonumber\\
&&=\left(C_A+\frac{C_F+C_F}{2}\right)|M_0|^2-C_F\left(C_F\left(A_1^2+A_2^2\right)+\left(-\frac{1}{2N_c}\right)2A_1A_2^*\right)\nonumber\\
&&=C_A|M_0|^2\ .
\end{eqnarray}
Finally, at the leading double logarithmic level, the soft gluon radiation contributes to
\begin{equation}
-\frac{\alpha_s}{2\pi}C_A\ln^2\left(\frac{P_\perp^2R_\perp^2}{c_0^2}\right) \ ,
\end{equation}
which is the same as that for Higgs boson production.

\subsection{$gg\to gg$}

\begin{figure}[tbp]
\begin{center}
\includegraphics[width=12cm]{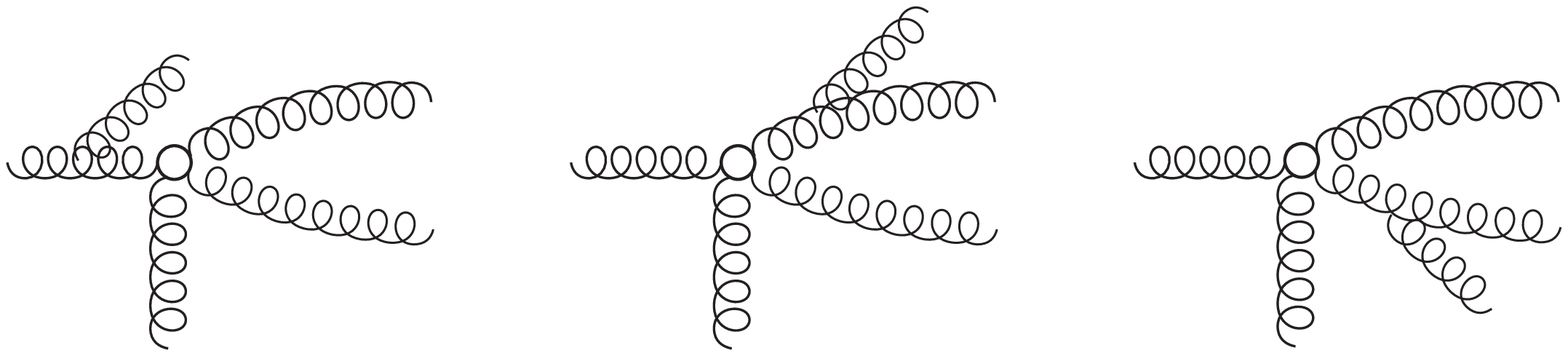}
\end{center}
\caption[*]{Soft gluon radiation in $gg\to gg$ process.}
\label{fig:gggg}
\end{figure}

For $gg\to gg$ channel, we can simply write down the following decomposition,
\begin{equation}
A_1f_{abe}f_{cde}+A_2f_{ace}f_{bde}+A_3f_{ade}f_{bce} \ ,
\end{equation}
where $a,b,c,d$ are color indices for the gluons.
The leading order amplitude squared can be written as
\begin{equation}
|M_0|^2=\left(A_1^2+A_2^2+A_3^2+A_1A_2^*-A_1A_3^*+A_2A_3^*\right) \ .
\end{equation}
One soft gluon radiation of Fig.~\ref{fig:gggg} can take the form,
\begin{eqnarray}
&&\frac{2k_1^\mu}{2k_1\cdot k_g}f_{gcf}\left[A_1f_{abe}f_{fde}+A_2f_{afe}f_{bde}+A_3f_{ade}f_{bfe}\right] \nonumber\\
&+&\frac{2k_2^\mu}{2k_2\cdot k_g}f_{gdf}\left[A_1f_{abe}f_{cfe}+A_2f_{ace}f_{bfe}+A_3f_{afe}f_{bce}\right] \nonumber\\
&+&\frac{2p_1^\mu}{2p_1\cdot k_g}f_{gaf}\left[A_1f_{fbe}f_{cde}+A_2f_{fce}f_{bde}+A_3f_{fde}f_{bce}\right] \ .
\end{eqnarray} 
The amplitude squared of the above radiation can be written as,
\begin{eqnarray}
&&k_1^\mu k_1^\nu\Rightarrow C_A |M_0|^2\ ,\nonumber\\
&&k_2^\mu k_2^\nu\Rightarrow C_A |M_0|^2\ ,\nonumber\\
&&p_1^\mu p_1^\nu\Rightarrow C_A |M_0|^2\ ,\nonumber\\
&&k_1^\mu k_2^\nu\Rightarrow\left[-\frac{N_c}{2}A_1^2-\frac{N_c}{4}\left(A_2^2+A_3^2+2A_1A_2^*-2A_1A_3^*\right)\right] \ ,\nonumber\\
&&k_1^\mu p_1^\nu\Rightarrow  \left[-\frac{N_c}{2}A_2^2-\frac{N_c}{4}\left(A_1^2+A_3^2+2A_1A_2^*+2A_2A_3^*\right)\right] \ ,\nonumber\\
&&k_2^\mu p_1^\nu\Rightarrow\left[-\frac{N_c}{2}A_3^2-\frac{N_c}{4}\left(A_1^2+A_2^2+2A_2A_3^*-2A_1A_3^*\right)\right] \ .
\end{eqnarray}
Adding them together, we have 
\begin{eqnarray}
&&\left(C_A+\frac{C_A+C_A}{2}\right)|M_0|^2+\left[-\frac{N_c}{2}A_1^2-\frac{N_c}{4}\left(A_2^2+A_3^2+2A_1A_2^*-2A_1A_3^*\right)\right] \nonumber\\
&&+ \left[-\frac{N_c}{2}A_2^2-\frac{N_c}{4}\left(A_1^2+A_3^2+2A_1A_2^*+2A_2A_3^*\right)\right] \nonumber\\
&&+\left[-\frac{N_c}{2}A_3^2-\frac{N_c}{4}\left(A_1^2+A_2^2+2A_2A_3^*-2A_1A_3^*\right)\right] \nonumber\\
&&=\left(C_A+C_A\right)|M_0|^2-N_c\left(A_1^2+A_2^2+A_3^2+A_1A_2^*-A_1A_3^*+A_2A_3^*\right)\nonumber\\
&&=C_A|M_0|^2\ .
\end{eqnarray}
Again, this leads to a leading double logarithmic contribution as,
\begin{equation}
-\frac{\alpha_s}{2\pi}C_A\ln^2\left(\frac{P_\perp^2R_\perp^2}{c_0^2}\right) \ ,
\end{equation}
the same as that for Higgs boson production.

\section{Double logarithms in jet-photon production in pA collisions}

In this section, we present an analysis on the Sudakov double logarithms 
in small-$x$ calculations, by extending our previous calculation of Higgs boson
production to the photon-jet production in $pA$ collisions. We will demonstrate
that for one-gluon radiation, the soft gluon contributes to the Sudakov double
logarithms, whereas the collinear gluon contributes to the small-$x$ 
evolution (in this case, it is the BK evolution). These two contributions
are well separated in the phase space of the radiated gluon, and also
by different diagrams. Once we have shown this example, we will 
carry out the calculations of leading double logs for other hard processes.

\subsection{Generic Arguments}
\begin{figure}[tbp]
\begin{center}
\includegraphics[width=7cm]{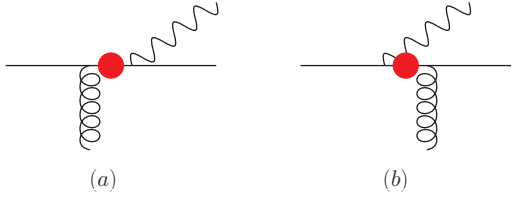}
\end{center}
\caption[*]{The leading order amplitudes of jet-photon pair production in $pA$ collisions. }
\label{LOp}
\end{figure}

The leading order cross section for 
real-photon and associate jet productions in pA collisions in the dipole model, as shown in Fig.~\ref{LOp}, 
can be written as 
\begin{eqnarray}
\frac{\text{d}\sigma^{pA\to q\gamma X}}{\text{d} \cal{P}.\cal{S}.} &=& \sum_f \alpha_{\text{e.m.}} e_f^2 
x_p q_f(x_p) 2\left[1+(1-z)^2\right] (1-z)  \notag \\
&&\times \int \frac{\text{d}^2 u_\perp \text{d}^2 v_\perp 
\text{d}^2 u_\perp^{\prime} \text{d}^2 v_\perp^\prime }{(2\pi)^6} 
e^{-iq_\perp\cdot (v_\perp- v_\perp^\prime)-iP_\perp\cdot (u_\perp- u_\perp^\prime)}\frac{u_\perp\cdot u_\perp^\prime}{u_\perp^2u_\perp^{\prime 2}}\nonumber \\
&& \times  
 \left[S^{(2)} \left(b_\perp, b_\perp^\prime \right)+S^{(2)} \left(v_\perp, v_\perp^\prime \right)
 -S^{(2)} \left(v_\perp, b_\perp^\prime \right)-S^{(2)} \left(b_\perp, v_\perp^\prime \right)\right] , \label{logamma}
\end{eqnarray}
where $v_\perp=zx_\perp+(1-z)b_\perp$ and $u_\perp=x_\perp-b_\perp$
 with $x_\perp$ and $b_\perp$ being the coordinates of the produced real 
 photon and quark, respectively. $z\equiv\frac{k^+_\gamma}{p^+}$ is defined as the longitudinal momentum fraction of the photon with respect to the incoming quark from the proton projectile. 

\begin{figure}[tbp]
\begin{center}
\includegraphics[width=10cm]{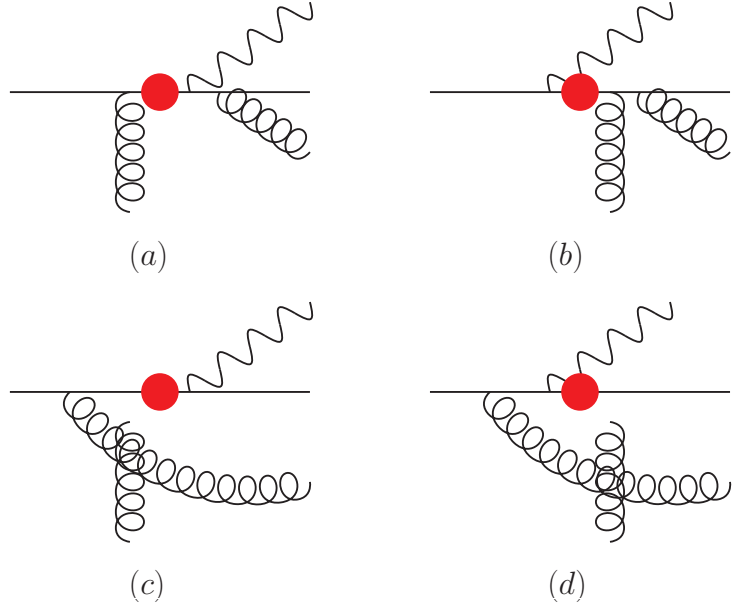}
\end{center}
\caption[*]{Four one loop real emission graphs which contribute to the 
leading power amplitude for the Sudakov factor, while two other graphs 
which have the radiated gluon attaching to the red dots are neglected. }
\label{one-loop-real}
\end{figure}

It straightforward to see that those four dipole scattering amplitudes correspond 
to the four different graphs after squaring the LO amplitudes as shown in 
Fig.~\ref{LOp}. The red dots in Fig.~\ref{LOp} indicate the highly virtual quark 
propagators in the dijet correlation limit ($P_\perp \gg q_\perp$). Simple power 
counting analysis shows that the LO cross section is proportional to 
$q_\perp^4/P_\perp^4$ as a result of the product of two quark propagators 
(as indicated by the red dots) which is proportional to $1/P_\perp^2$. This result 
is explicitly shown in Ref.~\cite{Dominguez:2011wm}. Therefore, in the leading 
power approximation at one loop order, one can treat this highly virtual quark 
propagator as an effective vertex, namely, any additional gluon attachment to 
the vertex, which brings another power of $1/P_\perp^2$, is power suppressed, and
therefore can be neglected. This power counting analysis only works 
when the radiated gluon is not collinear to the incoming target. Namely, if the 
longitudinal momentum of the radiated gluon vanishes, which indicates that the 
gluon is collinear to the nucleus target and generates the rapidity divergence, the 
radiated gluon can attach to the red vertex without being power suppressed. In this 
region, as we have expected, the rapidity divergence can be absorbed into the 
corresponding BK equation. Therefore, in fact, the four graphs which contribute to 
the Sudakov double logarithm which is shown in Fig.~\ref{one-loop-real} (graph (a) 
and (d) do not contribute to the BK evolution), are not the same as the graphs which 
contribute to the BK evolution (only graph (b) and (c) together with two other graphs 
which are not shown here contribute to the BK evolution) of the relevant dipole amplitudes. 
We have also explicitly worked out the derivation of the BK equation at one-loop level. 

For the sake of simplicity, with the cancelation of the $\frac{1}{\epsilon^2}$ 
term in mind, we can obtain the Sudakov factor from the real graphs only 
by choosing $\mu^2=P_\perp^2$ without dealing with the virtual graphs.
As far as the Sudakov double logarithm is concerned, 
there are only 4 real graphs at the amplitude level as shown in 
Fig.~\ref{one-loop-real} (16 in total at the amplitude squared level) 
which contribute to the leading power amplitude. 

According to what we learnt in previous sections for other processes and following the same procedure, we can easily obtain the contributions to the Sudakov double logarithms from those graphs in Fig.~\ref{one-loop-real} as follows. 
\begin{itemize}
\item $(a)^2+(b)^2+2(a)\times (b)$ $\Rightarrow$ $-\frac{\alpha_s 
\mathcal{C}_1}{2\pi} \ln^2\frac{P_\perp^2 R_\perp^2}{c_0^2}$ with 
$\mathcal{C}_1=\frac{1}{2} C_F$ where the factor $\frac{1}{2}$ arises 
due to the same reason as we have discussed in the DIS dijet production, while $C_F$ is just the usual quark-gluon color factor. Since the additional gluon considered here is radiated from the final state quark with large $P_\perp$, the resulting Sudakov double logarithmic contribution should always contain a factor of $\frac{1}{2}$ as we have shown in previous DIS dijet calculation. 
\item $(c)^2+(d)^2+2(c)\times (d)$ 
$\Rightarrow$ $-\frac{\alpha_s \mathcal{C}_2}{2\pi} \ln^2\frac{P_\perp^2 R_\perp^2}{c_0^2}$ 
with $\mathcal{C}_2=C_F$. Here since the gluon is radiated from 
initial state quark without large transverse momentum, there is no additional 
factor of $\frac{1}{2}$ as compared to the previous case. 
\item $2\left[(a)+(b)\right]\times \left[(c)+(d)\right]$ 
$\Rightarrow$ $-\frac{\alpha_s \mathcal{C}_3}{2\pi} \ln^2\frac{P_\perp^2 R_\perp^2}{c_0^2}$ 
with $\mathcal{C}_3=2\times \frac{1}{2}\frac{1}{2N_c}$ where the factor 
$2$ simply comes from the interference. For these eight diagrams, the leading 
$N_c$ contribution does not yield any Sudakov double logarithms, while the 
sub-$N_c$ correction, which is proportional to $-\frac{1}{2N_c}$, does contribute. Therefore, after factorizing out the leading order cross section, we find that these diagrams give
\begin{equation}
-2\times \frac{1}{2N_c} \times 4\alpha_s \int \frac{d\xi}{\xi} \int \frac{d^2 l_\perp}{(2\pi)^2} \frac{l_\perp \cdot (l_\perp +\xi P_\perp)}{l_\perp^2  (l_\perp +\xi P_\perp)^2} 
e^{-il_\perp \cdot R_\perp},
\end{equation}
which eventually gives the effective color factor $\mathcal{C}_3=\frac{1}{2N_c}$. 
In this case, similar to the DIS dijet calcultion, we find that there is an additional factor of $\frac{1}{2}$ coming from the $\xi$ integration as well. 
\end{itemize}
Therefore, by summing up all the above contributions, one find that the Sudakov factor
for the process $q+g \to q+\gamma$ in $pA$ collisions is $-\frac{\alpha_s \mathcal{C}}{2\pi} \ln^2\frac{P_\perp^2 R_\perp^2}{c_0^2}$ with the total color factor $\mathcal{C}=\frac{N_c}{2}+\frac{C_F}{2}$.

\subsection{BK Evolution}

First, let us transform the soft gluon radiation amplitude into more specific 
form, by applying the physical polarization for the radiated gluon, for which
we choose,
\begin{equation}
\epsilon(k_g)\cdot k_g=0,~~~\epsilon(k_g)\cdot p_2=0 \ ,
\end{equation} 
which leads to $\epsilon\cdot p_1=\vec{\epsilon}_\perp\cdot \vec{k}_{g\perp}p_1\cdot p_2/k_g\cdot p_2$.
For example, the gluon radiation from the initial state, we have
\begin{eqnarray}
\epsilon^\mu(k_g)\frac{2p_1^\mu}{2p_1\cdot k_g}=\frac{\epsilon\cdot p_1}{p_1\cdot k_g}=\frac{\vec{\epsilon}_\perp\cdot \vec{k}_{g\perp}p_1\cdot p_2}{p_1\cdot k_gp_2\cdot k_g} =\frac{2\vec{\epsilon}_\perp\cdot \vec{k}_{g\perp}}{k_{g\perp}^2}\ .
\end{eqnarray}
Similarly, we can rewrite 
\begin{eqnarray}
\frac{2k_1^\mu}{2k_1\cdot k_g}&\Rightarrow&\frac{2(k_{g\perp}-\xi_1k_{1\perp})^\mu}{(k_{g\perp}-\xi_1k_{1\perp})^2}  ,\nonumber\\
\frac{2k_2^\mu}{2k_2\cdot k_g}&\Rightarrow&\frac{2(k_{g\perp}-\xi_2k_{2\perp})^\mu}{(k_{g\perp}-\xi_2k_{2\perp})^2}  ,
\end{eqnarray}
where $\xi_i=k_g\cdot p_2/k_i\cdot p_2$. All the integral of the phase space resulting into
the leading double logs demonstrated in the previous section holds in the above functional
forms as well.

\begin{figure}[tbp]
\begin{center}
\includegraphics[width=8cm]{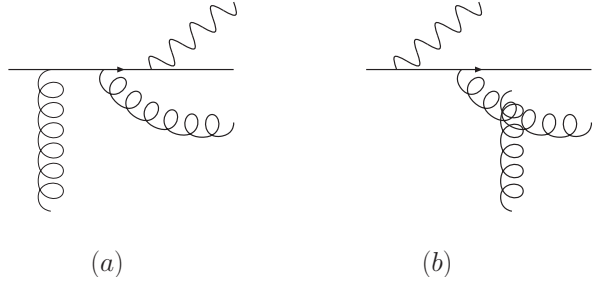}
\end{center}
\caption[*]{Real gluon radiation diagrams which only contribute to the BK-evolution, but not 
to the leading double logarithms.}
\label{one-loop-real-BK}
\end{figure}

In the analysis, we focus on two different regions of the radiated gluon: (1)
soft gluon, where $\alpha_g\sim\beta_g\ll 1$; (2) collinear to the momentum
of the nucleus, where $\alpha_g\ll 1$ but $\beta_g\sim 1$. The region (1) contribute
to the Sudakov double logarithms, whereas the region (2) contributes to 
the small-$x$ evolution for the unintegrated gluon distribution associated
with the nucleus. In both case, $\xi_i\to 0$ limit will be taken in the analysis
of their contributions. For the small-$x$ evolution, we require additional 
$k_g\cdot p_1\sim p_2\cdot p_1\gg (k_1+k_2)\cdot p_1\sim P_\perp^2$. These two regions will be
main focus to be analyzed in the gluon radiation to the
$qg\to q\gamma$ hard process.

Again, the leading order diagram has been shown in Fig.~\ref{LOp}, and can be summarized
into the following form,
\begin{equation}
A_0\sim \left[\frac{k_{2\perp}^\beta}{k_{2\perp}^2}-\frac{k_{2\perp}^\beta-(1-z)q_\perp^\beta}{(k_{2\perp}-(1-z)q_\perp)^2}\right]
\int d^2x_\perp e^{iq_\perp\cdot x_\perp}U(x_\perp) \ ,
\end{equation}
where $k_{1\perp}$, $k_{2\perp}$ represent the momenta for the 
final state quark and photon, $q_\perp=k_{1\perp}+k_{2\perp}$.
In the correlation limit, we will find out that the non-zero contribution
comes from the derivative of the Wilson line $U(x_\perp)$, which 
leads to the cross section is proportional to the dipole gluon distribution. 
 
Let us first analyze the small-$x$ evolution contribution, for which we focus 
on collinear gluon radiation parallel to the nucleus momentum. Some of the 
diagrams are straightforward, whereas some are non-trivial.
The diagram Fig.~\ref{one-loop-real-BK} (a) is, in particular, interesting, because it only contributes
to the small-$x$ evolution, not to the soft gluon double logarithms. 
The propagator goes as,
\begin{equation}
\frac{\!\!\not k\gamma^\mu(\!\!\not k+\!\!\not k_g)}{k^2(k+k_g)^2} \ ,
\end{equation}
where $k=k_1+k_2$. Since $k$ is far off-shell, $k^2\sim P_\perp^2$,
$(k+k_g)^2$ will be far off-shell as well if $k_g$ is soft. However,
it does have contribution in the collinear limit. We can work out the 
explicit dependence in the above expression,
\begin{equation}
\approx \frac{2\epsilon\cdot k_{g\perp}}{k_{g\perp}^2\left(1+\frac{k\cdot p_1}{k_g\cdot p_1}\right)} \ .
\end{equation}
In the soft gluon limit, $\alpha_g\sim\beta_g\ll 1$, $k\cdot p_1\sim p_2\cdot p_1\gg k_g\cdot p_1$.
Therefore, it is power suppressed in the soft gluon limit. However, in the collinear limit,
because $k_g\cdot p_1\gg k\cdot p_1$, there is a leading power contribution from
the above expression. Carrying out the rest of the amplitude, we find out
this diagram contributes to the small-$x$ evolution,
\begin{equation}
\frac{2k_{g\perp}^\mu}{k_{g\perp}^2}\frac{(k_{2\perp}-(1-z)q_\perp)^\beta}{(k_{2\perp}-(1-z)q_\perp)^2}
\int d^2x_\perp e^{i(q_\perp+k_{g\perp})\cdot x_\perp}\left[U(x_\perp)T^a\right] \ .
\end{equation}

On the other hand, both Fig.~\ref{one-loop-real}(a) and \ref{one-loop-real}(b) contribute to soft gluon radiation, but
only diagram Fig.~\ref{one-loop-real}(b) contributes to the small-$x$ evolution. In the soft gluon
limit, the propagators denoted by the red dots are in the same order. However,
in the collinear limit (small-$x$ from nucleus side), the propagator 
in Fig.~\ref{one-loop-real}(a) is in order of $(p_1+p_2)^2$, while that of Fig.~\ref{one-loop-real}(b) in order
of $(p_1-k_2)^2\sim P_\perp^2$. As we discussed above, for small-$x$ evolution,
we take the limit $(p_1+p_2)^2\gg P_\perp^2$. 
Therefore, Fig.~\ref{one-loop-real}(a) is power suppressed as compared to 
Fig.~\ref{one-loop-real}(b). Working out Fig.~\ref{one-loop-real}(b), we find that, it contributes,
\begin{equation}
-\frac{2k_{g\perp}^\mu}{k_{g\perp}^2}\frac{k_{2\perp}^\beta}{k_{2\perp}^2}\int d^2x_\perp e^{i(q_\perp+k_{g\perp})\cdot x_\perp}\left[T^aU(x_\perp)\right] \ .
\end{equation}
There is a relative sign difference between the above two contributions.

Similarly, Fig.~\ref{one-loop-real}(d) only contributes to the soft gluon radiation, by the same reason that for 
Fig.~\ref{one-loop-real}(a). There is an additional diagram, plotted as Fig.~\ref{one-loop-real-BK}(b) contributing to the small-$x$ evolution, which does not 
contribution to the soft gluon double logarithms. This diagram can be easily calculated as
\begin{equation}
\frac{2(k_{g\perp}-k_{g2\perp})^\mu}{(k_{g\perp}-k_{g2\perp})^2}\frac{k_{2\perp}^\beta}{k_{2\perp}^2}\int
d^2x_1d^2x_2e^{ik_{g1\perp}\cdot x_1+ik_{g2\perp}\cdot x_2}\left[U(x_1)U^\dagger(x_2)T^aU(x_2)\right] \ ,
\end{equation}
where $k_{g1}+k_{g2}=q_\perp+k_{g\perp}$. Fig.~\ref{one-loop-real}(c) can also be easily calculated, and we obtain
\begin{equation}
-\frac{2(k_{g\perp}-k_{g2\perp})^\mu}{(k_{g\perp}-k_{g2\perp})^2}\frac{(k_{2\perp}-(1-z)q_\perp)^\beta}{(k_{2\perp}-(1-z)q_\perp)^2}\int
d^2x_1d^2x_2e^{ik_{g1\perp}\cdot x_1+ik_{g2\perp}\cdot x_2}\left[U(x_1)U^\dagger(x_2)T^aU(x_2)\right] \ .
\end{equation}
Again, there is a relative sign difference between the above two terms.

By adding the above four terms together, we can write down the small-$x$ evolution 
coming from the following terms,
\begin{eqnarray}
&&\left(\frac{2(k_{g\perp}-k_{g2\perp})^\mu}{(k_{g\perp}-k_{g2\perp})^2}-\frac{2k_{g\perp}^\mu}{k_{g\perp}^2}\right)
\left(\frac{k_{2\perp}^\beta}{k_{2\perp}^2}-\frac{(k_{2\perp}-(1-z)q_\perp)^\beta}{(k_{2\perp}-(1-z)q_\perp)^2}\right)\nonumber\\
&&\times \int
d^2x_1d^2x_2e^{ik_{g1\perp}\cdot x_1+ik_{g2\perp}\cdot x_2}\left[U(x_1)U^\dagger(x_2)T^aU(x_2)\right] \ .
\end{eqnarray}
The above result leads to the real diagram contributions to the small-$x$ evolution of the dipole
gluon distribution from the nucleus.

\subsection{Detailed analysis of double logarithms of the soft gluon radiation}

To derive the soft gluon radiation contribution to the double logarithms, we will take the correlation
limit, i.e., $k_{1\perp}\sim k_{2\perp}\gg q_\perp=k_{1\perp}+k_{2\perp}$. Under this limit, the
leading order Born diagram can be simplified as,
\begin{equation}
A_0\approx \Gamma^\beta (k_{1\perp})\int d^2x_\perp e^{iq_\perp\cdot x_\perp}\partial_\perp^\beta U(x_\perp) \ .
\end{equation}
Clearly, the hard part $\Gamma^\beta$ is decouple from the Wilson line, as 
a result of the correlation limit and the effective $k_t$-factorization.
The above will result into the cross section proportional to the dipole gluon distribution $xG^{(1)}\sim \langle \partial _\perp U^\dagger(x_\perp) \partial_\perp U(x_\perp)\rangle$,
\begin{equation}
|A_0|^2=\Gamma^\beta(k_{1\perp})\Gamma^{\beta'}(k_{1\perp})\int d^2x_\perp d^2y_\perp e^{iq_\perp(x_\perp-y_\perp)}
 \langle \partial _\perp^\beta U^\dagger(y_\perp) \partial_\perp^{\beta\prime} U(x_\perp)\rangle\ .
 \end{equation}

The soft gluon radiation comes from initial and final state radiation. The diagrams have 
been shown in Fig.~\ref{one-loop-real}. The final state radiation (\ref{one-loop-real}(a) and \ref{one-loop-real}(b)) is easy to calculate, since
they do not involve multiple interactions with the nucleus. The total 
contribution from Figs.~\ref{one-loop-real}(a) and (b) can be written as
\begin{equation}
A_1=\frac{2(k_{g\perp}-\xi_1k_{1\perp})^\mu}{(k_{g\perp}-\xi_1k_{1\perp})^2}
\Gamma^\beta(k_{1\perp})\int d^2x_\perp e^{i(k_{g\perp}+q_\perp)\cdot x_\perp}\left[T^a \partial_\perp^\beta U(x_\perp)\right] \ ,
\end{equation}
where $a$ represents the color index for the radiated gluon and $\mu$ for
its polarization vector.
The evaluation of Figs.~\ref{one-loop-real}(c ) and (d) is a little involved. After some derivations,
we will find out that, they can be written as,
\begin{equation}
A_2=\frac{2(k_{g\perp}-k_{g2\perp})^\mu}{(k_{g\perp}-k_{g2\perp})^2}
\Gamma^\beta(k_{1\perp})\int d^2x_{1\perp} d^2x_{2\perp} e^{i(k_{g1\perp}\cdot x_{1\perp}+k_{g2\perp}\cdot x_{2\perp})}
\left[\partial_\perp^\beta U(x_{1\perp})U^\dagger(x_{2\perp})T^aU(x_{2\perp})\right] \ ,
\end{equation}
where $k_{g1\perp}+k_{g2\perp}=q_\perp+k_{g\perp}$. 

The contribution from the amplitude squared of $A_1$ can be written
as 
\begin{eqnarray}
|A_1|^2&=&\Gamma^\beta(k_{1\perp})\Gamma^{\beta'}(k_{1\perp})
\int d^2x_\perp d^2y_\perp e^{iq_\perp(x_\perp-y_\perp)}
 \langle \partial _\perp U^\dagger(y_\perp) \partial_\perp U(x_\perp)\rangle\nonumber\\
&&\times C_F \int \frac{d^3 k_{g}}{2E_{k_g}(2\pi)^2}e^{ik_{g\perp}\cdot (x_\perp-y_\perp)}\frac{1}{(k_{g\perp}-\xi_1k_{1\perp})^2} \ .
 \end{eqnarray}
Working out the integral as in the previous section, we will find that
\begin{eqnarray}
|A_1|^2&=&\int d^2x_\perp d^2 y_\perp e^{iq_\perp(x_\perp-y_\perp)}|\tilde A_0|^2 \left(-\frac{\alpha_s}{2\pi}\frac{C_F}{2}\right) 
\ln^2\left(\frac{P_\perp^2(x_\perp-y_\perp)^2}{c_0^2}\right)\ ,
\end{eqnarray}
where $|\tilde{A}_0|^2$ is the leading order amplitude squared
in the coordinate space only depending on $x_\perp-y_\perp$ and hard momentum
scale $P_\perp$. This contributes to $C_F/2$ factor for the leading double logarithms. 
Similarly, $|A_2|^2$ contribution,
\begin{eqnarray}
|A_2|^2&=&\int d^2x_\perp d^2 y_\perp e^{iq_\perp(x_\perp-y_\perp)}|\tilde A_0|^2 \left(-\frac{\alpha_s}{2\pi}C_F\right) 
\ln^2\left(\frac{P_\perp^2(x_\perp-y_\perp)^2}{c_0^2}\right)\ ,
\end{eqnarray}
where we have only kept the leading double logarithmic terms and neglected sub-leading
contributions. The interference between the above two terms leads to the following
contribution,
\begin{eqnarray}
2A_1A_2^*&=&-\frac{4(k_{g\perp}-\xi_1k_{1\perp})\cdot (k_{g\perp}-k_{g2\perp})}{(k_{g\perp}-\xi_1k_{1\perp})^2(k_{g\perp}-k_{g2\perp})^2}
\Gamma^\beta(k_{1\perp})\Gamma^{\beta'}(k_{1\perp})\nonumber\\
&&\times \int d^2x_{1\perp}d^2x_{2\perp}d^2x_\perp'
e^{i(k_{g1\perp}\cdot x_{1\perp}+k_{g2\perp}\cdot x_{2\perp})}e^{-i(k_{g\perp}+q_\perp)\cdot x_\perp'}\nonumber\\
&&\times {\rm Tr}\left[\partial_\perp^\beta U(x_{1\perp})U^\dagger(x_{2\perp})T^aU(x_{2\perp})\ \partial_\perp^{\beta'} U^\dagger(x_\perp') T^a \right] \ .
\end{eqnarray}
The leading double logarithmic contribution from the above equation is large $N_c$
suppressed,
\begin{eqnarray}
2A_1A_2^*&\approx &-\frac{4(k_{g\perp}-\xi_1k_{1\perp})\cdot (k_{g\perp})}{(k_{g\perp}-\xi_1k_{1\perp})^2(k_{g\perp})^2}
\Gamma^\beta(k_{1\perp})\Gamma^{\beta'}(k_{1\perp})\nonumber\\
&&\times \int d^2x_{1\perp}d^2x_{2\perp}d^2x_\perp'
e^{i(k_{g1\perp}\cdot x_{1\perp}+k_{g2\perp}\cdot x_{2\perp})}e^{-i(k_{g\perp}+q_\perp)\cdot x_\perp'}\nonumber\\
&&\times {\rm Tr}\left[\partial_\perp^\beta U(x_{1\perp})U^\dagger(x_{2\perp})T^aU(x_{2\perp})\ \partial_\perp^{\beta'} U^\dagger(x_\perp') T^a \right] \nonumber\\
&=&-\frac{4(k_{g\perp}-\xi_1k_{1\perp})\cdot (k_{g\perp})}{(k_{g\perp}-\xi_1k_{1\perp})^2(k_{g\perp})^2}
\Gamma^\beta(k_{1\perp})\Gamma^{\beta'}(k_{1\perp})\nonumber\\
&&\times \int d^2x_{1\perp}d^2x_{2\perp}d^2x_\perp'
e^{i(k_{g1\perp}\cdot x_{1\perp}+k_{g2\perp}\cdot x_{2\perp})}e^{-i(k_{g\perp}+q_\perp)\cdot x_\perp'}\delta^{(2)}(x_{1\perp}-x_{2\perp})\nonumber\\
&&\times {\rm Tr}\left[\partial_\perp^\beta U(x_{1\perp})U^\dagger(x_{2\perp})T^aU(x_{2\perp})\ \partial_\perp^{\beta'} U^\dagger(x_\perp') T^a \right] \ .\nonumber\\
&=&\Gamma^\beta(k_{1\perp})\Gamma^{\beta'}(k_{1\perp})
\int d^2x_\perp d^2y_\perp e^{iq_\perp(x_\perp-y_\perp)}
 \langle \partial _\perp U^\dagger(y_\perp) \partial_\perp U(x_\perp)\rangle\nonumber\\
&&\times \frac{1}{2N_c} \int \frac{d^3 k_{g}}{2E_{k_g}(2\pi)^2}e^{ik_{g\perp}\cdot (x_\perp-y_\perp)}\frac{4k_{g\perp}\cdot (k_{g\perp}-\xi_1k_{1\perp})^2}{k_{g\perp}^2(k_{g\perp}-\xi_1k_{1\perp})^2} \nonumber\\
&=&\int d^2x_\perp d^2 y_\perp e^{iq_\perp(x_\perp-y_\perp)}|\tilde A_0|^2 \left(-\frac{\alpha_s}{2\pi}\frac{1}{2N_c}\right) 
\ln^2\left(\frac{P_\perp^2(x_\perp-y_\perp)^2}{c_0^2}\right)\ ,
\end{eqnarray}
where the leading $N_c$ contribution in the trace vanishes because of ${\rm Tr}\left[\partial U(x_\perp)U^\dagger (x_\perp)\right]=0$.

Adding the above contributions together, we will find out the total contribution is
proportional to 
\begin{eqnarray}
|A_1+A_2|^2&=&\int d^2x_\perp d^2 y_\perp e^{iq_\perp(x_\perp-y_\perp)}|\tilde A_0|^2 \left(-\frac{\alpha_s}{2\pi}\frac{C_F+C_A}{2}\right) 
\ln^2\left(\frac{P_\perp^2(x_\perp-y_\perp)^2}{c_0^2}\right)\ ,
\end{eqnarray}
which is the same as that in the analysis of the collinear factorization calculations.
Therefore, the coefficient ${\cal C}$ for this process is identified as
${\cal C}_{qg\to q\gamma}=\left(C_F+C_A\right)/2$.

\subsection{Heuristic arguments}

\begin{figure}[tbp]
\begin{center}
\includegraphics[width=12cm]{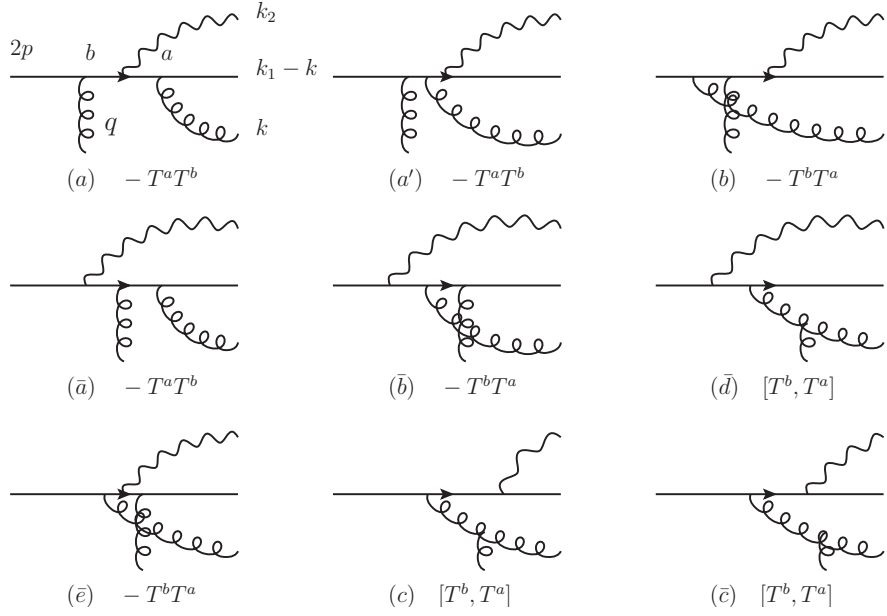}
\end{center}
\caption[*]{Soft gluon radiation in $qg\to q \gamma$ dijet process in the light cone perturbation formalism. All the labels and colour factors for each diagrams are provided below. The difference between diagram $(c)$ and $(\bar c)$ is just whether the $q\to q \gamma$ splitting occurs after or before the vertical gluon insertion.}
\label{pglight}
\end{figure}

In this part of the discussion, we employ the light-cone perturbation theory to analyse the real diagrams and show the Sudakov double logarithm for the $\gamma q$ dijet productions in $pA$ collisions. With the assumption of single scattering approximation, we find that there are nine real diagrams as shown in Fig.~\ref{pglight}. As we have learnt above, multiple scattering does not modify the Sudakov double logarithm, therefore single scattering calculation is sufficient for us to obtain the correct Sudakov factor in this process. 

To simplify the calculation, we set the kinematics as follows
\begin{eqnarray}
2p &=& (2p^+, 0,0), \quad \quad  
k_1-k = (p^+-k^+,\frac{(P_\perp-k_\perp)^2}{2(p^+-k^+)}, P_\perp-k_\perp),\\
k_2&=&(p^+,\frac{(P_\perp-q_\perp)^2}{2p^+}, q_\perp-P_\perp),\quad \quad 
k=(k^+, \frac{k_\perp^2}{2k^+},k_\perp).
\end{eqnarray}
The energy-momentum conservation constraint implies that the vertical small-$x$ gluon which comes from the target nucleus has a non-vanishing minus component $q^-=k_1^-+k^-_2+k^-$ and transverse component $q_\perp$. Since only the double logarithmic term is being considered, $z\equiv \frac{k^+}{p^+} \ll 1$ is always assumed. Without the gluon radiation, the photon ($\gamma$) and quark jets have almost back-to-back momenta $P_\perp\simeq \frac{1}{2}|k_{2\perp}-k_{1\perp}|$ and $q_\perp=|k_{1\perp}+k_{2\perp}|$ with $\frac{q_\perp}{P_\perp}\ll 1$. In the following calculation, we will neglect the transverse momentum $q_\perp$, since we know that the real diagrams will be approximately cancelled by the virtual diagrams in the region $k_\perp < q_\perp$. Namely, this region does not contribute to the Sudakov double logarithm. Therefore, only the region where $k_\perp >q_\perp$ is considered and $q_\perp$ plays a role as a natural lower cut-off for our calculation. On the other hand, the transverse momentum of the radiated gluon $k_\perp$ should not exceed the jet transverse momentum $P_\perp$, otherwise the corresponding diagram is suppressed. 

As a result, after dropping factors of $q_\perp$ in the light-cone perturbation theory, the denominators for the above nine graphs are
\begin{eqnarray}
D_a &=& \frac{-1}{\frac{(k_\perp-zP_\perp)^2}{2k^+}\left(\frac{P_\perp^2}{p^+}+\frac{k_\perp^2}{2k^+}\right)},  \quad D_{a^\prime}=  \frac{-1}{\frac{P_\perp^2}{p^+}\left(\frac{P_\perp^2}{p^+}+\frac{k_\perp^2}{2k^+}\right)} , \\
D_b &=& D_c = \frac{1}{\frac{P_\perp^2}{p^+}\frac{k_\perp^2}{2k^+}},  \quad D_{\bar c} =D_{\bar e}= \frac{-1}{\left(\frac{P_\perp^2}{p^+}+\frac{k_\perp^2}{2k^+}\right)\frac{k_\perp^2}{2k^+}}, \\
D_{\bar a} &=& \frac{1}{\frac{(k_\perp-zP_\perp)^2}{2k^+} \frac{P_\perp^2}{p^+}},  \quad D_{\bar b}=D_{\bar d}=\frac{-1}{\frac{P_\perp^2}{p^+}\left(\frac{P_\perp^2}{p^+}+\frac{k_\perp^2}{2k^+}\right)}.
\end{eqnarray}
To be consistent with the power counting analysis used throughout this paper, we neglect all the power suppressed terms. Taking the color factors into account, it is straightforward to find that the leading power contribution is cancelled completely
between all these nine diagrams which indicates
\begin{equation}
(a+a^\prime +b+c+\bar c)|_{q_\perp \to 0} = -(\bar a +\bar b+\bar d+\bar e)|_{q_\perp \to 0} . \label{cancellation}
\end{equation}
The above cancellation does not occur if we keep the incoming gluon transverse momentum $q_\perp$ finite. Instead, we would obtain the same leading power contribution which is proportional to $\frac{q_\perp^2}{P_\perp^2}$ as compared to the leading order $\gamma+q$-jets contribution calculated from the two diagrams as shown in Fig.~\ref{LOp}. To calculate the Sudakov double logarithmic term, the diagrams $(a)$, $(a^\prime)$, $(b)$ and $(c)$ are grouped together as the one loop correction to the Fig.~\ref{LOp} (a), since the vertical gluon insertion occurs before the $q\to q \gamma$ splitting in all of these diagrams. While $(\bar a)$,  $(\bar b)$, $(\bar d)$ and $(\bar e)$ are combined together as the one-loop correction to the Fig.~\ref{LOp}(b) since the vertical gluon insertion happens after the $q\to q \gamma$ splitting in all of these diagrams. It appears that there is an ambiguity for the last graph $(\bar c)$. If the vertical gluon transverse momentum $q_\perp$ were kept everywhere in the above calculation, there would be no ambiguity, but the calculation would be very tedious. Nevertheless, this problem goes away if we perform the same calculation using Feynman propagators instead of light-cone propagators where the diagrams $(c)$ and $(\bar c)$ become one single diagram in the Feynman propagator calculation. We obtain exactly the same result as the Feynman calculation if we combine $D_{c}$ and $D_{\bar c}$ together. Therefore, we combine the diagram $(\bar c)$ with the diagrams $(a)$, $(a^\prime)$, $(b)$ and $(c)$ as shown on the left hand side of the Eq.~(\ref{cancellation}). 

Now we are ready to evaluate the above diagrams and compute the Sudakov double logarithmic term. Here in this part, we do not include the jet decay double logarithmic term which depends on the jet-cone parameters, since this belongs to different type of physics. It is straightforward to find 
\begin{equation}
(a+a^\prime +b+c+\bar c)=\frac{T^aT^b}{\frac{(k_\perp-zP_\perp)^2}{2k^+}\left(\frac{P_\perp^2}{p^+}+\frac{k_\perp^2}{2k^+}\right)}- \frac{T^bT^a}{\left(\frac{P_\perp^2}{p^+}+\frac{k_\perp^2}{2k^+}\right)\frac{k_\perp^2}{2k^+}},
\end{equation}
which gives
\begin{equation}
(a+a^\prime +b+c+\bar c)\simeq\frac{1}{\frac{k_\perp^2}{2k^+}\frac{P_\perp^2}{p^+}}\left(\left.T^aT^b\right|_{\frac{k_\perp^2}{P_\perp^2}p^+<k^+<\frac{k_\perp}{P_\perp}p^+}^{\textrm{I}}-\left.T^bT^a\right|_{\frac{k_\perp^2}{P_\perp^2}p^+<k^+<p^+}^{\textrm{II}}\right). \label{sum}
\end{equation}
To arrive at the above expression, we have employed the knowledge that the region with $0<k^+ < \frac{k_\perp^2}{P_\perp^2}p^+$ does not contribute to the Sudakov factor at all since it gives rise to the BFKL type of single logarithmic term and therefore should be associated with the small-$x$ evolution of the dipole amplitudes. Now according to Eq.~(\ref{sum}), we can compute the leading Sudakov factor for Fig.~\ref{LOp}(a) term by term as follows
\begin{eqnarray}
\left.\textrm{Sudakov}\right|^{q\to q\gamma}_{\textrm{I}\times \textrm{I}^{\ast}}&=&-\frac{\alpha_s C_F}{\pi}\int_{q_\perp^2}^{P_\perp^2} \frac{\textrm{d} k_\perp^2}{k_\perp^2} \int_{\frac{k_\perp^2}{P_\perp^2}p^+}^{\frac{k_\perp}{P_\perp}p^+} \frac{\textrm{d}k^+}{k^+}=-\frac{\alpha_s C_F}{4\pi}\ln^2\frac{P_\perp^2}{q_\perp^2}, \\
\left.\textrm{Sudakov}\right|^{q\to q\gamma}_{\textrm{II}\times \textrm{II}^{\ast}}&=&-\frac{\alpha_s C_F}{\pi}\int_{q_\perp^2}^{P_\perp^2} \frac{\textrm{d} k_\perp^2}{k_\perp^2} \int_{\frac{k_\perp^2}{P_\perp^2}p^+}^{p^+} \frac{\textrm{d}k^+}{k^+}=-\frac{\alpha_s C_F}{2\pi}\ln^2\frac{P_\perp^2}{q_\perp^2}, \\
\left.\textrm{Sudakov}\right|^{q\to q\gamma}_{2\times \textrm{I}\times \textrm{II}^{\ast}}&=&-\frac{\alpha_s }{\pi N_c}\int_{q_\perp^2}^{P_\perp^2} \frac{\textrm{d} k_\perp^2}{k_\perp^2} \int_{\frac{k_\perp^2}{P_\perp^2}p^+}^{\frac{k_\perp}{P_\perp}p^+} \frac{\textrm{d}k^+}{k^+}=-\frac{\alpha_s }{4\pi N_c}\ln^2\frac{P_\perp^2}{q_\perp^2}, 
\end{eqnarray}
where the last line comes from the interference term with the colour factor $\frac{1}{N_c}\textrm{tr}\left[T^aT^bT^aT^b\right]=-\frac{C_F}{2N_c}$. Also one should note that a factor of $C_F$ should be factorized out for the leading order dijet cross section. Summing over all the contributions, one eventually obtains the total Sudakov double logarithmic term as follows 
\begin{equation}
\left.\textrm{Sudakov}\right|^{q\to q\gamma}=-\frac{\alpha_s}{2\pi}\left(\frac{C_F}{2}+\frac{N_c}{2}\right)\ln^2\frac{P_\perp^2}{q_\perp^2}.
\end{equation}
This result is in complete agreement with our earlier calculation with the identification of $q_\perp \sim \frac{1}{R_\perp}$. 

It is interesting to note that actually only diagrams $(a)$ and $(b)$ contribute to the Sudakov double logarithm when we take into account the constraint $k^+ > \frac{k_\perp^2}{P_\perp^2}p^+$, since $(a^\prime)$ and $(c)+(\bar c)$ are power suppressed in the corresponding region. This conclusion also agrees with our previous analysis. The graph $(a)$ represents the final state gluon radiation, while the graph $(b)$ corresponds to the initial state gluon radiation, which yield the contributions with colour factors $\frac{1}{2} C_F$ and $C_F$, respectively. Their interference graphs give a contribution with the colour factor $\frac{1}{2N_c}$. The Sudakov double logarithm can be viewed as the probabilistic correction to the back-to-back dijet configuration. If we define the probability of generating the dijet configuration at leading order as unity, then the Sudakov factor represents the probability that such configuration is destroyed by one-gluon radiation at the loop order. It is then straightforward to see that the initial state gluon radiation can destroy the desired dijet configuration by radiating a gluon with $k_\perp >q_\perp$. However, the case for the final state gluon radiation is slightly more complicated. Final state gluon radiation not only has to satisfy the requirement $k_\perp >q_\perp$, but also should have a large enough angle so that $\frac{k_\perp}{k^+}>\theta\equiv \frac{P_\perp}{p^+}$, in order to be distinguishable from the dijet configuration. Another way to view this additional requirement is to say that the rapidity of the emitted gluon $\ln\frac{k^+}{k_\perp}$ should not be greater than that of the original quark which is $\ln \frac{p^+}{P_\perp}$. This is a 'dead-zone' effect similar to what happens in the gluon radiation from a heavy quark. This naturally explains the above calculation and the different Sudakov contributions between the final state and initial state gluon radiations. 

Furthermore, as a consistency check, it is easy to find that the combination of graph $(\bar a)$,  $(\bar b)$, $(\bar d)$ and $(\bar e)$ yields a similar conclusion and the same Sudakov factor as shown above for Fig.~\ref{LOp}(b).  

In summary, as we have demonstrated above by employing three different analysis, we have found the Sudakov double logarithm for the photon-jet back-to-back correlations in $pA$ collisions. The effective color factor $\mathcal{C}$ is found to be $\frac{C_F}{2}+\frac{N_c}{2}$, which is the same as the one in the collinear factorization. 

\section{Double Logs in Dijet production in pA collisions}

In this section, we will extend the discussions in the previous section to 
the general dijet production in $pA$ collisions. We only focus on the 
Sudakov double logarithms. We expect the small-$x$ evolution
will follow the same as that for the photon-jet production process.
We will derive the double logs for all the hard processes relevant for
dijet production. The basic idea is to identify the initial and final state radiations,
and calculate the associated contributions in the leading double logarithmic 
approximation.

\subsection{$gg \to q\bar q$}

From Ref.\cite{Dominguez:2011wm}, we can write down the leading Born amplitude
for $q\bar q$ pair production,
\begin{equation}
A_0=\int d^2x_\perp e^{iq_\perp\cdot x_\perp} \Gamma^\beta(k_{1\perp})\left[(1-z)\partial_\perp U(x_\perp)T^aU^\dagger(x_\perp)
-zU(x_\perp)T^a\partial_\perp U^\dagger(x_\perp)\right]_{ij}\ ,
\end{equation}
where again we have taken the correlation limit. The amplitude squared
will be,
\begin{eqnarray}
|A_0|^2&=&\int d^2x_\perp y_\perp e^{iq_\perp\cdot (x_\perp-y_\perp)}
\Gamma^\beta(k_{1\perp}))\Gamma^{\beta'}(k_{1\perp}))\nonumber\\
&&\times \frac{1}{2}\left\{\left((1-z)^2+z^2\right) {\rm Tr}\left[U(x_\perp)U^\dagger(y_\perp)\right]
{\rm Tr}\left[\partial_\perp U(x_\perp)\partial_\perp U^\dagger(y_\perp')\right]\right.\nonumber\\
&&\left. -2z(1-z)
{\rm Tr}\left[U(x_\perp)\partial_\perp U^\dagger(y_\perp)\right]{\rm Tr}
\left[ U(x_\perp)\partial_\perp U^\dagger(y_\perp')\right]\right\} \ .
\end{eqnarray} 
This is consistent with what we have found in Ref.~\cite{Dominguez:2011wm}.

\begin{figure}[tbp]
\begin{center}
\includegraphics[width=9cm]{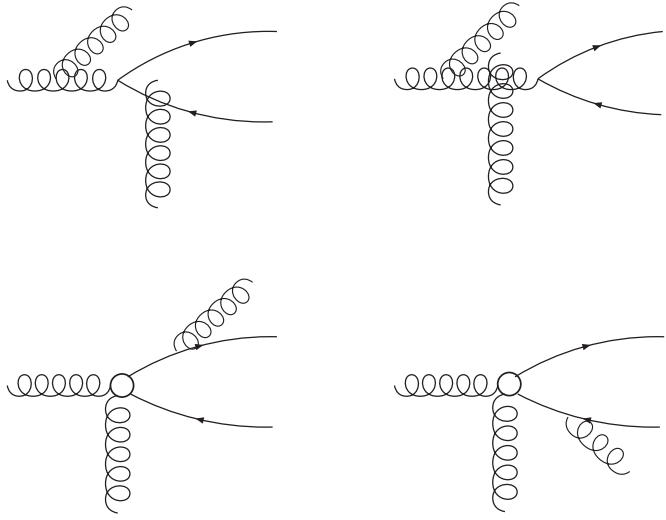}
\end{center}
\caption[*]{Soft gluon radiation in $gg\to q\bar q$ process in the saturation formalism. 
The blobs in the lower two diagrams represent the multiple gluon interaction with nucleus
formulated in Ref.~\cite{Dominguez:2011wm} in the correlation limit.}
\label{cgc-qq}
\end{figure}

The initial state radiation contribution of Fig.~\ref{cgc-qq} can be written as
\begin{eqnarray}
A_1&=&\frac{2(k_{g\perp}-k_{g2\perp})^\mu}{(k_{g\perp}-k_{g2\perp})^2}
\Gamma^\beta(k_{1\perp})\frac{1}{N_F}{\rm Tr}[T^bU(x_2)T^cU^\dagger(x_2)]
\left[(1-z)\partial_\perp U(x_\perp)[T^a,T^c]U^\dagger(x_\perp)\right.\nonumber\\
&&\left.-zU(x_\perp)[T^a,T^c]\partial_\perp U^\dagger(x_\perp)\right]_{ij} \ ,
\end{eqnarray}
where $a$ represents the color index for  incoming gluon, $b$ for radiated
gluon, $ij$ for the final state quark pair. Gluon radiation from the quark and
antiquark lines of Fig.~\ref{cgc-qq} can be written as
\begin{eqnarray}
A_2&=&\frac{2(k_{g\perp}-\xi_1k_{1\perp})^\mu}{(k_{g\perp}-\xi_1k_{1\perp})^2}
\Gamma^\beta(k_{1\perp})
\left[(1-z)T^b\partial_\perp U(x_\perp)T^aU^\dagger(x_\perp)-zT^bU(x_\perp)T^a\partial_\perp U^\dagger(x_\perp)\right]_{ij} \ ,
\nonumber\\
A_3&=&-\frac{2(k_{g\perp}-\xi_2k_{2\perp})^\mu}{(k_{g\perp}-\xi_2k_{2\perp})^2}
\Gamma^\beta(k_{1\perp})
\left[(1-z)\partial_\perp U(x_\perp)T^aU^\dagger(x_\perp)T^b-zU(x_\perp)T^a\partial_\perp U^\dagger(x_\perp)T^b\right]_{ij} \ .\nonumber
\\
\end{eqnarray}
The amplitude squared of the above terms can be easily calculated, following
the example in previous section,
\begin{equation}
|A_1|^2=C_A|A_0|^2,~~|A_2|^2=\frac{C_F}{2}|A_0|^2,~~|A_3|^2=\frac{C_F}{2}|A_0|^2 \ .
\end{equation}
The interference between them is a little involved, but also straightforward.
For example, for $2A_1A_2^*$, we have, for the Wilson line part,
\begin{eqnarray}
2A_1A_2^*&=&\frac{1}{N_F}{\rm Tr}[T^bU(x_2)T^cU^\dagger(x_2)]\nonumber\\
&&\times {\rm Tr}\left[
\left((1-z)\partial_\perp U(x_\perp)[T^a,T^c]U^\dagger(x_\perp)-zU(x_\perp)[T^a,T^c]\partial_\perp U^\dagger(x_\perp)\right)
\right.\nonumber\\
&&\left.\left((1-z)U(y_\perp)T^a\partial_\perp U^\dagger (y_\perp)T^b-z\partial_\perp U(y_\perp)T^a U^\dagger(y_\perp)T^b\right)\right] \ .
\end{eqnarray}
Again, taking the leading double logarithmic approximation, we will have $\delta^{(2)}(x_\perp-x_{2\perp})$,
which will simplify the above expression, and we shall obtain,
\begin{eqnarray}
2A_1A_2^*&=&-\frac{N_c}{2}\frac{z^2}{2} {\rm Tr}\left[U(x_\perp)U^\dagger(y_\perp)\right]
{\rm Tr}\left[\partial_\perp U(x_\perp)\partial_\perp U^\dagger(y_\perp')\right]\nonumber\\
&&+\frac{N_c}{2}\frac{z(1-z)}{2}{\rm Tr}\left[U(x_\perp)\partial_\perp U^\dagger(y_\perp)\right]{\rm Tr}
\left[ U(x_\perp)\partial_\perp U^\dagger(y_\perp')\right]\ .
\end{eqnarray}
Similarly, following the same procedure, we have 
\begin{eqnarray}
2A_1A_3^*&=&\frac{1}{N_F}{\rm Tr}[T^bU(x_2)T^cU^\dagger(x_2)]\nonumber\\
&&\times {\rm Tr}\left[
\left((1-z)\partial_\perp U(x_\perp)[T^a,T^c]U^\dagger(x_\perp)-zU(x_\perp)[T^a,T^c]\partial_\perp U^\dagger(x_\perp)\right)
\right.\nonumber\\
&&\left.\left((1-z)T^bU(y_\perp)T^a\partial_\perp U^\dagger (y_\perp)-zT^b\partial_\perp U(y_\perp)T^a U^\dagger(y_\perp)\right)\right] \nonumber\\
&&\Rightarrow \frac{N_c}{2}\frac{(1-z)^2}{2} {\rm Tr}\left[U(x_\perp)U^\dagger(y_\perp)\right]
{\rm Tr}\left[\partial_\perp U(x_\perp)\partial_\perp U^\dagger(y_\perp')\right]\nonumber\\
&&~~~-\frac{N_c}{2}\frac{z(1-z)}{2}{\rm Tr}\left[U(x_\perp)\partial_\perp U^\dagger(y_\perp)\right]{\rm Tr}
\left[ U(x_\perp)\partial_\perp U^\dagger(y_\perp')\right]\ . 
\end{eqnarray}
On the other hand, the interference between $A_2$ and $A_3$ is large $N_c$ suppressed.

By adding the above two equations together, we find that
the interference terms will contribute to a factor of $N_c/2$ to the leading double logarithms.
Therefore, the total contribution will be
\begin{eqnarray}
|A_1+A_2+A_3|^2&=&\int d^2x_\perp d^2 y_\perp e^{iq_\perp(x_\perp-y_\perp)}|\tilde A_0|^2 \left(-\frac{\alpha_s}{2\pi}N_c\right) 
\ln^2\left(\frac{P_\perp^2(x_\perp-y_\perp)^2}{c_0^2}\right)\ ,
\end{eqnarray}
in the large $N_c$ limit.

\begin{figure}[tbp]
\begin{center}
\includegraphics[width=11cm]{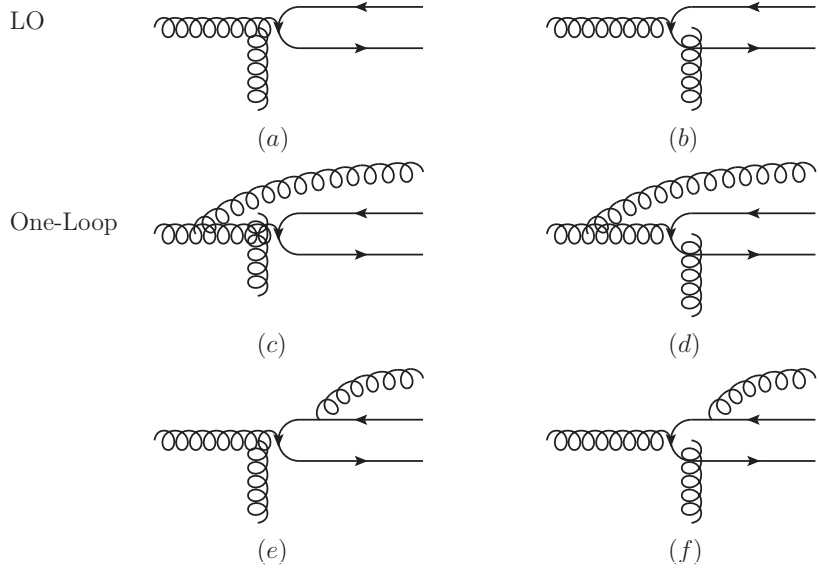}
\end{center}
\caption[*]{Relevant graphs in the $g+g\to q \bar q$ process. }
\label{gqqbar}
\end{figure}

Similar to the above calculation, one can also obtain the color factor of the Sudakov factor for the $g+g\to q \bar q$ channel as illustrated in Fig.~\ref{gqqbar} as follows.
\begin{itemize}
\item $(a)^2+(b)^2+2(a)\times (b)$ $\Rightarrow$ the LO cross section which sets the baseline for the extraction of the Sudakov factor from the one-loop calculation.
\item $(c)^2+(d)^2+2(c)\times (d)$ $\Rightarrow$ $-\frac{\alpha_s \mathcal{C}_1}{2\pi} \ln^2\frac{P_\perp^2 R_\perp^2}{c_0^2}$ with $\mathcal{C}_1=N_c$. 
\item $(e)^2+(f)^2+2(e)\times (f)$ $\Rightarrow$ $-\frac{\alpha_s \mathcal{C}_2}{2\pi} \ln^2\frac{P_\perp^2 R_\perp^2}{c_0^2}$ with $\mathcal{C}_2=\frac{1}{2} \times 2 (C_F+\frac{1}{2N_c})=\frac{1}{2}N_c$. Here since the gluon is radiated from the final state quark with large transverse momentum $P_\perp$, there is an additional factor of $\frac{1}{2}$ which is compensated by a factor of $2$ since we have two quark lines in this case. Here we have also taken the sub-$N_c$ correction into account. The color factor of the contribution is identical to the one in the DIS dijet case. 
\item $2\left[(c)+(d)\right]\times \left[(e)+(f)\right]$ $\Rightarrow$ $-\frac{\alpha_s \mathcal{C}_3}{2\pi} \ln^2\frac{P_\perp^2 R_\perp^2}{c_0^2}$ with $\mathcal{C}_3=-2\times \frac{1}{2}\frac{N_c}{2}$ where the factor $2$ simply comes from the interference. This part of the result is different from the $q+g\to q +\gamma$ process due to different color structures, although these two processes look similar in terms of the dipole amplitudes. Here we believe that the non-linear term, which is leading $N_c$ contribution, do contribute to the Sudakov factor. 
\end{itemize}

Let us compare the calculation in detail for the $q+g\to q +\gamma$ process with the $g+g\to q +\bar q$ process, which can help us understand their differences for the interference terms as we mentioned above. 
In the case of $q+g\to q +\gamma$, the last term of the LO cross section as shown in Eq.~(\ref{logamma}) reduces to $\partial_v \partial_{v^\prime} S^{(2)}(v,v^\prime) $ in the dijet correlation limit. This is the exact reason that this particular process is measuring the dipole gluon distributio (which is proportional to $q^2 F_{x_g}(q)$ in the momentum space) in the dijet limit. Now we consider the interference term ($2\left[(a)+(b)\right]\times \left[(c)+(d)\right]$ in Fig~(\ref{one-loop-real})), the leading $N_c$ term is non-linear which is proportional to 
\begin{equation}
\partial_v \partial_{v^\prime} \left[S^{(2)}(v,z)S^{(2)}(z,v^\prime)\right],
\end{equation}
where $z$ is the coordinate of the radiated gluon, while $v$ and $v^\prime$ are the coordinates of the incoming quark. It is important to note that the above expression vanishes in the limit $z\to v$ or $z\to v^\prime$. Therefore, only the sub-$N_c$ correction contributes to the Sudakov factor in this case. 

On the other hand, for the $g+g\to q +\bar q$ channel in the large $N_c$ limit, the leading order amplitude is proportional to
\begin{equation}
\left[\alpha^2+(1-\alpha)^2\right]S^{(2)}(v^\prime,v)\partial_v \partial_{v^\prime} \left[S^{(2)}(v,v^\prime)\right] -2\alpha (1-\alpha) \partial_v  S^{(2)}(v^\prime,v)\partial_{v^\prime} S^{(2)}(v,v^\prime) \label{LOgqqbar}
\end{equation} 
where $\alpha$ is the longitudinal momentum fraction of the produced quark with respect to the incoming gluon. At the one-loop order, we find that the interference term (from $2\left[(c)+(d)\right]\times \left[(e)+(f)\right]$) yields
\begin{eqnarray}
&&\left[\alpha^2+(1-\alpha)^2\right]S^{(2)}(v,z)S^{(2)}(z,v^\prime)\partial_v \partial_{v^\prime} \left[S^{(2)}(v,v^\prime) \right]\notag \\
&+&\left[\alpha^2+(1-\alpha)^2\right]S^{(2)}(v,v^\prime)\partial_v \partial_{v^\prime} \left[S^{(2)}(v,z)S^{(2)}(z,v^\prime) \right]\notag\\
&-& 2\alpha (1-\alpha) \left[\partial_v  S^{(2)}(v^\prime,z)S^{(2)}(z,v)\right]\partial_{v^\prime} S^{(2)}(v,v^\prime)\notag\\
&-& 2\alpha (1-\alpha) \partial_v  S^{(2)}(v^\prime,v)\partial_{v^\prime} \left[S^{(2)}(v,z)S^{(2)}(z,v^\prime)\right]
\end{eqnarray}
in the large $N_c$ limit. In the region where $z$ is close to $v$ or $v^\prime$, the above expression does not vanish and reduces to the LO expression as shown in Eq.~(\ref{LOgqqbar}) in this region. As a result, the integration over $z$ in the vicinity of $v$ and $v^\prime$ could generate a Sudakov double logarithm with a color factor $-\frac{N_c}{2}$. In the large $N_c$ limit, in which one can approximate a gluon to a pair of quark-antiquark lines in terms of the color structure, we find that the extra antiquark line in this channel as compared to the $q+g\to q +\gamma$ channel cause the different outcomes for the interference contributions in these two channels. 
 
Therefore, by summing up all the contributions, one find the Sudakov factor for the $g+g \to q+\bar q$ channel  is $-\frac{\alpha_s \mathcal{C}}{2\pi} \ln^2\frac{P_\perp^2 R_\perp^2}{c_0^2}$ with the effective color factor $\mathcal{C}_{gg\to q\bar q}=N_c$. For heavy quark pair productions in $pA$ collisions considered in a recent study\cite{Fujii:2013}, we believe that there should also be an associated Sudakov factor with the same effective color factor. 
\subsection{$qg\to qg$ channel}

Tthe leading Born amplitude for $qg\to qg$ channel can be written as,
\begin{equation}
A_0=\int d^2x_\perp e^{iq_\perp\cdot x_\perp} \Gamma^\beta(k_{1\perp})\left[\partial_\perp U(x_\perp)U^\dagger(x_\perp)T^aU(x_\perp)
-zT^a\partial_\perp U(x_\perp)\right]_{ij}\ ,
\end{equation}
where again we have taken the correlation limit. The amplitude squared
will be,
\begin{eqnarray}
|A_0|^2&=&\int d^2x_\perp d^2y_\perp e^{iq_\perp\cdot (x_\perp-y_\perp)}
\Gamma^\beta(k_{1\perp}))\Gamma^{\beta'}(k_{1\perp}))\nonumber\\
&&\times\left\{\frac{1}{2} {\rm Tr}\left[U(x_\perp)U^\dagger(y_\perp)\right]
{\rm Tr}\left[\partial_\perp U(x_\perp)U^\dagger(x_\perp)\partial_\perp U(y_\perp)U^\dagger(y_\perp)\right]\right.\nonumber\\
&&\left. +z^2 C_F\left[\partial_\perp U(x_\perp)\partial_\perp U^\dagger(y_\perp)\right]\right\} \ .
\end{eqnarray} 
This is consistent with what we have found in Ref.~\cite{Dominguez:2011wm}.

\begin{figure}[tbp]
\begin{center}
\includegraphics[width=9cm]{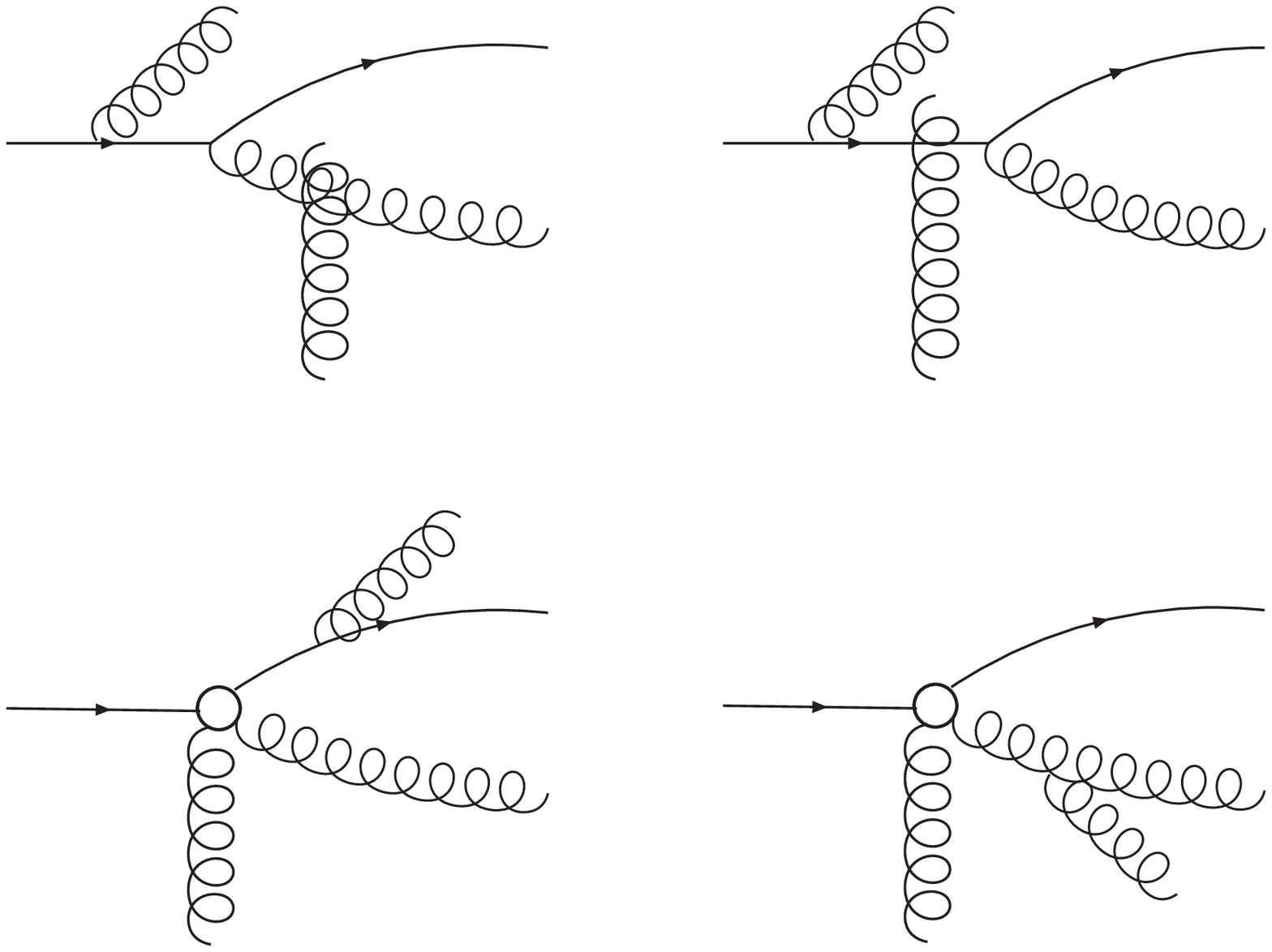}
\end{center}
\caption[*]{Same as Fig.~\ref{cgc-qq} for $qg\to qg$ process.}
\label{cgc-qg}
\end{figure}

The initial state radiation contribution of Fig.~\ref{cgc-qg} can be written as
\begin{eqnarray}
A_1&=&\frac{2(k_{g\perp}-k_{g2\perp})^\mu}{(k_{g\perp}-k_{g2\perp})^2}
\Gamma^\beta(k_{1\perp})\left[-\partial_\perp U(x_\perp)U^\dagger(x_\perp)T^a U(x_\perp)U^\dagger(x_{2\perp})T^bU(x_{2\perp})
\right.\nonumber\\
&&\left.+zT^a\partial_\perp U(x_\perp)U^\dagger (x_{2\perp})T^bU(x_{2\perp})\right]_{ij} \ ,
\end{eqnarray}
where $a$ represents the color index for out going final state gluon, $b$ for radiated
gluon, $ij$ for the initial and final state quarks. Gluon radiation from the quark and gluon
lines of Fig.~\ref{cgc-qg} can be written as
\begin{eqnarray}
A_2&=&\frac{2(k_{g\perp}-\xi_1k_{1\perp})^\mu}{(k_{g\perp}-\xi_1k_{1\perp})^2}
\Gamma^\beta(k_{1\perp})
\left[-T^b\partial_\perp U(x_\perp)U^\dagger(x_\perp)T^aU(x_\perp)+zT^bT^a\partial_\perp U(x_\perp)\right]_{ij} \ ,
\nonumber\\
A_3&=&-\frac{2(k_{g\perp}-\xi_2k_{2\perp})^\mu}{(k_{g\perp}-\xi_2k_{2\perp})^2}
\Gamma^\beta(k_{1\perp})
\left[-\partial_\perp U(x_\perp)U^\dagger(x_\perp)[T^a,T^b]U(x_\perp)+z[T^a,T^b]\partial_\perp U(x_\perp)\right]_{ij} \ .\nonumber
\\
\end{eqnarray}
The amplitude squared of the above terms can be easily calculated, following
the example in previous section,
\begin{equation}
|A_1|^2=C_F|A_0|^2,~~|A_2|^2=\frac{C_F}{2}|A_0|^2,~~|A_3|^2=\frac{C_A}{2}|A_0|^2 \ .
\end{equation}
For the interference between them $2A_2A_3^*$, we have, for the Wilson line part,
\begin{eqnarray}
2A_2A_3^*&=&{\rm Tr}\left[\partial_\perp U(x_\perp)U^\dagger(x_\perp)[T^a,T^b]U(x_\perp)U^\dagger(y_\perp)T^aU(y_\perp)\partial_\perp
U^\dagger(y_\perp)T^b\right.\nonumber\\
&&\left.+z^2[T^a,T^b]\partial_\perp U(x_\perp)\partial_\perp U^\dagger(y_\perp)T^aT^b\right] \ ,
\end{eqnarray}
where we will find out that the first term is large $N_c$ suppressed and the second term
can be calculated easily. Finally, we have
\begin{eqnarray}
2A_2A_3^*&=&z^2\left(\frac{-C_A}{2}\right)C_F{\rm Tr}\left[\partial_\perp U(x_\perp)\partial_\perp U^\dagger(y_\perp)\right] \ .
\end{eqnarray}
Similarly, following the same procedure, we have 
\begin{eqnarray}
2A_1A_3^*&=& {\rm Tr}\left[\partial_\perp U(x_\perp)U^\dagger(x_\perp)T^aU(x_\perp)
U^\dagger(x_{2\perp})T^bU(x_{2\perp})U^\dagger(y_\perp)[T^b,T^a]U(y_\perp)\partial_\perp 
U^\dagger(y_\perp)\right.\nonumber\\
&&\left.+z^2T^a\partial_\perp U(x_\perp) U^\dagger(x_{2\perp})T^bU(x_{2\perp})\partial_\perp 
U(y_\perp)[T^b,T^a]\right]\ . 
\end{eqnarray}
By applying the Delta function of $\delta^{(2)}(x_\perp-x_{2\perp})$, we will find out the leading $N_c$ 
contribution from the second term vanishes. Therefore, we will only have the first term
contribution,
\begin{eqnarray}
2A_1A_3^*&=& \frac{N_c}{2}\frac{1}{2}{\rm Tr}\left[U(x_\perp)U^\dagger(y_\perp)\right]{\rm Tr}
\left[\partial_\perp U(x_\perp)U^\dagger(x_\perp)U(y_\perp)\partial_\perp U^\dagger(y_\perp)\right] \ .
\end{eqnarray}
On the other hand, the interference between $A_1$ and $A_2$ is large $N_c$ suppressed.

By adding the above two equations together, we find that
the interference terms will contribute to a factor of $N_c/2$ to the leading double logarithms.
Therefore, the total contribution will be
\begin{eqnarray}
|A_1+A_2+A_3|^2&=&\int d^2x_\perp d^2 y_\perp e^{iq_\perp(x_\perp-y_\perp)}|\tilde A_0|^2 \left(-\frac{\alpha_s}{2\pi}\frac{C_A+C_F}{2}\right) 
\ln^2\left(\frac{P_\perp^2(x_\perp-y_\perp)^2}{c_0^2}\right)\ ,
\end{eqnarray}
in the large $N_c$ limit. Therefore, the ${\cal C}$ coefficient for this process is
identified as ${\cal C}_{qg\to qg}=\left(C_F+C_A\right)/2$.

\subsection{$gg\to gg$ channel}

In the correlation limit and leading $N_c$, the Born diagram reads as,
\begin{eqnarray}
A_0&=&\int d^2x_\perp e^{iq_\perp\cdot x_\perp} \Gamma^\beta(k_{1\perp})\nonumber\\
&&\times \frac{1}{N_F}
{\rm Tr}\left[zT^a[U^\dagger(x_\perp)T^bU(x_\perp),\partial_\perp U^\dagger(x_\perp)T^cU(x_\perp)+
U^\dagger(x_\perp)T^c\partial_\perp U(x_\perp)]\right.\nonumber\\
&&\left.-(1-z)T^a[\partial_\perp U^\dagger(x_\perp)T^bU(x_\perp)+
U^\dagger(x_\perp)T^b\partial_\perp U(x_\perp),U^\dagger(x_\perp)T^cU(x_\perp)]\right]_{ij}\ ,
\end{eqnarray}
where $a,b,c$ are color indices for incoming, and outgoing gluon lines.
The amplitude squared will be,
\begin{eqnarray}
|A_0|^2&=&\int d^2x_\perp d^2y_\perp e^{iq_\perp\cdot (x_\perp-y_\perp)}
\Gamma^\beta(k_{1\perp}))\Gamma^{\beta'}(k_{1\perp}))\nonumber\\
&&\times \left[\left(z+(1-z)^2\right)F_{g}^{(1)}-2z(1-z)F_g^{(2)}+F_g^{(3)}\right]\ ,
\end{eqnarray}
Where $F_g^{(1,2,3)}$ represent the following Wilson lines,
\begin{eqnarray}
F_g^{(1)}&=&N_c{\rm Tr}[\partial_\perp U(x_\perp)\partial_\perp U^\dagger(y_\perp)]{\rm Tr}[U(x_\perp)U^\dagger(y_\perp)] \ ,\nonumber\\
F_g^{(2)}&=&N_c{\rm Tr}[\partial_\perp U(x_\perp) U^\dagger(y_\perp)]{\rm Tr}[U(x_\perp)\partial_\perp U^\dagger(y_\perp)] \ ,\nonumber\\
F_g^{(3)}&=&{\rm Tr}[\partial_\perp U(x_\perp)U^\dagger(x_\perp)\partial_\perp U(y_\perp)U^\dagger(y_\perp)]{\rm Tr}[U(x_\perp)U^\dagger(y_\perp)]{\rm Tr}[U(x_\perp)U^\dagger(y_\perp)] \ .
\end{eqnarray}
With these expressions, we will obtain the same differential cross sections as that 
in Ref.~\cite{Dominguez:2011wm}.

\begin{figure}[tbp]
\begin{center}
\includegraphics[width=9cm]{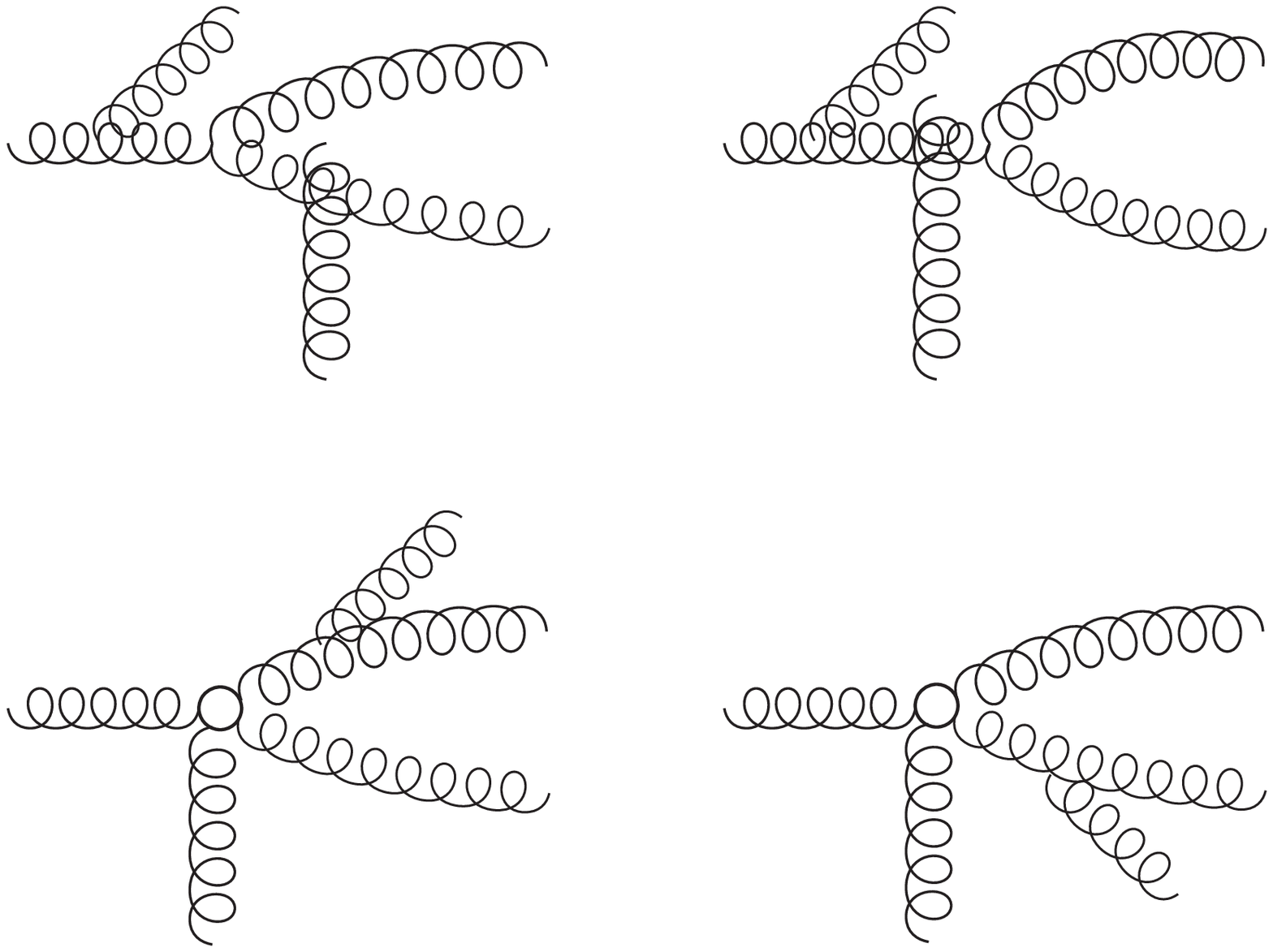}
\end{center}
\caption[*]{Same as Fig.~\ref{cgc-qq} for $gg\to gg$ process.}
\label{cgc-gg}
\end{figure}

The initial and final state radiation contributions of Fig.~\ref{cgc-gg} can be written as
\begin{eqnarray}
A_1&=&\frac{2(k_{g\perp}-k_{g2\perp})^\mu}{(k_{g\perp}-k_{g2\perp})^2}\Gamma^\beta(k_{1\perp})
\frac{1}{N_F}{\rm Tr}[T^dU(x_2)T^eU^\dagger(x_2)]\nonumber\\
&&\times \frac{1}{N_F}
{\rm Tr}\left[z[T^d,T^a][U^\dagger(x_\perp)T^bU(x_\perp),\partial_\perp U^\dagger(x_\perp)T^cU(x_\perp)+
U^\dagger(x_\perp)T^c\partial_\perp U(x_\perp)]\right.\nonumber\\
&&\left.-(1-z)[T^d,T^a][\partial_\perp U^\dagger(x_\perp)T^bU(x_\perp)+
U^\dagger(x_\perp)T^b\partial_\perp U(x_\perp),U^\dagger(x_\perp)T^cU(x_\perp)]\right]_{ij}\ ,\nonumber\\ 
A_2&=&\frac{2(k_{g\perp}-\xi_1k_{1\perp})^\mu}{(k_{g\perp}-\xi_1k_{1\perp})^2}\Gamma^\beta(k_{1\perp})\nonumber\\
&&\times \frac{1}{N_F}
{\rm Tr}\left[zT^a[U^\dagger(x_\perp)[T^d,T^b]U(x_\perp),\partial_\perp U^\dagger(x_\perp)T^cU(x_\perp)+
U^\dagger(x_\perp)T^c\partial_\perp U(x_\perp)]\right.\nonumber\\
&&\left.-(1-z)T^a[\partial_\perp U^\dagger(x_\perp)T^bU(x_\perp)+
U^\dagger(x_\perp)[T^d,T^b]\partial_\perp U(x_\perp),U^\dagger(x_\perp)T^cU(x_\perp)]\right]_{ij}\nonumber\ ,\\
A_3&=&\frac{2(k_{g\perp}-\xi_2k_{2\perp})^\mu}{(k_{g\perp}-\xi_2k_{2\perp})^2}\Gamma^\beta(k_{1\perp})\nonumber\\
&&\times \frac{1}{N_F}
{\rm Tr}\left[zT^a[U^\dagger(x_\perp)T^bU(x_\perp),\partial_\perp U^\dagger(x_\perp)[T^d,T^c]U(x_\perp)+
U^\dagger(x_\perp)T^c\partial_\perp U(x_\perp)]\right.\nonumber\\
&&\left.-(1-z)T^a[\partial_\perp U^\dagger(x_\perp)T^bU(x_\perp)+
U^\dagger(x_\perp)T^b\partial_\perp U(x_\perp),U^\dagger(x_\perp)[T^d,T^c]U(x_\perp)]\right]_{ij}\ ,
\end{eqnarray}
where $d$ is the color index for the radiated gluon.

The amplitude squared of the above three terms can be easily calculated,
\begin{equation}
|A_1|^2=C_A|A_0|^2,~~|A_2|^2=\frac{C_A}{2}|A_0|^2,~~|A_3|^2=\frac{C_A}{2}|A_0|^2 \ .
\end{equation}
Although it is tedious, we can calculate the interference between the above amplitude
straightforwardly. For the interference between $A_2$ and $A_3$, we have
\begin{equation}
2A_2A_3^*=-\frac{N_c}{2}\left[\left(z^2+(1-z)^2\right)F_g^{(1)}-2z(1-z)F_g^{(2)} \right] \ .
\end{equation}
The interference between $A_1$ and $A_2$ is,
\begin{equation}
2A_1A_2^*=-\frac{N_c}{2}\left[F_g^{(3)}+z^2F_g^{(1)}-z(1-z)F_g^{(2)}\right]\ .
\end{equation}
Similarly, for the interference between $A_1$ and $A_3$,
\begin{equation}
2A_1A_3^*=-\frac{N_c}{2}\left[F_g^{(3)}+(1-z)^2F_g^{(1)}-z(1-z)F_g^{(2)}\right]\ .
\end{equation}
Adding them together, we will find that the interference contributions is proportional
to a factor of $N_c$. Therefore, the total contributions 
take the following leading double logarithms,
\begin{eqnarray}
|A_1+A_2+A_3|^2&=&\int d^2x_\perp d^2 y_\perp e^{iq_\perp(x_\perp-y_\perp)}|\tilde A_0|^2 \left(-\frac{\alpha_s}{2\pi}N_c\right) 
\ln^2\left(\frac{P_\perp^2(x_\perp-y_\perp)^2}{c_0^2}\right)\ ,
\end{eqnarray}
in the large $N_c$ limit. Therefore, the ${\cal C}$ coefficient for this 
process is identified as ${\cal C}_{gg\to gg}=N_c$.

\section{Conclusion}
In summary, through the detailed one-loop calculation in coordinate space together with the heuristic discussion, we have computed the Sudakov factor, including both the double and single logarithmic terms, for the Higgs production in $pA$ collisions with the leading power approximation. The collinear divergence and the rapidity divergence, which appear in the one-loop calculation, have been absorbed into the corresponding DGLAP evolution of the collinear gluon distribution for the proton and the corresponding small-$x$ evolution equation for the WW gluon distribution for the target nucleus, respectively. We can also reproduce the QCD renormalization group equation for the running coupling $\alpha_s$ when we remove the $\textrm{UV}$ divergence by redefining the coupling constant. This demonstrates that one can perform the Sudakov resummation consistently in the small-$x$ formalism, when the small-$x$ type of large logarithms are also resumed.

Furthermore, we believe that this calculation can be generalized to the leading power contributions of the one-loop calculation for Higgs production in nucleus-nucleus collisions as long as the Higgs mass $M$ is much greater than its transverse momentum $k_\perp$. One should be able to obtain the same Sudakov double logarithmic term. Here in this case, one should use the WW unintegrated gluon distribution for both of the incoming gluons, and should also take into account the linearly polarized component of the WW gluon distribution\cite{Metz:2011wb}. In general, the genuine $k_t$ factorization does not apply unless the final state observed particles are color neutral which is the exact reason why the Higgs production process is peculiar as compared to other processes. With the help of the leading $\frac{k_\perp^2}{M^2}$ power approximation, we expect that the true $k_t$ factorization for Higgs production can be demonstrated to hold at least for the leading power contribution up to one-loop order. 

In addition, we have considered the processes for heavy quark pair productions and dijet productions in DIS, as well as various channels of dijet production processes in $pA$ collisions. Due to the complexity of these processes, we can only obtain the coefficient of the Sudakov double logarithmic term, which is the most important one-loop correction, by using the technique developed in the calculation of the Higgs production. We find that the Sudakov factor is process dependent, and the coefficient $\mathcal{C}$ can be determined from the color of the incoming partons. 

As derived in this manuscript, the Sudakov factor appears not only in Higgs productions in $pA$ collisions, but also in various back-to-back dijet processes and other interesting processes, such as heavy quark pair or heavy quarkonium productions and two-photon productions. The phenomenological application of this theoretical study will play an important role in the search of experimental evidences for saturation phenomena at RHIC and the LHC as well as future EIC. Last but not least, the azimuthal angle correlation of the dijet productions in heavy ion collisions\cite{Aad:2010bu} may also contain similar Sudakov suppression effect. There are many further relevant studies which should be considered and completed in the future. 

\begin{acknowledgments}
We thank J. W. Qiu for useful comments and discussions. This work was supported in part by the U.S. Department of Energy under the contracts DE-AC02-05CH11231. B.X. wishes to thank Dr. S. Munier, Dr. B. Pire and the CPHT at the Ecole Polytechnique for hospitality and support during his visit when this work is finalized. 
\end{acknowledgments}
\appendix
\section{Evaluation of Several Integrals in Dimensional Regularization}
In this part of the manuscript, we provide some more technique details for the evaluation of several integrals used in the derivation in dimensional regularization. The basics, conventions and usual tricks involved in dimensional regularization can be found in Refs.~\cite{Collins:1984xc, Collins:2011zzd}. 

To evaluate Eq.~(\ref{realsudakov}), one needs to first use the following identity \cite{Collins:2011zzd}:
\begin{eqnarray}
\int \frac{d^{2-2\epsilon} q_\perp}{(q_\perp^2)^\alpha} e^{-iq_\perp \cdot R_\perp} =\frac{\pi^{1-\epsilon}\Gamma (1-\alpha-\epsilon)}{\Gamma (\alpha)}\left(\frac{R_\perp^2}{4}\right)^{\alpha +\epsilon  -1}, \label{qtint}
\end{eqnarray}
which can be easily proved by rewriting $(q_\perp^2)^{-\alpha}$ as follows
\begin{equation}
(q_\perp^2)^{-\alpha} =\frac{1}{\Gamma (\alpha)}\int_0^\infty dx \, x^{\alpha -1}e^{-xq_\perp^2}.
\end{equation}
By setting $\alpha =1+a$, one can easily find
\begin{equation}
\mu^{2\epsilon}\int \frac{d^{2-2\epsilon} q_\perp}{(2\pi)^{2-2\epsilon}q_\perp^2}\left(\frac{M^2}{q_\perp^2}\right)^{a} e^{-iq_\perp \cdot R_\perp} =\frac{1}{(4\pi)^{1-\epsilon}}\left(\frac{R_\perp^2 \mu^2}{4}\right)^{\epsilon} \frac{\Gamma (-a-\epsilon)}{\Gamma (1+a)}\left(\frac{R_\perp^2M^2}{4}\right)^{a}.
\end{equation}
By differentiating over $a$ on both sides of the above equation and setting $a=0$ afterwards, one can obtain
\begin{eqnarray}
&&\mu^{2\epsilon}\int \frac{d^{2-2\epsilon} q_\perp}{(2\pi)^{2-2\epsilon}q_\perp^2}\ln\left(\frac{M^2}{q_\perp^2}\right) e^{-iq_\perp \cdot R_\perp} \notag \\
&=&\frac{1}{(4\pi)^{1-\epsilon}}\left(\frac{R_\perp^2 \mu^2}{4}\right)^{\epsilon} \left[\gamma_E+\ln \frac{R_\perp^2M^2}{4}-\psi^{(0)}(-\epsilon)\right]\Gamma (-\epsilon),
\end{eqnarray}
where $\psi^{(0)}(x)=\frac{\Gamma^\prime (x)}{\Gamma (x)}$ is the zeroth order polygamma function. In the convention of $\overline{\textrm{MS}}$ subtraction scheme, we should also multiply a factor of $S^{-1}_{\epsilon}=(4\pi e^{-\gamma_E})^{-\epsilon}$ to the right hand side of the above results to convert from $\textrm{MS}$ scheme to $\overline{\textrm{MS}}$ scheme. At last, by expanding $\epsilon$ around $0$, we can find the result in Eq.~(\ref{realc}). 

Using Eq.~(\ref{qtint}) and the convention in the $\overline{\textrm{MS}}$ subtraction scheme, one can easily compute the collinear singularity from the following integral
\begin{equation}
\mu^{2\epsilon}\int\frac{\text{d}^{2-2\epsilon}q_{\perp}}{(2\pi)^{2-2\epsilon}}e^{-iq_{\perp
}\cdot R_\perp}\frac{1}{q_{\perp}^{ 2}} =\frac{1}{4\pi}\left(-\frac{1}{\epsilon}+\ln\frac{c_0^2}{\mu^2 R_\perp^2}\right). \label{colm}
\end{equation}
To derive the convention and method for calculations in coordinate space for dimensional regularization, we can compute the corresponding integral of Eq.~(\ref{colm}) in the coordinate space. Through comparison between these two different calculation, we should be able to extract the connection of computations between the momentum space and the coordinate space in dimensional regularization. Using the following identity
\begin{equation}
\int \textrm{d}^2u_\perp \frac{ i u_\perp }{u_\perp^2} e^{-ik_\perp \cdot u_\perp} = 2\pi \frac{k_\perp}{k_\perp^2}, \label{utok}
\end{equation}
it is straightforward to find 
\begin{equation}
\int\frac{\text{d}^{2}q_{\perp}}{(2\pi)^{2}}e^{-iq_{\perp}\cdot R_\perp}\frac{1}{q_{\perp}^{ 2}} =\int\frac{\text{d}^{2}u_{\perp}}{(2\pi)^{2}}\frac{u_\perp \cdot \left(u_\perp +R_\perp\right)}{u_{\perp}^{ 2} \left(u_\perp +R_\perp\right)^2}.
\end{equation}
Thus, the corresponding integral in dimensional regularization gives 
\begin{equation}
\bar \mu^2 \int\frac{\text{d}^{2+2\epsilon}u_{\perp}}{(2\pi)^{2+2\epsilon}}\frac{u_\perp \cdot \left(u_\perp +R_\perp\right)}{u_{\perp}^{ 2} \left(u_\perp +R_\perp\right)^2} = \frac{1}{4\pi} \left(-\frac{1}{\epsilon}+\ln\frac{1}{\bar \mu^2 R_\perp^2}\right), \label{colc}
\end{equation} 
where we have multiplied a factor of $S_{\epsilon}=(4\pi e^{-\gamma_E})^{\epsilon}$ for the $\overline{\textrm{MS}}$ scheme. By comparing Eq.~(\ref{colc}) to Eq.~(\ref{colm}), we find that one can always convert the coordinate space results to the momentum space ones by setting $\bar\mu^2=\frac{\mu^2}{c_0^2}$ in dimensional regularization.

There are two ways to evaluate Eq.~(\ref{virtualsud}). In coordinate space, following the same convention summarized above, together with the identity
\begin{eqnarray}
a\int_0^\infty du \, u^{2\epsilon} \textrm{K}_1 (a u)&=& \frac{1}{2} \left(\frac{4}{a^2}\right)^{\epsilon} \Gamma (\epsilon) \Gamma (1+\epsilon) \\
a^2\int_0^\infty du \, u^{1+2\epsilon} \textrm{K}_0 (a u)&=& \left(\frac{4}{a^2}\right)^{\epsilon} \Gamma^2 (1+\epsilon),
\end{eqnarray}
it is straightforward to find that Eq.~(\ref{virtualsud}) gives
\begin{equation}
\frac{1}{2\pi}\left( \frac{4e^{-\gamma_E} \bar \mu^2}{-M^2}\right)^{\epsilon}\frac{\Gamma(\epsilon)\Gamma^2(-\epsilon)}{\Gamma(-2\epsilon)}=\frac{1}{\pi}\left(-\frac{1}{\epsilon^2}+\frac{1}{\epsilon}\ln \frac{M^2}{c_0^2\bar\mu^2}-\frac{1}{2}\ln^2\frac{M^2}{c_0^2\bar\mu^2} +\frac{\pi^2}{2}+\frac{\pi^2}{12}\right). \label{coov}
\end{equation}
It is interesting to notice that the contribution from $\textrm{K}_0 (\epsilon_f^\prime u_\perp)$ term is equal to $\epsilon$ times the contribution from $\textrm{K}_1 (\epsilon_f^\prime u_\perp)$ term, therefore cancels the $\epsilon$ dependence in the $\frac{2}{d}$ factor with $d=2(1+\epsilon)$. If we change $d$ back to $2$, then we only need to evaluate the $\textrm{K}_1 (\epsilon_f^\prime u_\perp)$ part. 

To evaluate Eq.~(\ref{virtualsud}) in the momentum space, we use the following identities
\begin{eqnarray}
\frac{1}{2\pi} K_0(au_\perp)& =&\int \frac{ \textrm{d}^2k_\perp}{(2\pi)^2}\frac{e^{-i k_\perp \cdot u_\perp}}{k_\perp^2 +a^2} \\
\frac{1}{2\pi} au_\perp K_1(au_\perp)& =&\int\textrm{d}^2k_\perp \frac{ ik_\perp \cdot u_\perp}{(2\pi)^2}\frac{e^{-i k_\perp \cdot u_\perp}}{k_\perp^2 +a^2},
\end{eqnarray}
together with Eq.~(\ref{utok})
to cast Eq.~(\ref{virtualsud}) into 
\begin{equation}
\frac{1}{1+\epsilon}\int_{0}^1 \frac{\text{d}\xi}{\xi (1-\xi)}\int\frac{\text{d}^{2-2\epsilon}k_{\perp}}{(2\pi)^{2-2\epsilon}} \left[-\frac{2\epsilon _{f}^{\prime 2}}{\left(k_\perp^2+\epsilon _{f}^{\prime 2}\right)k_\perp^2}+(2\pi)\delta^{(2)}(k_\perp)\frac{\epsilon _{f}^{\prime 2}}{k_\perp^2+\epsilon _{f}^{\prime 2}}\right], 
\end{equation}
where we have changed the dimension of the $k_\perp$ integration to $2-2\epsilon$. Using the $\overline{\textrm{MS}}$ conventional factors $S_\epsilon^{-1}$, the above momentum space integral yields 
\begin{equation}
\frac{1}{\pi} \left(-\frac{1}{\epsilon^2}+\frac{1}{\epsilon}\ln \frac{M^2}{\mu^2}-\frac{1}{2}\ln^2\frac{M^2}{\mu^2} +\frac{\pi^2}{2}+\frac{\pi^2}{12}\right). \label{momv}
\end{equation}
Again, by identifying $c_0^2\bar \mu^2$ with $\mu^2$, Eq.~(\ref{coov}) and Eq.~(\ref{momv}) give the same results.

\end{document}